%% file: main.tex
\documentclass[3p,times,preprint]{elsarticle}

\usepackage{MyArxiv}

\volume{00}

\firstpage{1}

\journalname{Computer Methods in Applied Mechanics and Engineering}

\runauth{C.F. Christensen et al.}

\jid{CMAME}

\jnltitlelogo{\normalsize{Computer Methods in Applied Mechanics and Engineering}}

\usepackage{amssymb}
\usepackage{amsmath}

\usepackage{lineno}

\usepackage[figuresright]{rotating}

\usepackage{subcaption}

\usepackage{multirow}

\usepackage{import}

\usepackage{xcolor}

\usepackage{siunitx}

\usepackage{tikz}

\begin{document}

\begin{frontmatter}



\dochead{}

\title{Topology Optimization of Multiscale Structures Considering Local and Global Buckling Response}



\author{Christoffer Fyllgraf Christensen\corref{cor1}}
\ead{chrify@mek.dtu.dk}
\author{Fengwen Wang}
\author{Ole Sigmund}

\address{Department of Civil and Mechanical Engineering, Technical University of Denmark, Nils Koppels Allé, Building 404, 2800 Kgs. Lyngby, Denmark}

\cortext[cor1]{Corresponding author.}

\begin{abstract}
Much work has been done in {topology optimization of multiscale structures} for maximum stiffness or minimum compliance design. Such approaches date back to the original homogenization-based work by Bendsøe and Kikuchi from 1988, which lately has been revived due to advances in manufacturing methods like additive manufacturing. Orthotropic microstructures locally oriented in principal stress directions provide for highly efficient stiffness optimal designs, whereas for the pure stiffness objective, porous isotropic microstructures are sub-optimal and hence not useful. It has, however, been postulated and exemplified that isotropic microstructures (infill) may enhance structural buckling stability but this has yet to be directly proven and optimized. In this work, we optimize buckling stability of multiscale structures with isotropic porous infill. To do this, we establish local density dependent Willam-Warnke yield surfaces based on {local} buckling estimates from Bloch-Floquet-based cell analysis to predict local instability of the homogenized materials. These local buckling-based stress constraints are combined with a global buckling criterion to obtain topology optimized designs that take both local and global buckling stability into account. De-homogenized structures with small and large cell sizes confirm validity of the approach and demonstrate huge structural gains as well as time savings compared to standard singlescale approaches.
\end{abstract}


\end{frontmatter}


\section{Introduction}
\label{sec:introduction}
\input{Introduction.tex}

\section{Buckling Optimization of Multiscale Structures}
\label{sec:method}
\input{Method.tex}

\section{Numerical Results}
\label{sec:results}
\input{Results.tex}

\section{Conclusion}
\label{sec:conclusion}
\input{Conclusion.tex}

\section*{Acknowledgements}{
The authors acknowledge funding from the Villum Fonden through the Villum Investigator Project “InnoTop”.}

\appendix

\section{Constraint Definition}
\label{app:Constraint}
\input{APP_Constraint}

\section{Sensitivity Analysis}
\label{app:Sensitivity}
\input{APP_Sensitivity}



\bibliographystyle{elsarticle-num}
\bibliography{main.bib}







\end{document}

%% file: Introduction.tex
In recent years advances in additive manufacturing have facilitated the fabrication of multiscale or infill structures \cite{Zheng2016}, which in turn has further promoted topology optimization {of multiscale structures}. The fact that structures with tailored microstructures can be manufactured means that multiscale structures are no longer restricted to theoretical research but can actually be utilized in real designs. A lot of research considering  multiscale structural optimization following the original work of Bendsøe and Kikuchi \cite{BENDSOE1988197} has been done in recent years as reviewed by Wu, Sigmund and Groen \cite{Wu2021}. Work in \cite{pantz2008a} showed that multiscale designs can be projected to singlescale using an implicit geometry description. This work was elaborated in \cite{Groen2018, Allaire2019} for 2D problems and \cite{Groen2020,Groen2021} for 3D. 

Previous multiscale structural optimization has mainly focused on compliance minimization without considering structural stability. Recently, singlescale studies of buckling optimization has become more popular. Work in \cite{Ferrari2020a} focused on the the use of linear buckling with the scope of solving large scale topology optimization. An extension of this is the work in \cite{Russ2021}, where buckling resistance and local ductile failure constraints are combined. Furthermore, the work in \cite{Mendes2022} considers singlescale buckling optimization, using the Topology Optimization of Binary Structures (TOBS) method, subject to design dependent pressure loads. This method is limited to singlescale optimization due to the binary nature of the TOBS method. Recent work in \cite{ISHIDA2022} focused on singlescale buckling optimization using the level set method. Work on multiscale stability optimization is only sparsely covered in literature. Recent work in \cite{Wang2021} focuses on using filleted lattice structures considering a simplified local buckling formulation and yield stress constraints for compliance optimization. The method lacks the opportunity to generate true void and solid due to the local stress constraints and strut radii bounds used in the microstructure. As a result microstructures appear in the entire design domain even though global buckling is not considered and multiscale isotropic material is sub-optimal for compliance optimized designs~\cite{Sigmund2016}.

Earlier work \cite{Clausen2016} has shown that isotropic infill may increase structural buckling resistance at the price of a small stiffness reduction even with a predefined constant infill density. To fully explore the potential buckling resistance, this study suggests to freely optimize the isotropic microstructure infill densities in the entire design domain via topology optimization. Previous studies have shown that microstructure buckling strength can be evaluated using Bloch-Floquet wave theory \cite{Triantafyllidis1998,thomsen2018a}. Based on buckling stability analysis of a chosen microstructure evaluated at different densities it is possible to formulate interpolation functions for local buckling stress constraints. Combining this with a global buckling optimization provides a way of performing buckling optimization {of multiscale structures}.  In this work, the global buckling load is estimated using the linearized buckling analysis framework presented in \cite{ferrari2021a}, as it increases computational efficiency, while producing sufficiently accurate buckling estimates as discussed in \cite{Ferrari2020a,ferrari2019a}. {Furthermore, the robust formulation by \cite{Wang2011} and two-field formulation by \cite{Giele2021} are used to allow control over the minimum density, thus allowing true void in the design.}

\begin{figure}
     \centering
     \begin{tabular}{cc}
     \begin{subfigure}[b]{0.3\textwidth}
         \centering
         \includegraphics[width=4.77cm]{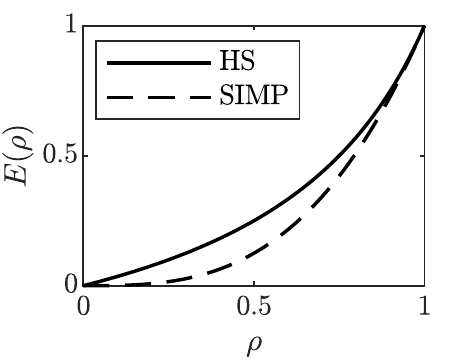}
         \caption{}
         \label{fig:SIMPvsHS:Intro:1}
     \end{subfigure}
     &
     \begin{subfigure}[b]{0.4\textwidth}
     	\centering
		  \scalebox{1}{\includegraphics[scale=1]{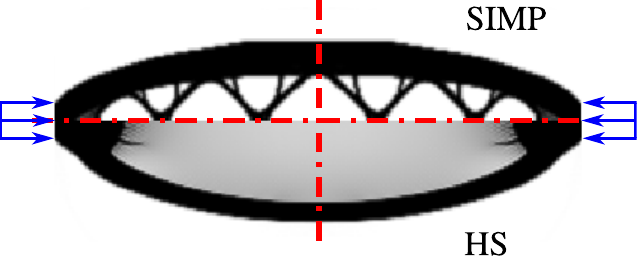}}
		\vspace{0.72cm}         
         \caption{}
         \label{fig:SIMPvsHS:Intro:2}
     \end{subfigure}
     \\
     \multicolumn{2}{c}{
     \begin{subfigure}[b]{0.73\textwidth}
         \centering
		\vspace{-0.4cm}
         \scalebox{1}{\includegraphics[scale=1]{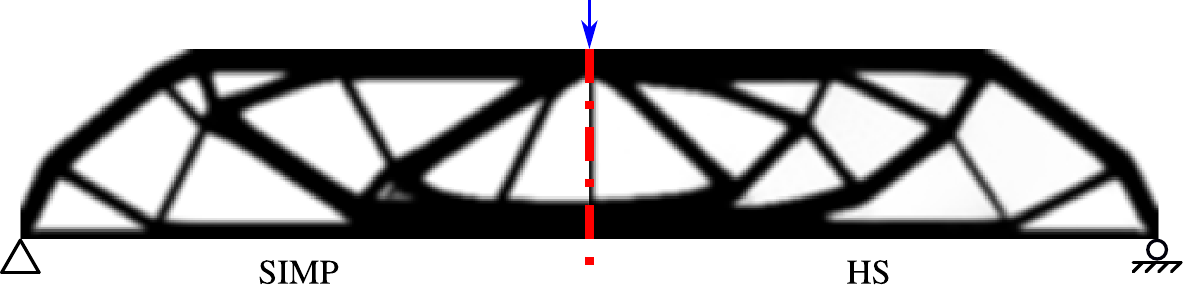}}
         \caption{}
         \label{fig:SIMPvsHS:Intro:3}
     \end{subfigure}
     }
     \end{tabular}
     \vspace{-0.3cm}
        \caption{Comparison of HS and SIMP stiffness interpolations: (\subref{fig:SIMPvsHS:Intro:1}) Stiffness interpolation with $p=3$ in SIMP, (\subref{fig:SIMPvsHS:Intro:2}) Buckling load maximization where the top half uses SIMP ($BLF=36.26$) and the bottom half uses HS ($BLF=41.56$), (\subref{fig:SIMPvsHS:Intro:3}) Compliance minimization where the left half uses SIMP ($C=13.95  \cdot 10^{-4}$) and the right half uses HS ($C=13.29 \cdot 10^{-4}$).}
        \label{fig:SIMPvsHS:Intro}
\end{figure}

{We consider local and global buckling topology optimization of multiscale structures composed of a fixed isotropic triangular microstructure with spatially varying density. The stiffness of the isotropic triangular microstructure is very close to the theoretical maximum provided by the Hashin-Shtrikman (HS) upper bound \cite{Hashin1963} as found in e.g. \cite{Traff2019}. Thus, instead of performing a formal homogenization approach to compute the effective properties of the triangular microstructure, we can simply use the bulk and shear modulus provided by the HS upper bound to model its elastic properties \cite{Sigmund2016,Bendsoe1999,Sigmund2000}. For pure single-load compliance minimization problems, the use of the isotropic (near-optimal) microstructure is suboptimal and results in 0--1, non-composite solutions, c.f.~\cite{Sigmund2016}. For such problems, orthotropic rank-2 or rectangular hole microstructures are optimal, however, such microstructures have low shear stiffness and thus perform badly in terms of buckling stability \cite{Bluhm2020}. For this reason, we here opt for the stiffness sub-optimal but buckling-superior triangular microstructure. An added benefit of the stiffness isotropic triangular microstructure is that its buckling response is rather isotropic as well, thus making the establishing of a buckling ``yield surface'' considerably simpler.

A comparison of the stiffness interpolations using HS and standard SIMP (p=3) interpolations is visualized in Figure~\ref{fig:SIMPvsHS:Intro:1}.}
{The difference between the two interpolation schemes is significant for lower densities and explains the advantage of using microstructures when considering buckling. Using the two schemes for pure compliance minimization confirms, contrary to optimally oriented rank-2 laminates or rectangular hole cells, that there is no advantage in using isotropic infill for compliance (C), see Figure~\ref{fig:SIMPvsHS:Intro:3}, c.f.~\cite{Sigmund2016}. The small difference in compliances is simply a result of the different stiffness interpolations acting in the intermediate density regions that occur due to the density filter. However, for buckling load maximization large differences are visible in both design and performance as measured by the Buckling Load Factor (BLF). The design obtained using SIMP (seen in the upper half of Figure~\ref{fig:SIMPvsHS:Intro:2}) is black and white as intermediate densities are penalized. The multiscale design, based on the HS interpolation (seen in the lower part of Figure~\ref{fig:SIMPvsHS:Intro:2}), shows a large region of intermediate densities. The explanation lies in the higher stiffness for intermediate densities provided by the HS interpolation compared to SIMP as seen in Figure~\ref{fig:SIMPvsHS:Intro:1}. Comparing the performance of the two designs shows a 14.62\% improvement in global buckling stability for the multiscale design. This indicates that isotropic microstructures are superior for buckling objectives. Of course this is only true if the structure does not reach local buckling of the infill before global buckling is reached. Therefore, local buckling stability must be taken into account when optimizing against buckling using topology optimization of multiscale structures.}

\begin{figure}
   \centering
   \includegraphics[scale=1]{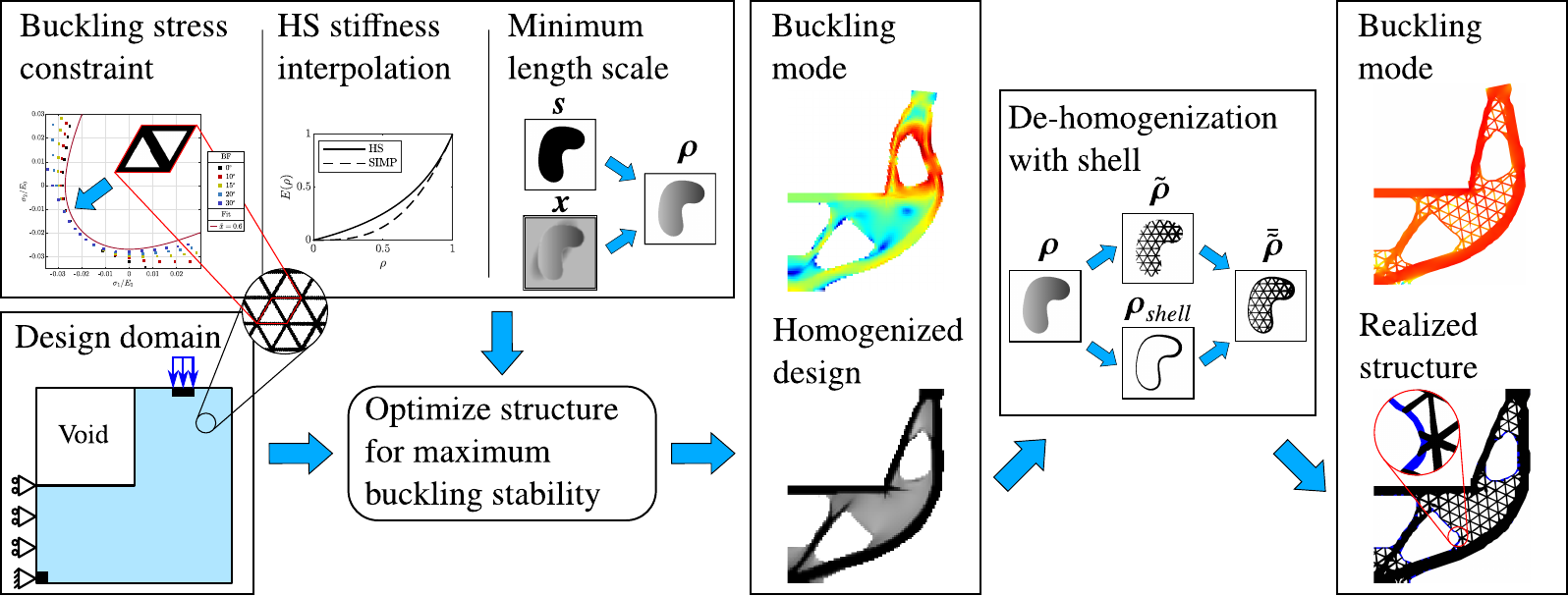}
   \caption{Flowchart illustrating the work flow of the method presented in this paper. The upper left box shows the main building blocks to set up the homogenized optimization problem. The lower left box shows an example of a design domain. The center column show a optimized homogenized design and its critical buckling mode. The second box from the right shows the de-homogenization method. Finally, the de-homogenized structure with its critical buckling mode is shown in the right most box.}
   \label{fig:FlowChart}
\end{figure}

{This paper will present the work flow for performing topology optimization of multiscale structures while preventing buckling on both local and  global scale. This is done according to the work flow illustrated in Figure~\ref{fig:FlowChart}. The} paper is organized as follows: In Section~\ref{sec:method} the optimization problem {for multiscale structures} is presented. This includes a multifield method, interpolations and buckling stress constraint. Section~\ref{sec:results} presents the results of two numerical examples demonstrating the advantage of the method. Finally, Section~\ref{sec:conclusion} concludes the work presented in this paper.

%% file: Method.tex
This study aims to optimize global structural stability while preventing local microstructure buckling via buckling stress constraints using a robust linearized buckling framework. The optimization problem using the robust formulation \cite{Wang2011} and a two-field formulation that allows control over the minimum density \cite{Giele2021} is written as
\begin{equation} \label{eq:optProblem}
\begin{array}{rll}
\min\limits_{\mathbf{x},\mathbf{s}}  &: \quad \max\limits_{m}\left(g_{\lambda}(\boldsymbol{\rho}^m) = {\dfrac{J^{KS}\left(\gamma_i(\boldsymbol{\rho}^m)\right)}{J^{KS}_0}}\right),  &m\in\{e,b,d\}, \ \gamma_i\in\mathcal{B}  \\  
\textrm{s.t.} &: \quad g_c(\boldsymbol{\rho}^e) = \dfrac{\boldsymbol{f}^T \boldsymbol{u}(\boldsymbol{\rho}^e)}{C_e^*}-1 \leq 0 , &\\
	&: \quad {g_s(\boldsymbol{\rho}^m,\tilde{\boldsymbol{x}},\gamma_i(\boldsymbol{\rho}^m)) 
	\leq 0} , &m\in\{e,b,d\} , \ \gamma_i\in\mathcal{B}\\
	&: \quad g_V(\boldsymbol{\rho}^d) = \dfrac{\sum_j v_j \rho_j^d}{V_d^* V_{\Omega}} - 1 \leq 0, & \\
    &: \quad \rho_j^m = \tilde{x}_j \bar{\tilde{s}}_j^m  , &m\in\{e,b,d\}  \\
    &: \quad  x_{min} \leq x_j \leq1, & \forall j    \\
    &: \quad 0\leq s_j \leq1.  &\forall j  \\
\end{array}    
\end{equation}
The meaning of each term in \eqref{eq:optProblem} will be explained over the next several pages and subsections. {First, $g_{\lambda}(\boldsymbol{\rho}^m)$ represents the normalized aggregated inverse global stability objectives for $m\in\{e,b,d\}$ being the \textit{eroded}, \textit{blue print} and  \textit{dilated} designs. The normalization factor $J^{KS}_0$ is simply the value of $J^{KS}\left(\gamma_i(\boldsymbol{\rho}^m)\right)$ for the initial design.} {$g_{c}$ and $g_{V}$ represent structural compliance and volume constraints, respectively. $g_{s}$ is the local buckling stress constraint, which is dependent on the global buckling load. The exact definition of $g_{s}$ is presented in the following subsections.} $v_j$ and $V_{\Omega}$ are the elemental and total structural volumes.  The physical design field $\boldsymbol{\rho}^m$ is constructed using a filtered density field  $\tilde{\boldsymbol{x}}$ and a void indicator field $ \bar{\tilde{\boldsymbol{s}}}$ following the method by \cite{Giele2021}. This provides control over the minimum density while allowing void regions in the design. The detailed design procedure will be presented in the following subsection.

The compliance measure $\boldsymbol{f}^T \boldsymbol{u}$ depends on the external load $\boldsymbol{f}$ and the displacements $\boldsymbol{u}(\boldsymbol{\rho}^m)$ for the design, $m$. The displacements are calculated by solving
\begin{equation} \label{eq:compliance}
	\boldsymbol{K}(\boldsymbol{\rho}^m) \boldsymbol{u} = \boldsymbol{f},
\end{equation}
where $\boldsymbol{K}(\boldsymbol{\rho}^m)$ is the linear symmetric global stiffness matrix for the considered design $\boldsymbol{\rho}^m$. The constant $C^*_e$ is the compliance upper limit for the \textit{eroded} design.  The maximum allowed volume fraction is enforced through the \textit{dilated} design using $V_d^*$. The value of $V_d^*$ is continuously updated to ensure that the target volume $V_b^*$ is enforced on the \textit{blue print} design. { This strategy is used to eliminate numerical artifacts and stabilize convergence of the optimization problem \cite{Wang2011}.}

The critical buckling load is approximated using linearized buckling analysis (see \cite{cook2002a}) based on the equilibrum in \eqref{eq:compliance}, defined as
\begin{equation}\label{eq:ModifiedLinearBucklingEquation}
	\left(\boldsymbol{G}_\sigma(\boldsymbol{\rho}^m,\boldsymbol{u}) - \gamma_i \boldsymbol{K}(\boldsymbol{\rho}^m) \right) \boldsymbol{\varphi}_i = 0. 
\end{equation}
{where the eigenpairs $(\gamma_i, \ \boldsymbol{\varphi}_i)$ represent the eigenvalue and buckling mode vector, respectively. The actual Buckling Load Factors (BLF) $\lambda_i$ are recovered through the eigenvalues $\gamma_i$ as $\lambda_i = -1/ \gamma_i$, ordered such that $\lambda_i \leq \lambda_{i+1}$.} The global stress stiffness matrix $\boldsymbol{G}_\sigma(\boldsymbol{\rho}^m,\boldsymbol{u})$ is built considering the design $\boldsymbol{\rho}^m$ and the current stress state resulting from the displacements $\boldsymbol{u}(\boldsymbol{\rho}^m)$. A Matlab implementation of \eqref{eq:ModifiedLinearBucklingEquation} is presented in \cite{ferrari2021a}.

There exist as many eigenpairs $(\gamma_i, \ \boldsymbol{\varphi}_i), \ i \in \mathcal{B}_0$ as there are DOFs in the system. Only a subset of eigenpairs $\mathcal{B} \subset \mathcal{B}_0$ is included in the analysis. This subset should be large enough to  capture all relevant buckling modes but not so large that the problem becomes computationally infeasible~(\cite{Bruyneel2008,Dunning2016}).  The critical buckling load is determined by the fundamental BLF $\lambda_1$  through $\boldsymbol{f}_{cr} = \lambda_1\boldsymbol{f}$.

The Kreisselmeier–Steinhauser (KS) function~\cite{Kreisselmeier1979} approximates the minimum structural buckling factor by aggregating the considered eigenvalues into one differentiable objective for each design $m$ in the robust formulation. { The version of the KS function used here differs from the one in \cite{Kreisselmeier1979} by normalizing the eigenvalues. This provides better scaling which improves stability of the optimization problem. The definition of the modified KS function is}
\begin{equation} \label{eq:KSFunction}
	{J^{KS}\left(\gamma_i(\boldsymbol{\rho}^m)\right)_{i\in\mathcal{B}} = \gamma_0
	+ \dfrac{\gamma_0}{P} \ln \left(
	\sum\limits_{i\in \mathcal{B}}
	e ^ {P  \left(
		\dfrac{\gamma_i(\boldsymbol{\rho}^m)}{\gamma_0}-1
		\right)}
	\right),}
\end{equation}
where $P$ is the aggregation parameter {and $\gamma_0$ is equal to the fundamental eigenvalue $\gamma_1$ from the initial design}. Here \eqref{eq:KSFunction} provides a smooth upper bound approximation of $|\gamma_1| =  \max\limits_{i \in \mathcal{B}} | \gamma_i|$. The result is a smooth lower bound for $\lambda_1$.

\subsection{Multifield Method and Material Interpolation}
\label{sec:Multi_Material_Formulation}
The multifield method by \cite{Giele2021} is used in this work to generate the physical design field $\boldsymbol{\rho}^m$. It provides control over the minimum density of porous regions while also allowing fully void regions in the design. This is essential as both the buckling resistance and buckling stress limit goes to zero for $\rho \rightarrow 0$. If nothing is done to deal with this issue the optimization will always require some amount of material in every elements to comply with the constraint. This essentially means that the method will not allow topology changes and is only useful for infill optimization of already known geometries. However, using the multifield method by \cite{Giele2021} solves this issue as the stress limits are interpolated using the density field $\tilde{\boldsymbol{x}}$ while the stiffness interpolation is performed on the physical design field $\boldsymbol{\rho}^m$.

\begin{figure}
	\centering
	\includegraphics[scale=1]{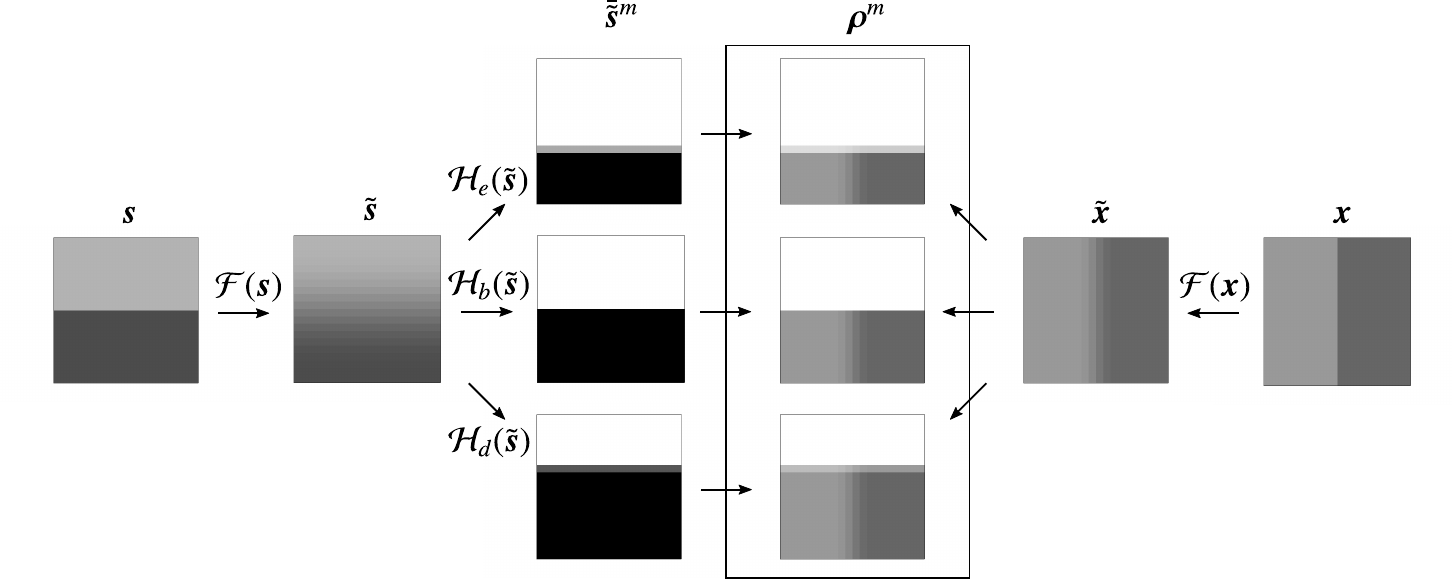}
	\caption{Illustration of how the physical design field $\boldsymbol{\rho}^m$ is constructed from the density field $\boldsymbol{x}$ and the void indicator field $\boldsymbol{s}$.}
	\label{fig:MultifieldIllustration}
\end{figure}

The construction procedure for the physical design field $\boldsymbol{\rho}^m$ is illustrated in Figure~\ref{fig:MultifieldIllustration}. 
It shows how the indicator field, $\boldsymbol{s}$, is filtered through $\mathcal{F}(\boldsymbol{s})$ then projected using a Heaviside function $\mathcal{H}_m(\tilde{\boldsymbol{s}})$ at three different threshold levels. The density field is only filtered through $\mathcal{F}(\boldsymbol{x})$. The product of these two fields form the physical design field $\boldsymbol{\rho}^m$.

{The purpose of the density filter is to eliminate numerical artifacts such as checker boarding and make the design mesh-independent \cite{Bourdin2001}.} 
The definition of the density filter is \cite{Bourdin2001,Bruns2001}
\begin{equation}
	\tilde{y}_j = \dfrac{\sum_{i\in N_{j,i}} w(\boldsymbol{n}_i) y_i}{\sum_{i\in N_{j,i}} w(\boldsymbol{n}_i)} \Rightarrow \tilde{\boldsymbol{y}} = \mathcal{F} (\boldsymbol{y})
	, \quad \boldsymbol{y} \in \left\lbrace\boldsymbol{x},\boldsymbol{s}\right\rbrace,
\end{equation}
where $N_{j,i}$ is the set of neighborhood elements within the filter radii $R_s$ and $R_x$ for the indicator field $\boldsymbol{s}$ and density field $\boldsymbol{x}$ respectively. In this work $R_x = 1.5\Delta x$ and $R_s = 4.5\Delta x$ is used throughout. The densities are weighted through $w(\boldsymbol{n}_i)$, where $\boldsymbol{n}_i$ are the element center points, i.e.
\begin{equation}
	w(\boldsymbol{n}_i) = \max \left( 0, R_{s/x} - |\boldsymbol{n}_i - \boldsymbol{n}_j |\right).
\end{equation}

The {smoothed} Heaviside function $\mathcal{H}_m(\tilde{\boldsymbol{s}})$  (see \cite{Wang2011,Lazarov2016}) is defined as
\begin{equation}\label{eq:Heaviside}
	\bar{\tilde{\boldsymbol{s}}} = \mathcal{H}_m(\tilde{\boldsymbol{s}}) = 
	\dfrac{\tanh(\beta \eta_m) + \tanh \left(\beta (\tilde{\boldsymbol{s}} - \eta_m) \right) }
	{\tanh(\beta \eta_m) + \tanh \left(\beta (\boldsymbol{1} - \eta_m) \right)}
	, \quad m \in \{e,b,d\},
\end{equation}
where $\eta_m$ is the threshold value. In this work $\eta_b = 0.5$ and the \textit{eroded} and \textit{dilated} values are determined by $\eta_b \pm \Delta \eta${, where $\Delta \eta$ is specified explicitly for each of the numerical examples.} The steepness of the projection is determined by $\beta$. When $\beta$ increases the Heaviside function gets more and more non-linear. Therefore, a continuation method is used on $\beta$ implying that it is initialized as $\beta = 2$ and slowly increased throughout the optimization process to $\beta_{max} = 256$ to ensure a clear representation of void regions. {The specific continuation scheme used in this work is stated in Section~\ref{sec:results}.}

{As discussed in the introduction, the stiffness of the considered triangular microstructure is very close to the theoretical optimum provided by the HS bounds. Therefore, instead of performing numerical homogenization to compute its properties, we may with just small errors use the stiffness propeties provided by the bounds. Hence,} the HS upper bound is employed for elemental stiffness interpolation \cite{Hashin1963}, i.e.,
\begin{equation}\label{eq:stiffnessInterpolation}
	\begin{split}
			\mu ({\rho}_j^m)=&
		\rho_j^m \mu_0 +(1-\rho_j^m) \mu_{\min} -
		\dfrac{\rho_j^m (1-\rho_j^m)(\mu_0-\mu_{\min})^2}{(1-\rho_j^m)\mu_0 + \rho_j^m \mu_{\min} + \dfrac{\kappa_0 \mu_0}{\kappa_0+2 \mu_0}}, \\
		\kappa ({\rho}_j^m) =&
		\rho_j^m \kappa_0 +(1-\rho_j^m) \kappa_{\min} -
		\dfrac{\rho_j^m (1-\rho_j^m)(\kappa_0-\kappa_{\min})^2}{(1-\rho_j^m)\kappa_0 + \rho_j^m \kappa_{\min} + \mu_0},
		\end{split}
\end{equation}
where $\mu ({\rho}_j^m)$ and $\kappa({\rho}_j^m)$ are the effective shear and bulk moduli, respectively. The subscripts $0$ and $\min$ refer to the base material and void. The shear and bulk modulus for the base material are defined as $\mu_0 = E_0/(2(1+\nu_0))$ and $\kappa_0 = E_0/(2(1-\nu_0))$ and the stiffness of the void is determined similarly using $E_{min} = 10^{-6} E_0$. The interpolation functions are used to build both the linear stiffness and stress stiffness matrices as
\begin{equation}
\begin{split}
		\boldsymbol{K}^j({\rho}_j^m) = &
		\mu ({\rho}_j^m)
		\boldsymbol{K}^j_{0,\mu}+
		\kappa( {\rho}_j^m)
		\boldsymbol{K}^j_{0,\kappa},\\
		\boldsymbol{G}^j_\sigma({\rho}_j^m,\boldsymbol{u}_j) =& \mu ({\rho}_j^m) \boldsymbol{G}^j_{0,\mu}(\boldsymbol{u}_j) + \kappa({\rho}_j^m) \boldsymbol{G}^j_{0,\kappa}(\boldsymbol{u}_j).
	\end{split}
\end{equation}
Here $\boldsymbol{K}^j_{0,\mu}$ and $\boldsymbol{K}^j_{0,\kappa}$ are the finite elemental shear and bulk stiffness matrices for element $j$ with unit shear and bulk moduli. $\boldsymbol{G}^j_{0,\mu}(\boldsymbol{u}_j)$ and $\boldsymbol{G}^j_{0,\kappa}(\boldsymbol{u}_j)$ are the finite elemental base stress stiffness matrices for unit shear and bulk modulus respectively. { The definition of these matrices are based on  the choice of finite element, hence they are not defined here for generality. For the 4-node quadrilateral elements used here we refer to the public Matlab code provided by \cite{ferrari2021a}}. To avoid artificial buckling modes in the low stiffness regions, the stress stiffness matrix interpolation uses $\mu_{\min} = \kappa_{\min} = 0$ (see \cite{thomsen2018a,ferrari2019a,Gao2015}).

\subsection{Buckling Stress Constraint}
\label{sec:bucklingStressConstraint}
We employ a Willam-Warnke failure criterion to characterize the microstructure buckling failure under macroscopic stress situations in each structural element. The Willam-Warnke failure criterion, known from failure prediction of concrete, usually has a relatively small limit in tension but allows larger stresses in compression \cite{Giraldo-Londono2020}. In the case considered here, where buckling of microstructure should be prevented, the purpose of the Willam-Warnke failure surface is flipped to make it more suitable to characterize the buckling failure  with limited compression stresses but allowing tension. The failure surface is determined by the stress limits in uniaxial tension $\sigma_t$, uniaxial compression $\sigma_c$ and equibiaxial compression $\sigma_b$. 

\subsubsection{Microstructure Buckling Stress Limits}\label{sec:stressLimits}
\begin{figure}
	\centering
	\begin{subfigure}[b]{0.25\textwidth}
		\centering
		\includegraphics[scale=1]{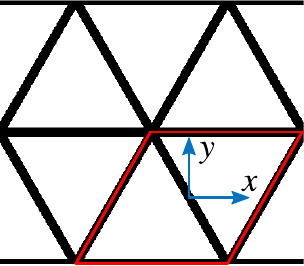}
		\caption{$\tilde{x} = 0.2$}
		\label{fig:triangularMicrostructure:1}
	\end{subfigure}
	\hspace{0.75cm}
	\begin{subfigure}[b]{0.25\textwidth}
		\centering
		\includegraphics[scale=1,angle =90]{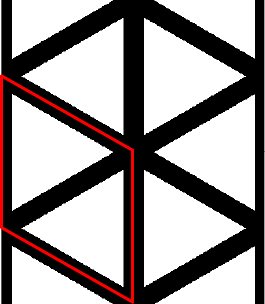}
		\caption{$\tilde{x} = 0.4$}
		\label{fig:triangularMicrostructure:2}
	\end{subfigure}
	\hspace{0.75cm}
	\begin{subfigure}[b]{0.25\textwidth}
		\centering
		\includegraphics[scale=1,angle =90]{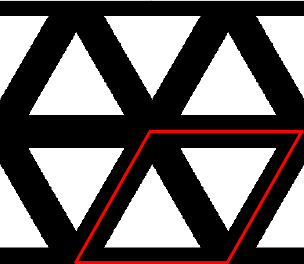}
		\caption{$\tilde{x} = 0.6$}
		\label{fig:triangularMicrostructure:3}
	\end{subfigure}
	\\
	\begin{subfigure}[]{0.55\textwidth}
		\centering
		\includegraphics[scale=1,trim={3.5cm 0 3.5cm 0},clip]{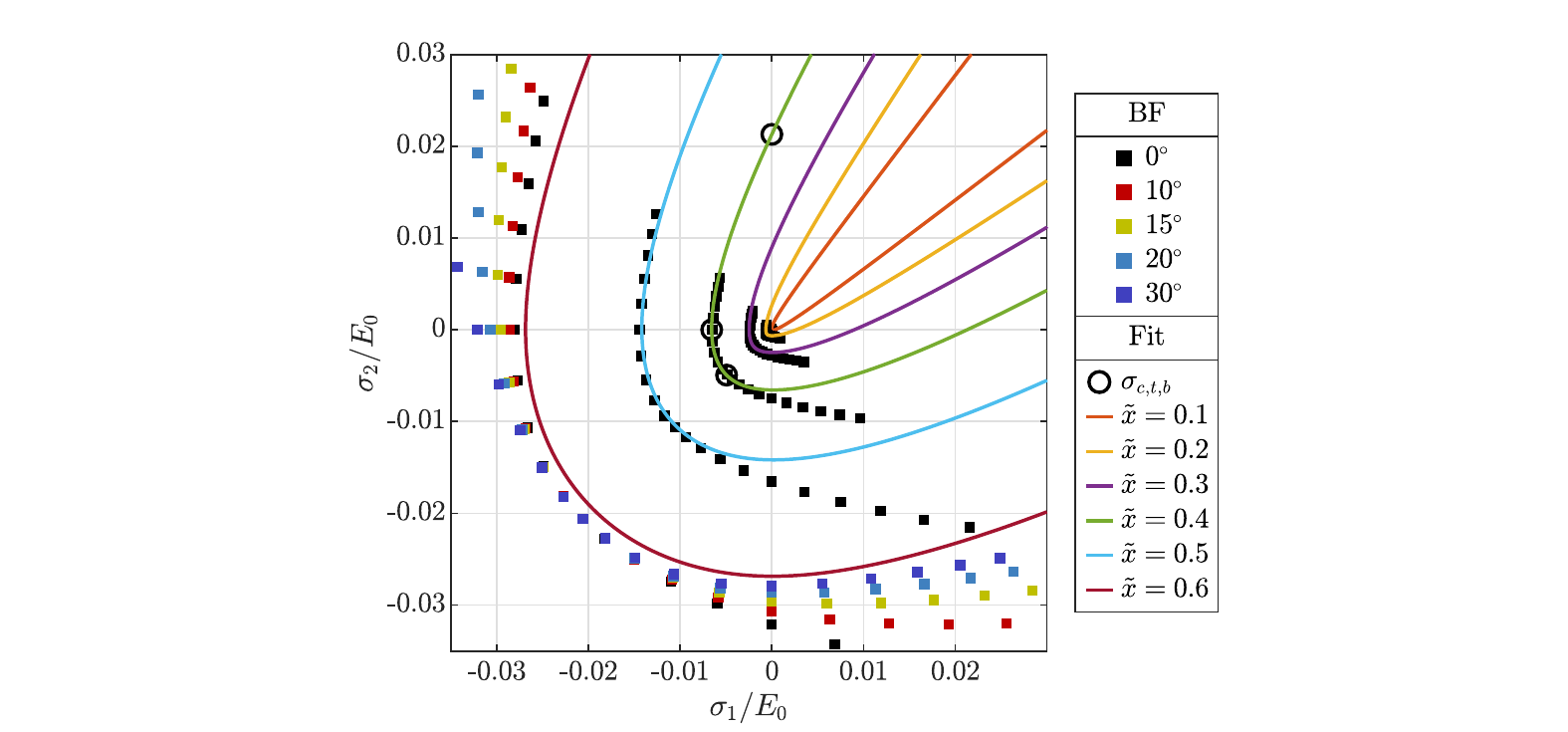}
		\caption{}
		\label{fig:FittedCurves:2}
	\end{subfigure}
	\hfill
	\begin{subfigure}[]{0.44\textwidth}
		\centering
		\includegraphics[scale=1,trim={0.1cm 0 0.6cm 0},clip]{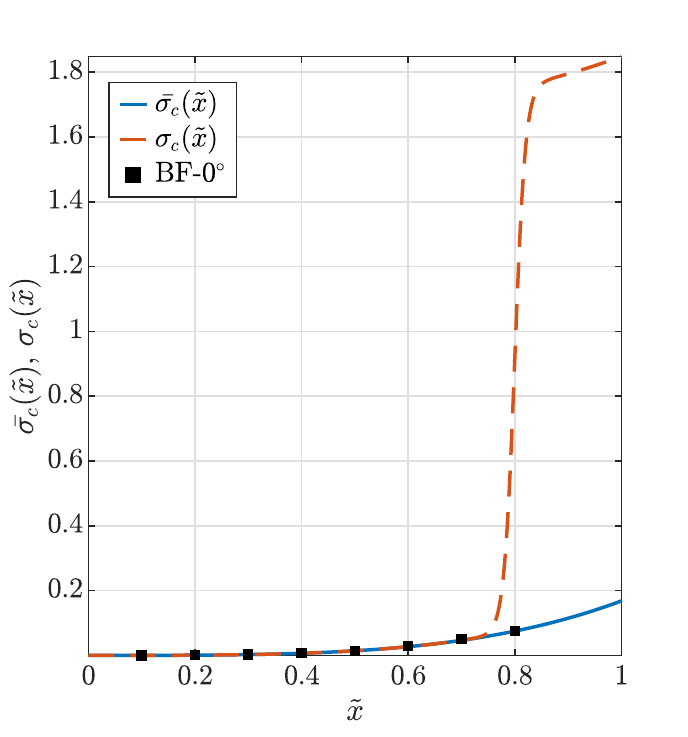}
		\caption{}
		\label{fig:FittedCurves:1}
	\end{subfigure}
	\caption{Illustration of triangular microstructures and stress limit fits:
	(\subref{fig:triangularMicrostructure:1} -- \subref{fig:triangularMicrostructure:3}) Triangular microstructures at different relative densities with red line indicating one unit cell { oriented at $0^{\circ}$ relative to the coordinate system in (\subref{fig:triangularMicrostructure:1})}.
	 (\subref{fig:FittedCurves:2}) Fitted buckling stress surfaces compared to {Block-Floquet-based cell analysis (BF)}. Squares are results from the cell analysis, black circles indicate the values of $\sigma_c$, $\sigma_t$ and $\sigma_b$ for $\tilde{x} = 0.4$. Finally, solid lines are the fitted surfaces at different densities $\tilde{x}$. 
		(\subref{fig:FittedCurves:1}) Buckling stress limit in compression $\bar{\sigma}_c$ defined by \eqref{eq:RelaxedStressLimit} marked by the dashed red line. The solid blue line is the fit of \eqref{eq:FitEquation} to the data points, marked by black points, obtained using Bloch-Floquet-based cell analysis.}
	\label{fig:FittedCurves}
\end{figure}

In order to determine the three stress limits mentioned above, we characterize the microstructure buckling strength under arbitrary loading using the material buckling strength evaluations presented in~\cite{Triantafyllidis1998,thomsen2018a}.  The material buckling strength under a given macroscopic stress situation is evaluated using linear buckling analysis together with Bloch-Floquet boundary conditions to capture all the possible buckling modes for an infinite periodic material. {In short, Bloch-Floquet analysis numerically computes the unit cell buckling problem for all possible wave vectors, i.e. for buckling modes with all possible periodicities from cell periodic to infinite. The lowest computed buckling mode is the most critical one and thus used to create the buckling yield surface.} The detailed formulations for the material buckling strength evaluation have been presented in~\cite{thomsen2018a} and are not repeated here for brevity. The general strength characteristics under arbitrary loading of a microstucture are evaluated by considering 21 evenly distributed macroscopic stress situations. {This is done by interpolating between external stress states $\boldsymbol{\sigma}_0=[\sigma_{xx},\sigma_{yy},\sigma_{xy}]$} ranging from $\boldsymbol{\sigma}_0=[-1,1,0]$ via $\boldsymbol{\sigma}_0=[-1,0,0]$, $\boldsymbol{\sigma}_0=[-1,-1,0]$ and $\boldsymbol{\sigma}_0=[0,-1,0]$ to {$\boldsymbol{\sigma}_0=[1,-1,0]$}. The considered isotropic stiffness-optimal microstructures are characterized (see Figure~\ref{fig:FittedCurves}) for  eight different relative densities (volume fractions), equally distributed in the interval $[0.1,0.8]$. 
{ Furthermore, the microstructure is evaluated at different angles to capture the worst case orientation. Due to symmetry in the microstructure 5 different orientation in the interval $[0^{\circ}-30^{\circ}]$ are deemed sufficient to capture all orientations.} The corresponding buckling strength surfaces are presented in Figure~\ref{fig:FittedCurves:2} by the square points. 
{ The data points for all orientation are only presented for the microstructure using $\tilde{x}=0.6$. It is shown that the symmetry in the microstructure provides equal but flipped surfaces for the $0^{\circ}$ and $30^{\circ}$ as well as $10^{\circ}$ and $20^{\circ}$ angles. The worst case orientations are $0^{\circ}$ and $30^{\circ}$, i.e. when one of the uniaxial load cases are aligned with one of the bars in the microstructure. Therefore, the Willam-Warnke surfaces have to be fitted accordingly. Only the data points for the $0^{\circ}$ angle are presented for the remaining densities, meaning that the fitted Willam-Warnke surfaces should match only for the upper left data points.}

{We remark here that the estimated cell buckling surface is scale independent (assuming infinite periodicity). This can a.o. be understood from considering the buckling yield stress of a pinned-pinned column, which is 
\begin{equation}
\sigma_c = \pi E I \dfrac{h^2}{L^2}.
\end{equation}
Thus, the estimated cell buckling values should be accurate for the cell size going to zero but be influenced by  boundary conditions for cell sizes approaching the macroscale.}

For each considered volume fraction, the buckling strengths under uniaxial ${\sigma_c}$  and biaxial compression ${\sigma_b}$ are directly obtained from the buckling surface. The tension strength, ${\sigma_t}$,  is estimated such that the failure surface has the best fit with the Willam-Warnke failure surface. Figure~\ref{fig:FittedCurves:2} shows the stress limits $\sigma_c$, $\sigma_t$ and $\sigma_b$ for $\tilde{x} = 0.4$ by circles and the solid lines represent the fitted surfaces.
{ As it is visible in the figure the fitted lines deviate more for the higher densities. This is a result of prioritizing accuracy for lower densities. However, the deviation results in a conservative approximation of the buckling stress limit and therefore still prevents local buckling for all densities.}

Similar to \cite{Andersen2021a}, two term interpolation schemes are employed to interpolate the buckling strength as function of the relative density from the eight evaluated microstructures. Thus the function to be fitted is given as  
\begin{equation}\label{eq:FitEquation}
	\bar{\sigma}_k(\tilde{x}_j) = E_0 \left(b_{0,k} \tilde{x}_j^{n_0} + b_{1,k} \tilde{x}_j^{n_0+1}\right), \quad k\in\{c,t,b\}
\end{equation}
where  $\tilde{x}_j$ is the elementwise relative density of the microstructures. The subscript  $k\in\{c,t,b\}$ indicates the strength for uniaxial compression, uniaxial tension and biaxial compression and the exponent $n_0=3$ is chosen to represent the buckling-to-relative-density exponent of a column for the microstructures at low density regimes.  The coefficients  $b_0$ and $b_1$  are obtained using curve fitting. Figure~\ref{fig:FittedCurves:1} shows the fitted curve (solid blue line) and data points (black points) for the buckling strengths under uniaxial compression.  All coefficients, $b_0$ and $b_1$, for all the buckling strengths are presented in Table~\ref{tab:FittedCoefficients}.
\begin{table}
	\centering
	\caption{Values of the fitted coefficients used to describe $\sigma_c(\rho)$, $\sigma_t(\rho)$ and $\sigma_b(\rho)$.}
	\label{tab:FittedCoefficients}
	\begin{tabular}{llll}
		\hline
		& $c$ & $t$ & $b$ \\
		\hline
		$b_{0,k}$ & $0.05882$ & $0.3327$ & $0.05079$\\
		$b_{1,k}$ & $0.1092$ & $3.301\mathrm{e}{-13}$ & $0.06596$ \\
		\hline
	\end{tabular}
\end{table}

Dense microstructures are more likely to fail due to plastic yielding than local buckling \cite{Timoshenko1989}. In this work we do not consider plastic yielding. Therefore, the buckling stress constraint in the high volume fraction limit is relaxed using a Heaviside function equivalent to \eqref{eq:Heaviside}
\begin{equation}\label{eq:RelaxedStressLimit}
	\sigma_k(\tilde{x}_j) = \bar{\sigma}_k(\tilde{x}_j) + 
	\psi \mathcal{H}(\tilde{x}_j,\eta,\bar{\beta}), \quad k\in\{c,t,b\}.
\end{equation}
Here $\psi$ is the relaxation parameter, which in this work is defined as 10 times the stress limit for a solid unit cell using \eqref{eq:FitEquation}, i.e. $\psi = 10 \bar{\sigma}_k(1)$. The threshold $\eta$ is chosen at the highest density of the known data points such that $\eta = 0.8$. Finally, a sharpness parameter of $\bar{\beta} = 50$ is used to get a relatively smooth yet steep relaxation at the threshold value. The relaxed buckling limit under uniaxial compression is shown by the dashed red line in Figure~\ref{fig:FittedCurves:1}. {The effect is that the buckling stress constraint is made inactive for high densities by overestimating the buckling stress value.}

\subsubsection{Willam-Warnke Failure Surface}\label{sec:WillamWarnkeModel}
Using the interpolated stress limits defined by \eqref{eq:FitEquation}, the density dependent  Willam-Warnke failure surface is defined using  the unified approach from \cite{Giraldo-Londono2020} in terms of the equivalent stress, $\sigma_{eq}(\rho_j^m,\tilde{x}_j)$,  given as
\begin{equation} \label{eq:sigmaEQ}
	\sigma_{eq}(\rho_j^m,\tilde{x}_j) = \alpha(\rho_j^m,\tilde{x}_j) \sqrt{3 J_2(\rho_j^m)} + G(I_1(\rho_j^m),\tilde{x}_j),
\end{equation}
where $J_2$ and $I_1$ are the second invariant of the Cauchy tensor $\mathbf{\sigma}$ and invariant of the deviatoric stress tensor $\mathbf{s}$, respectively. The term $G(I_1(\rho_j^m),\tilde{x}_j)$ is a function determined from the first stress invariant of the deviatoric stress. All of these are defined in \ref{app:Constraint}. Finally, the term $\alpha(\rho_j^m,\tilde{x}_j)$ defines the shape of the yield surface. It is defined as 
\begin{equation}\label{eq:alpha}
	\alpha(\rho_j^m,\tilde{x}_j) = \frac{A(\tilde{x}_j) \cos^2(\hat{\theta}(\rho_j^m))+ B(\tilde{x}_j)}{C(\tilde{x}_j) \cos(\hat{\theta}(\rho_j^m)) + \sqrt{D(\tilde{x}_j) \cos^2(\hat{\theta}(\rho_j^m)) + E(\tilde{x}_j)}},
\end{equation}
with $A(\tilde{x}_j)$, $B(\tilde{x}_j)$, $C(\tilde{x}_j)$, $D(\tilde{x}_j)$ and $E(\tilde{x}_j)$ determined from the stress limits and $\hat{\theta}(\rho_j^m)$ being the modified Lode angle. The definitions of all these are presented in \ref{app:Constraint}.

\subsubsection{Buckling Stress Constraint Formulation}\label{sec:ConstraintFormulation}
The \textit{p}-norm stress measure  proposed by \cite{Duysinx1998} is used to approximate the maximum equivalent stress as
\begin{equation}\label{eq:pNorm}
	\max_{\forall j}(\sigma_{eq}(\rho_j^m,\tilde{x}_j)) \approx ||\sigma_{eq}(\rho_j^m,\tilde{x}_j)||_p = \left(\sum_j \sigma_{eq}(\rho_j^m,\tilde{x}_j)^p \right)^{\frac{1}{p}} = \sigma_{PN}(\boldsymbol{\rho}^m,\tilde{\boldsymbol{x}}).
\end{equation}
The \textit{p}-norm approximation of the max-function depends on the value of $p$. As $p \rightarrow \infty$ the \textit{p}-norm approaches $\max\limits_{\forall j}(\sigma_{eq}(\rho_j^m,\tilde{x}_j))$ from above. This means that the \textit{p}-norm will always overestimate the actual maximum constraint value and therefore act as a conservative constraint. 

The actual buckling  stress constraint is enforced on the \textit{p}-norm stress measure under the critical  global buckling load. This results in the buckling stress constraint
\begin{equation}\label{eq:sigma_pn_tilde}
	\tilde{\sigma}_{PN}(\boldsymbol{\rho}^m,\tilde{\boldsymbol{x}}) = c\min\limits_{i\in \mathcal{B}}(\lambda_i)   \sigma_{PN}(\boldsymbol{\rho}^m,\tilde{\boldsymbol{x}})   -1 \approx  - \dfrac{c \sigma_{PN}(\boldsymbol{\rho}^m,\tilde{\boldsymbol{x}})}{J^{KS}(\gamma_i(\boldsymbol{\rho}^m))}  - 1 \leq 0,
\end{equation}
where $c$ is the correction presented by \cite{Le2010} for better approximation of  $\max\limits_{\forall j}(\sigma_{eq}(\rho_j^m,\tilde{x}_j))$. The correction is defined as
\begin{equation}
	c^n = \alpha^n \frac{\max\limits_{\forall j}(\sigma_{eq}^{n-1}(\rho_j^m,\tilde{x}_j))}{\sigma_{PN}^{n-1}(\boldsymbol{\rho}^m,\tilde{\boldsymbol{x}})} + \left(1-\alpha^n \right) c^{n-1},
\end{equation}
where $n$ is the design iteration number and $\alpha \in (0,1]$ controls the update speed between design iterations. In this work $\alpha = 0.1, \ \forall n,$ is used and the correction value is initialized as $c^0 = 1$. The final constraint is formulated by normalizing (\ref{eq:sigma_pn_tilde}) with the initial value at $n=0$, which yields
\begin{equation}\label{eq:StressConstraint}
	g_s(\boldsymbol{\rho}^m,\tilde{\boldsymbol{x}}) = \frac{\tilde{\sigma}_{PN}(\boldsymbol{\rho}^m,\tilde{\boldsymbol{x}})}{\tilde{\sigma}_{PN}^0},
\end{equation}
where $\tilde{\sigma}_{PN}^0 = -\sigma_{PN}^0 / J^{KS,0} $.

The optimization problem is solved using the method of moving asymptotes {(MMA)} \cite{Svanberg1987} based on the sensitivity of the objective and constraints presented in \ref{app:Sensitivity}.

%% file: Results.tex
The approach presented in this paper is verified and demonstrated on two examples. The results illustrate the superiority of multiscale structures using isotropic microstructures over single scale structures. Both examples use $E_0 = 1$ and $\nu_0 = \nu_{min} = 0.3$ and the coefficients from Table~\ref{tab:FittedCoefficients} to define the buckling failure surface. All examples use a $\beta$-continuation where $\beta$ is initialized as $\beta = 2$. After the first 125 iterations $\beta$ is doubled every 75 iterations until $\beta_{max} = 256$ is reached.

\subsection{Uni-axial Compression}
\label{sec:uniaxialComprresion}
\begin{figure}
\centering
\includegraphics[scale=1]{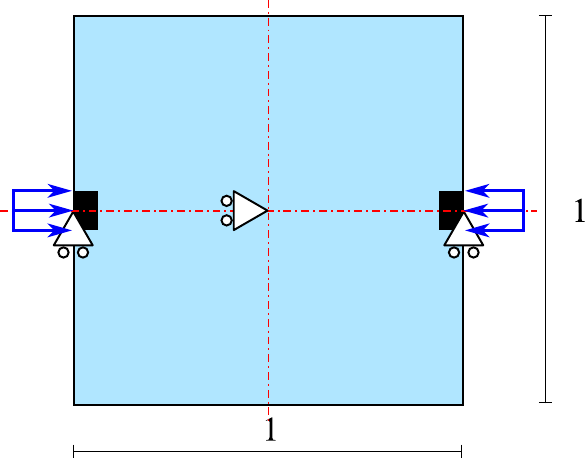}
\caption{Illustration of the design domain used for the uni-axial compression optimization.}
\label{fig:uniaxialDomain}
\end{figure}

\newcommand\lambdaSIMP{6.86} 
\newcommand\lambdaHS{10.81} 
\newcommand\lambdaHSS{9.24} 
\newcommand\lambdaSIMPOmega{6.17} 
\newcommand\lambdaHSOmega{6.68} 
\newcommand\lambdaHSSOmega{6.54} 
\newcommand\lambdaSIMPOmegaTwo{7.43}
\newcommand\lambdaHSOmegaTwo{6.82}
\newcommand\lambdaHSSOmegaTwo{6.63} 
\pgfmathsetmacro{\SIMPtoHS}{\lambdaHS/\lambdaSIMP*100-100} 
\pgfmathsetmacro{\SIMPtoHSS}{\lambdaHSS/\lambdaSIMP*100-100} 
\pgfmathsetmacro{\HSStoHS}{\SIMPtoHS-\SIMPtoHSS} 
\pgfmathsetmacro{\SIMPtoHSSOmega}{\lambdaHSSOmega/\lambdaSIMPOmega*100-100} 
The first example compares singlescale with multiscale optimization. Therefore three optimizations are performed. The first is a singlescale optimization using SIMP as stiffness interpolation, performed using the code from \cite{ferrari2021a}. The second is a multiscale optimization using HS for the stiffness interpolation. Finally, a multiscale optimization using the local buckling stress constraint {with HS stiffness  interpolation} is performed. This will be denoted HS-S where the added S indicates the use of the buckling stress constraint. All three optimizations are performed using the design domain in Figure~\ref{fig:uniaxialDomain} which is discretized using $200 \times 200$ bilinear quadrilateral elements. {The robust formulation is set up using $\Delta \eta = 0.25$ to generate the \textit{eroded} and \textit{dilated} designs.}
The center nodes on each side of the domain are restricted against vertical displacements. The node in the center of the domain is restricted against horizontal displacements. Two inward pointing distributed loads are applied around the horizontal symmetry axis. The widths of these loads are $L_{load} = 0.04$ with a magnitude of $10^{-3}$ at each load. { Passive solid regions are located at the loads with a width of $0.025$.} The design is forced to be symmetric around the horizontal and vertical center axes, { due to the symmetry in the design problem,} but no symmetry is used on the physics {to allow for both symmetric and unsymmetric buckling modes}. The objective function is a weighted average between compliance and buckling using weight $\{\omega \ | \ 0\leq \omega \leq 1\}$. This yields the optimization problem
\begin{equation} \label{eq:optProblem:uniaxial}
\begin{array}{rll}
\min\limits_{\mathbf{x},\mathbf{s}}  &: \quad \max\limits_m \left( g_{w}(\boldsymbol{\rho}^m) = \omega {\dfrac{C(\boldsymbol{\rho}^m)}{C_0}}  + (1- \omega) {\dfrac{J^{KS}\left(\gamma_i(\boldsymbol{\rho}^m)\right)}{J^{KS}_0}}\right),  &m\in\{e,b,d\}, \ \gamma_i\in\mathcal{B}, \ 0 \leq \omega \leq 1  \\
\textrm{s.t.} &: \quad g_V(\boldsymbol{\rho}^d) = \dfrac{\sum_j v_j \rho_j^d}{V_d^* V_{\Omega}} - 1 \leq 0, & \\
	&: \quad \color{gray}{
	{g_s(\boldsymbol{\rho}^m,\tilde{x},\gamma_i(\boldsymbol{\rho}^m)) 
	 \leq 0} ,} &\color{gray}{m\in\{e,b,d\}, \ \gamma_i\in\mathcal{B} } \\
    &: \quad \rho_j^m = \tilde{x}_j \bar{\tilde{s}}_j^m  , &m\in\{e,b,d\}  \\
    &: \quad  x_{min} \leq x_j \leq1, & \forall j    \\
    &: \quad 0\leq s_j \leq1.  &\forall j  \\
\end{array}    
\end{equation}
where $g_{w}(\boldsymbol{\rho}^m)$ is the weighted objective function. { The objective function consists of the compliance $C$ and aggregated eigenvalues $J^{KS}$ normalized using the values for the initial design, i.e. $C_0$ and $J^{KS}_0$.} Furthermore, a global volume constraint $g_V(\boldsymbol{\rho}^d)$ is used with a target volume on the \textit{blue print} design of $V^*_b = 0.15$. Finally, the local buckling stress constraint is used only on the last of the three multiscale optimizations, hence the gray color. 
%
{The optimization is performed by first optimizing the single- and multiscale problems for pure compliance i.e. $\omega = 1$. Then $\omega$ is lowered step by step using the optimized design from the previous $\omega$ until  $\omega = 0$ is reached. Finally, the multiscale problem with the active buckling stress constraint is optimized using the singlescale designs as initial guesses.}

\begin{figure}
	\begin{subfigure}[c]{0.3\textwidth}
		\centering
		\includegraphics[scale=0.5]{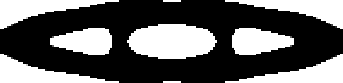}
		\caption{$C = 8.64\times 10^{-6}$}
		\label{fig:UniaxialResults:Omega05:1}
	\end{subfigure}
\hfill
	\begin{subfigure}[c]{0.3\textwidth}
		\centering
		\includegraphics[scale=0.5]{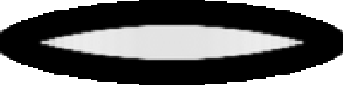}
		\caption{$C = 8.48\times 10^{-6}$}
		\label{fig:UniaxialResults:Omega05:2}
	\end{subfigure}
\hfill
	\begin{subfigure}[c]{0.3\textwidth}
		\centering
		\includegraphics[scale=0.5]{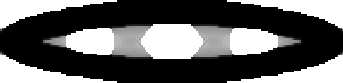}
		\caption{$C = 8.51\times 10^{-6}$}
		\label{fig:UniaxialResults:Omega05:3}
	\end{subfigure}
\hfill
	\begin{subfigure}[c]{0.08\textwidth}
		\vspace{-0.3cm}
		\includegraphics[scale=1]{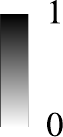}
	\end{subfigure}
	\\
	\begin{subfigure}[c]{0.3\textwidth}
		\centering
		\includegraphics[scale=0.405]{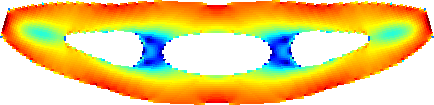}
		\caption{$\lambda_1 = \lambdaSIMPOmega$}
		\label{fig:UniaxialResults:Omega05:4}
	\end{subfigure}
\hfill
	\begin{subfigure}[c]{0.3\textwidth}
		\centering
		\scalebox{1}[-1]{\includegraphics[scale=0.405]{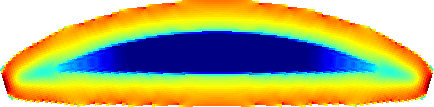}}
		\caption{$\lambda_1 = \lambdaHSOmega$}
		\label{fig:UniaxialResults:Omega05:5}
	\end{subfigure}
\hfill
	\begin{subfigure}[c]{0.3\textwidth}
		\centering
		\scalebox{1}[-1]{\includegraphics[scale=0.405]{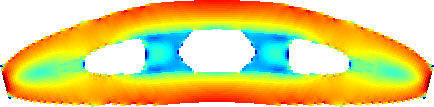}}
		\caption{$\lambda_1 = \lambdaHSSOmega$}
		\label{fig:UniaxialResults:Omega05:6}
	\end{subfigure}
\hfill
	\begin{subfigure}[c]{0.08\textwidth}
		\vspace{-0.3cm}
		\includegraphics[scale=1]{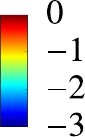}
	\end{subfigure}
	\\
	\begin{subfigure}[c]{0.3\textwidth}
		\centering
		\includegraphics[scale=0.405]{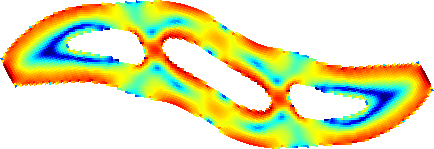}
		\caption{$\lambda_2 = \lambdaSIMPOmegaTwo$}
		\label{fig:UniaxialResults:Omega05:7}
	\end{subfigure}
\hfill
	\begin{subfigure}[c]{0.3\textwidth}
		\centering
		\scalebox{1}[-1]{\includegraphics[scale=0.405]{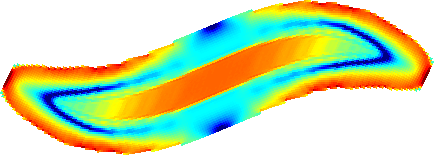}}
		\caption{$\lambda_2 = \lambdaHSOmegaTwo$}
		\label{fig:UniaxialResults:Omega05:8}
	\end{subfigure}
\hfill
	\begin{subfigure}[c]{0.3\textwidth}
		\centering
		\includegraphics[scale=0.405]{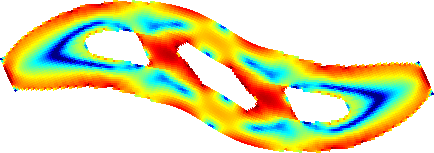}
		\caption{$\lambda_2 = \lambdaHSSOmegaTwo$}
		\label{fig:UniaxialResults:Omega05:9}
	\end{subfigure}
\hfill
	\begin{subfigure}[c]{0.08\textwidth}
		\vspace{-0.3cm}
		\includegraphics[scale=1]{ColorBarJetVert_converted.pdf}
	\end{subfigure}
	\caption{Optimized designs and the first two buckling modes using $\omega = 0.5$ and color indication of the normalized strain energy density $\log(\phi_e/\phi_{max})$:  
(\subref{fig:UniaxialResults:Omega05:1}) Singlescale design (SIMP),
(\subref{fig:UniaxialResults:Omega05:2}) Multiscale design (HS),
(\subref{fig:UniaxialResults:Omega05:3}) Design obtained using the multiscale formulation with local buckling stress constraint (HS-S),
(\subref{fig:UniaxialResults:Omega05:4}) First buckling mode for the SIMP design,
(\subref{fig:UniaxialResults:Omega05:5}) First buckling mode for the HS design,
(\subref{fig:UniaxialResults:Omega05:6}) First buckling mode for the HS-S design,
(\subref{fig:UniaxialResults:Omega05:7}) Second buckling mode for the SIMP design,
(\subref{fig:UniaxialResults:Omega05:8}) Second buckling mode for the HS design,
(\subref{fig:UniaxialResults:Omega05:9}) Second buckling mode for the HS-S design
.}
	\label{fig:UniaxialResults:Omega05}
\end{figure}

For the case using $\omega = 0.5$ the three designs in Figure~\ref{fig:UniaxialResults:Omega05} are obtained.
The figure also presents the first two mode shapes with the strain energy density calculated on element level by $\phi_j = \boldsymbol{\varphi}_{ij}^T \boldsymbol{k}_j  \boldsymbol{\varphi}_{ij}$ as indicated by the color scheme. It is seen how the three optimizations produce different designs performing at different levels for the BLF and almost similar level for compliance. All three designs have similar solid oval shells as the dominating part of the structures. These shells carry the majority of the load supported by an internal shear web that increases the buckling stability. The difference in the designs lie in the infill that connects the shells. The singlescale material produces a design which uses { thick bar} connections between the shell walls. These help stabilize the structure by restricting the shell from being pushed outward during compression. The disadvantage of these {bar} connections is that the shell is only stabilized discretely. Therefore the shells have to be thicker to prevent buckling between these connections. This results in a BLF of $\lambda_1 = \lambdaSIMPOmega$.

Looking at the multiscale design in Figure~\ref{fig:UniaxialResults:Omega05:2} it is seen that a more homogeneous infill with intermediate density is used. This means that the shells are connected on their entire inner surfaces. Therefore the shells are restricted from being pushed outwards and the shell itself only has to carry the horizontal load. The first two modes shapes are very similar to the singlescale modes. { The critical BLF has increased to $\lambda_1 = \lambdaHSOmega$ and the second mode is active at $\lambda_2 = \lambdaHSOmegaTwo$.} Based on this, the multiscale design is superior for buckling stability. However, using infill with an intermediate density introduces a risk of having local buckling before global buckling. This is not prevented in the simple multiscale design in Figure~\ref{fig:UniaxialResults:Omega05:2}.
{
The risk of having local buckling is tested using de-homogenization based on the approach described in \cite{Groen2020} which is briefly reviewed in Section~\ref{sec:LShape:De_Homogenization}. The first buckling mode is presented in Figure~\ref{fig:UniaxialResults:DeHomogenized:1}. Local buckling of the microstructure is clearly visible in the { right} side of the design. As a result the BLF is reduced to only $\lambda_1 = 0.96$, which is significantly lower than for the homogenized design. This BLF, together with the buckling mode, confirms that local buckling occurs if local stability is not taken into account during the optimization.}
\newcommand\MyscaleUni{0.53}
\begin{figure}
     \centering
     \begin{subfigure}[b]{0.29\textwidth}
         \centering
         \includegraphics[scale=\MyscaleUni]{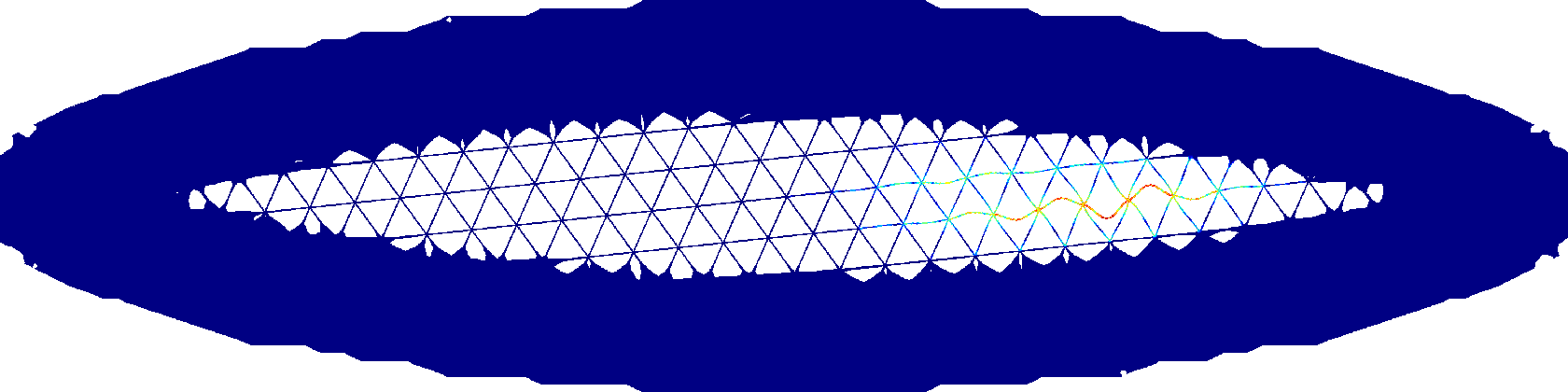}
         \vspace{0.27cm}
         \caption{$\lambda_1 = 0.96$}
         \label{fig:UniaxialResults:DeHomogenized:1}
     \end{subfigure}
     \hfill
     \begin{subfigure}[b]{0.29\textwidth}
         \centering
         \includegraphics[scale=\MyscaleUni]{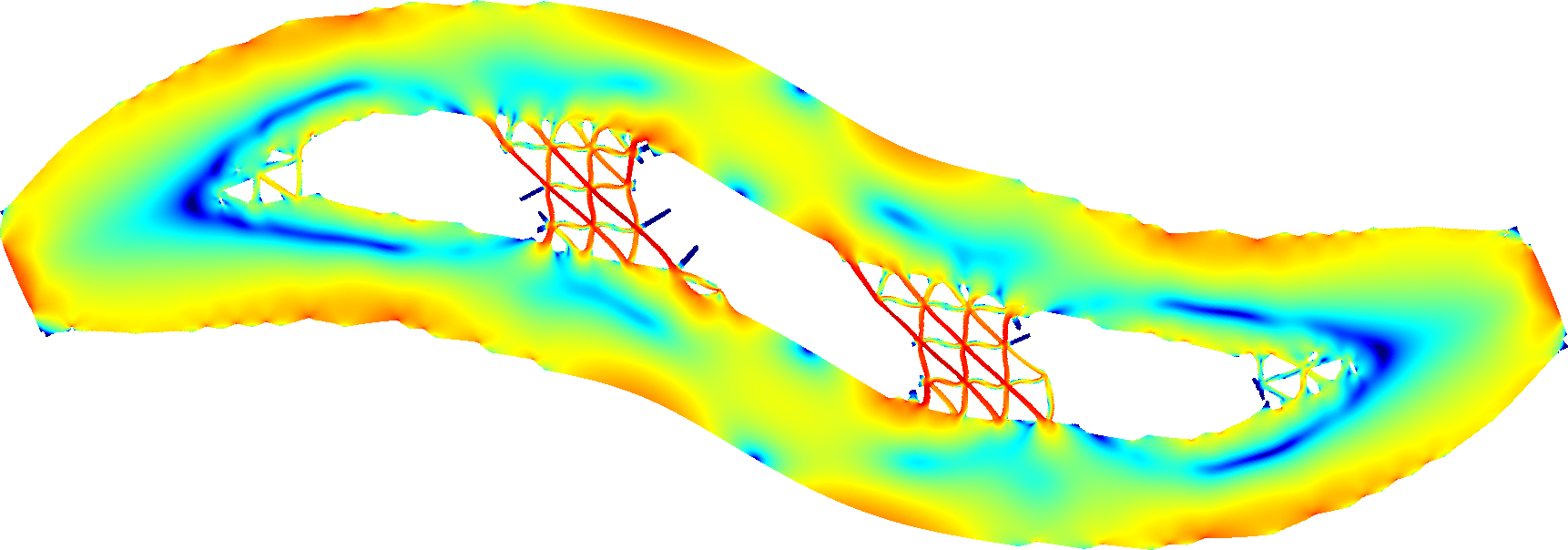}  
         \caption{$\lambda_1 = 6.47$}
         \label{fig:UniaxialResults:DeHomogenized:2}
     \end{subfigure}
     \hfill
     \begin{subfigure}[b]{0.29\textwidth}
         \centering
         \scalebox{1}[-1]{\includegraphics[scale=\MyscaleUni]{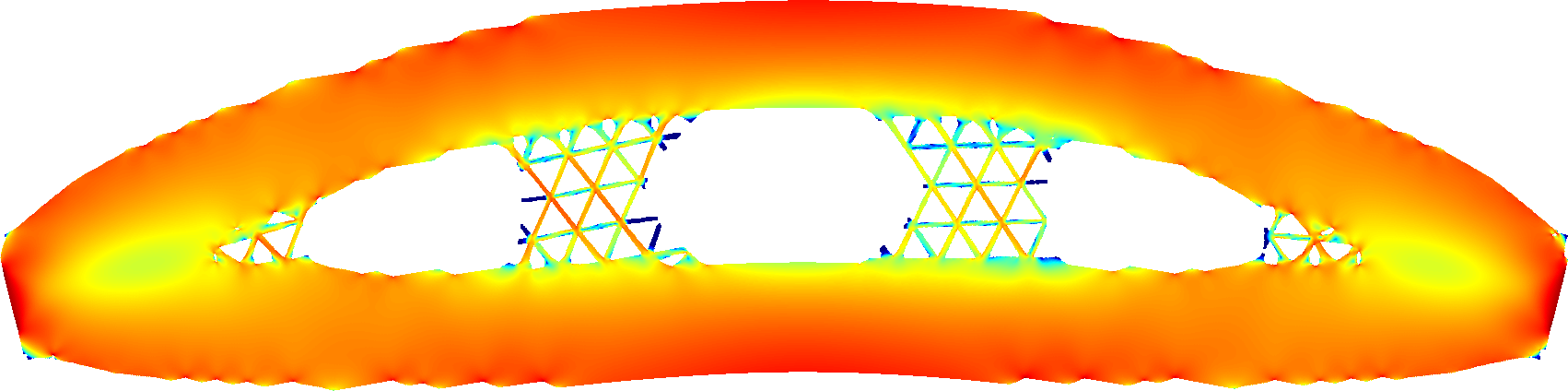}} 
         \vspace{-0.07cm}
         \caption{$\lambda_2 = 6.65$}
         \label{fig:UniaxialResults:DeHomogenized:3}
     \end{subfigure}
     \hfill
     \begin{subfigure}[c]{0.1\textwidth}
		\centering
		\vspace{-2cm}
		\includegraphics[scale=1]{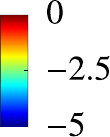}
	\end{subfigure}
        \caption{Comparison of de-homogenized results of the two multiscale designs 		from Figure~\ref{fig:UniaxialResults:Omega05}. The color scale illustrates 				the normalized strain energy density $\log(\phi_j/\phi_{max})$.
		(\subref{fig:UniaxialResults:DeHomogenized:1}) First buckling mode without buckling stress constraint (HS),
		(\subref{fig:UniaxialResults:DeHomogenized:2}) First buckling mode with buckling stress constraint (HS-S), 
		(\subref{fig:UniaxialResults:DeHomogenized:3}) Second buckling mode with buckling stress constraint (HS-S).}
        \label{fig:UniaxialResults:DeHomogenized}
\end{figure}

Finally, the multiscale design with the buckling stress constraint is presented in Figure~\ref{fig:UniaxialResults:Omega05:3}. The design is somewhat a mix of the multiscale design without the local buckling stress constraint and the singlescale design. The same {bar} structures as in the singlescale design are visible inside the shell. However, the {bars} are no longer solid, but instead they are wider and filled with material of intermediate density. This means that the distance between weak free hanging shells is reduced and therefore stability is increased. Looking at the first two buckling modes in Figure~\ref{fig:UniaxialResults:Omega05:6} and \ref{fig:UniaxialResults:Omega05:9} it is seen that they are similar to the modes from the singlescale and unconstrained multiscale designs. The eigenvalues of the two modes are $\lambda_1 = \lambdaHSSOmega$ and $\lambda_2 = \lambdaHSSOmegaTwo$ respectively, which means that both are likely to be active at the same time. The critical BLF is $\lambda_1 = \lambdaHSSOmega$ which is a slight  improvement of \pgfmathprintnumber[precision=2]{\SIMPtoHSSOmega}\% in buckling stability compared to the singlescale design.
{
To verify that the buckling stress constraint works as anticipated, the design is de-homogenized and the two first buckling modes are presented in Figure~\ref{fig:UniaxialResults:DeHomogenized:2} and \ref{fig:UniaxialResults:DeHomogenized:3}, respectively. The buckling modes are clearly global, which confirms that the buckling stress constraint prevents local buckling. The buckling modes are switched compared to the homogenized results. The reason is most likely found in the de-homogenization of the two thick bars with intermediate density. The buckling modes for the homogenized design shows that the strain energy density in the two bars is low for the first buckling mode, but large for the second buckling mode. This means that the second buckling mode is more sensitive to the size of the de-homogenized microstructure used in the two bars. Therefore, the BLF of the second buckling mode is likely to drop below the first buckling mode when de-homogenized. This is indeed the case with a BLF of $\lambda_1 = 6.47$ which is slightly lower than that of the homogenized design. The BLF of the second buckling mode is $\lambda_2 = 6.65$ which is actually a bit higher than the corresponding buckling mode for the homogenized design. Nevertheless, both buckling modes have BLF much closer to those of the homogenized design than it was the case when the buckling stress constraint was not included. The reason for the lower BLF value is examined thoroughly in the following section. Furthermore, a solution is presented to produce de-homogenized designs that perform at the same level or higher than the homogenized designs.}

\begin{figure}
\centering
\includegraphics[scale=1]{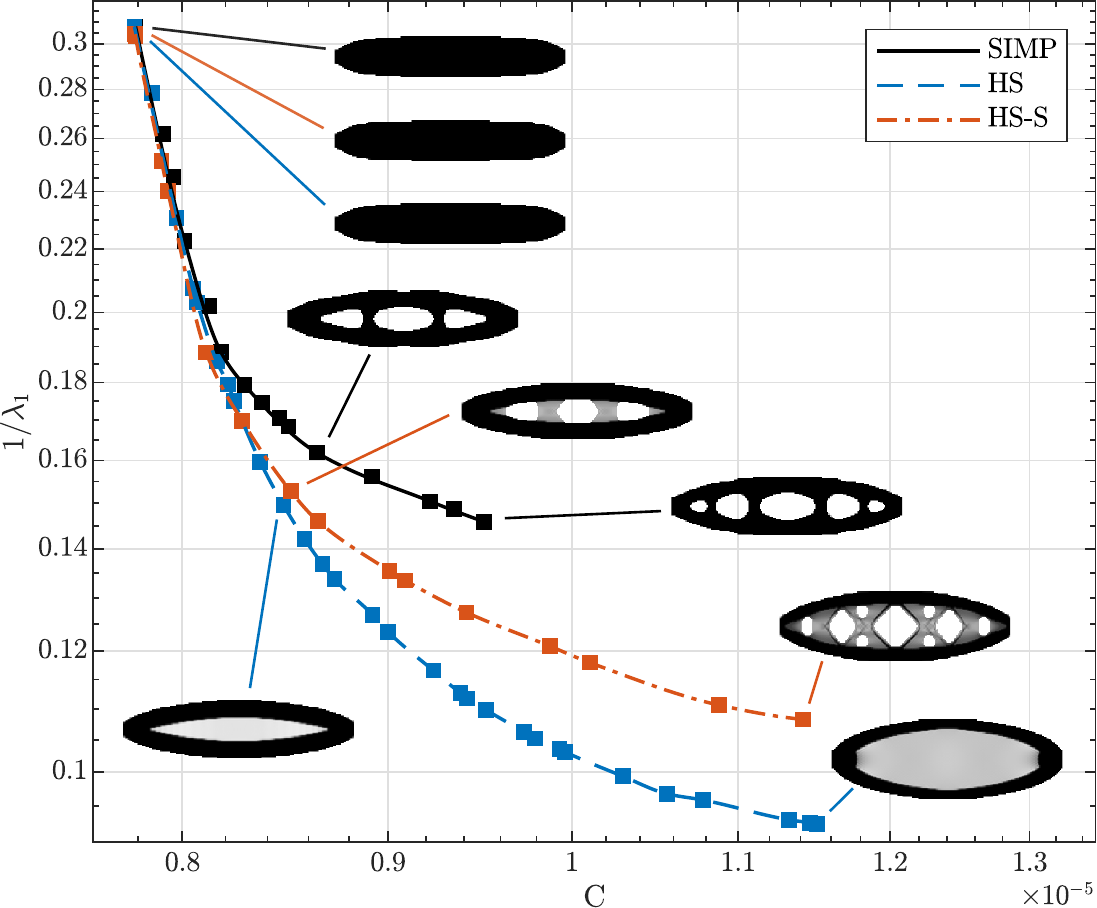}
\caption{Pareto fronts of the three uni-axial optimization cases, i.e. SIMP is single scale, HS is multiscale using Hashin-Shtrikman and HS-S is the same but with the buckling stress constraint included. Results are similar for compliance dominated problems. However, both multiscale optimization's show superior performance in buckling.}
\label{fig:uniaxialPareto}
\end{figure}

The results for all the optimization setups with all the values of $\omega$ are presented in Figure~\ref{fig:uniaxialPareto}.
A curve is fitted to the data points of each of the optimization setups to visualize the position of the Pareto front. Selected designs from all three test cases illustrating similarities and differences are also presented in the figure. The figure shows that for pure compliance optimization both single and multiscale optimizations reach similar designs with almost identical performance. Looking at the pure buckling case it is seen that the multiscale optimization without the local buckling constraint is the superior choice. However, the low density region in the center of that design is likely to buckle locally at significantly lower loads than the global buckling load at $\lambda_1 = \lambdaHS$. The design which takes local buckling stability into account shows a lower global buckling resistance at $\lambda_1 = \lambdaHSS$. But it is still better than the single scale design which experiences global buckling at $\lambda_1 = \lambdaSIMP$. 

To sum up, the Pareto fronts show a potential buckling stability increase of \pgfmathprintnumber[precision=1]{\SIMPtoHS}\% when using multiscale material with isotropic microstructures instead of single scale. When local buckling stability is taken into account the increased performance is reduced to \pgfmathprintnumber[precision=1]{\SIMPtoHSS}\% which is still a significant improvement compared to the singlescale design. It does however, leave a potential room for improvements of \pgfmathprintnumber[precision=1]{\HSStoHS} percentage points. This could possibly be exploited using multiscale microstructures with enhanced buckling stability \cite{thomsen2018a, Bluhm2020}.
\subsection{Modified L-Beam}
\label{sec:L_Beam}
\begin{figure}
\centering
\includegraphics[scale=1]{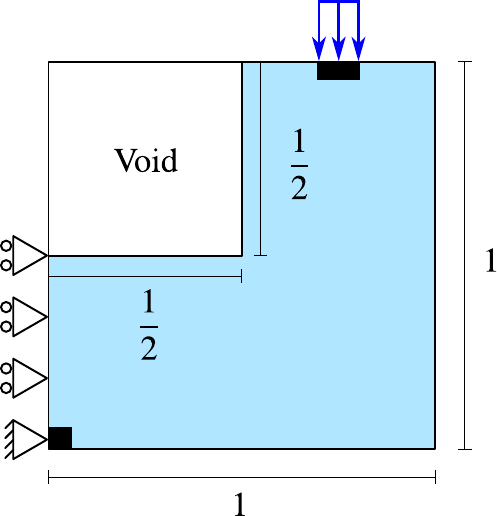}
\caption{Illustration of the design domain used for the L-beam optimization problem.}
\label{fig:LShapeDomain}
\end{figure}

%
\newcommand\lambdaSingleHomA{0.13}
\newcommand\lambdaSingleHomB{1.12}
\newcommand\lambdaSingleHomC{1.34}
\newcommand\lambdaSingleHomD{2.30}
\newcommand\CSingleHomB{5.33\times 10^{-5}}
\newcommand\CSingleHom{6.9\times 10^{-5},\ 5.33\times 10^{-5},\ 4.68\times 10^{-5}}
\newcommand\lambdaSingleBuckHomA{1.06}
\newcommand\lambdaSingleBuckHomB{1.75}
\newcommand\lambdaSingleBuckHomC{2.66}
\newcommand\lambdaSingleBuckHomD{3.31}
\newcommand\CSingleBuckHomB{6.36\times 10^{-5}}
\newcommand\CSingleBuckHom{6.97\times 10^{-5},\ 6.36\times 10^{-5},\ 5.77\times 10^{-5}}
\newcommand\lambdaMultiHomA{2.73}
\newcommand\lambdaMultiHomB{2.85}
\newcommand\lambdaMultiHomC{2.95}
\newcommand\lambdaMultiHomD{3.12}
\newcommand\CMultiBuckHomBVal{7.5120} 
\newcommand\CMultiBuckHomB{\CMultiBuckHomBVal \times 10^{4}}    
\newcommand\CMultiBuckHomBOriginal{6.66\times 10^{-5}}    
\newcommand\CMultiBuckHom{6.97\times 10^{-5},\ 6.66\times 10^{-5},\ 6.40\times 10^{-5}}
\pgfmathsetmacro{\SIMPtoHSHom}{\lambdaMultiHomA/\lambdaSingleHomA-1} 
\pgfmathsetmacro{\SIMPtoSIMPBuck}{\lambdaSingleBuckHomA/\lambdaSingleHomA*100-100} 
\pgfmathsetmacro{\HSBuckAtoHSBuckD}{\lambdaMultiHomC-\lambdaMultiHomA}
%
\newcommand\lambdaMultiDeHomAMin{2.76}
\newcommand\lambdaMultiDeHomAMax{2.82}
\newcommand\lambdaMultiDeHomAUser{2.78}
\newcommand\lambdaMultiDeHomAMean{2.78}
\newcommand\cMultiDeHomAMeanValOriginal{6.7430} 
\newcommand\cMultiDeHomAMeanVal{7.2452} 
\newcommand\cMultiDeHomAMean{\cMultiDeHomAMeanVal \times 10^{4}} 
\newcommand\volMultiDeHomAMeanVal{0.2047} 
\pgfmathsetmacro{\HomtoDehomA}{(\cMultiDeHomAMeanVal)/(\CMultiBuckHomBVal)*100-100} 
\newcommand\CellSizeA{0.01}
\newcommand\lambdaMultiDeHomCMin{1.71} 
\newcommand\lambdaMultiDeHomCMax{2.45}
\newcommand\lambdaMultiDeHomCUser{2.07}
\newcommand\CellSizeC{0.05} 
\newcommand\lambdaMultiDeHomBMin{1.33} 
\newcommand\lambdaMultiDeHomBMax{2.22}
\newcommand\lambdaMultiDeHomBUser{1.75}
\pgfmathsetmacro{\HomtoDehomB}{(\lambdaMultiDeHomBMax)/(\lambdaMultiHomA)*100-100}
\newcommand\CellSizeB{0.08}
\newcommand\lambdaMultiDeHomDMin{0.53} 
\newcommand\lambdaMultiDeHomDMax{2.}
\newcommand\lambdaMultiDeHomDUser{1.34}
\newcommand\CellSizeD{0.15}
\newcommand\volMultiDeHomFMeanVal{0.2018} 
\newcommand\ComplianceRelaxation{1}
%
\newcommand\lambdaMultiDeHomAShell{2.96} 
\newcommand\lambdaMultiDeHomAShellMean{2.96} 
\pgfmathsetmacro{\DeHomtoDehomShellA}{(\lambdaMultiDeHomAShell)/(\lambdaMultiDeHomAUser)*100-100} 
\pgfmathsetmacro{\DeHomtoDehomShellAMean}{(\lambdaMultiDeHomAShellMean)/(\lambdaMultiDeHomAMean)*100-100} 
\newcommand\cMultiDeHomAShellMeanVal{6.53} 
\newcommand\cMultiDeHomAShellMean{\cMultiDeHomAShellMeanVal \times 10^{-5}}
\pgfmathsetmacro{\cDehomtoDehomAShell}{(\cMultiDeHomAShellMeanVal)/(\cMultiDeHomAMeanValOriginal)*100-100} 
\newcommand\volMultiDeHomAShellMeanVal{0.2062} 
\newcommand\volMultiDeHomAShellMean{\volMultiDeHomAShellMeanVal}
\pgfmathsetmacro{\HomtoDehomAShell}{(\volMultiDeHomAShellMeanVal)/(0.2)*100-100} 
\pgfmathsetmacro{\volDehomtoDehomAShell}{(\volMultiDeHomAShellMeanVal)/(\volMultiDeHomAMeanVal)*100-100} 
\newcommand\CellSizeAShell{0.01}
\newcommand\lambdaMultiDeHomCShell{3.01} 
\newcommand\CellSizeCShell{0.05}
\newcommand\lambdaMultiDeHomBShell{3.05} 
\newcommand\CellSizeBShell{0.08}
\newcommand\lambdaMultiDeHomDShell{3.09} 
\pgfmathsetmacro{\DeHomtoDehomShellD}{(\lambdaMultiDeHomDShell)/(\lambdaMultiDeHomDMin)*100-100} 
\newcommand\CellSizeDShell{0.15}
\newcommand\lambdaMultiDeHomEShellMin{2.70} 
\newcommand\lambdaMultiDeHomEShellMax{3.25} 
\newcommand\volMultiDeHomEShellMeanVal{0.2252} 
\newcommand\volMultiDeHomEShellMean{\volMultiDeHomEShellMeanVal}
\pgfmathsetmacro{\ShellMinToMaxE}{(\lambdaMultiDeHomEShellMax)/(\lambdaMultiDeHomEShellMin)*100-100} 
\pgfmathsetmacro{\volDehomtoDehomEShell}{(\volMultiDeHomEShellMeanVal)/(\volMultiDeHomFMeanVal)*100-100} 
\newcommand\CellSizeEShell{0.2}
The second problem is a modified version of the frequently used benchmark L-beam problem. The domain is illustrated in Figure~\ref{fig:LShapeDomain} with the domain rotated $90^{\circ}$ counter clockwise and the load in the vertical downward pointing direction.
This change provides a more interesting buckling response as seen in the following. The width of the distributed load is $0.1$ and its center is located $0.25$ from the right side of the domain. The magnitude of the load is $10^{-3}$. The upper left quadrant of the domain is passive void. Furthermore, two passive solid regions are present in the domain. The first is located at the load with the same width as the load and a height of $0.05$. The second is positioned at the lower left corner with the dimensions $0.05 \times 0.05$. The left boundary of the active domain is simply supported in the horizontal direction. The part at the passive solid domain is also supported in the vertical direction. The domain is discretized by $100 \times 100$ bilinear quadrilateral elements. { The robust formulation is set up using $\Delta \eta = 0.1$ to generate the \textit{eroded} and \textit{dilated} designs. Furthermore, all designs are obtained from uniform initial guesses.} The problem is solved using both singlescale { material (SIMP) and multiscale material with the buckling stress constraint (HS-S)}. The optimized designs are post-processed using de-homogenization where the multiscale material is projected onto a fine mesh and interpreted as a triangular microstructure with a varying bar width $w$ using the method by \cite{Groen2018,Groen2020}. For this reason $x_{min} = 0.27$, corresponding to a slenderness ratio of $L/w = 10.3$, is used to ensure that individual bars in the microstructure can always be resolved using 3 elements across the width of the bar when projected onto the fine mesh. This is done to prevent node connections and reduce artificial stiffness through shear locking. { In reality the choice of $x_{min}$ should be based on the available manufacturing method and associated limitations.}

\begin{figure}
     \centering
     \begin{subfigure}[t]{0.29\textwidth}
         \centering
         \includegraphics[height=0.8\linewidth,angle=90,origin=c]{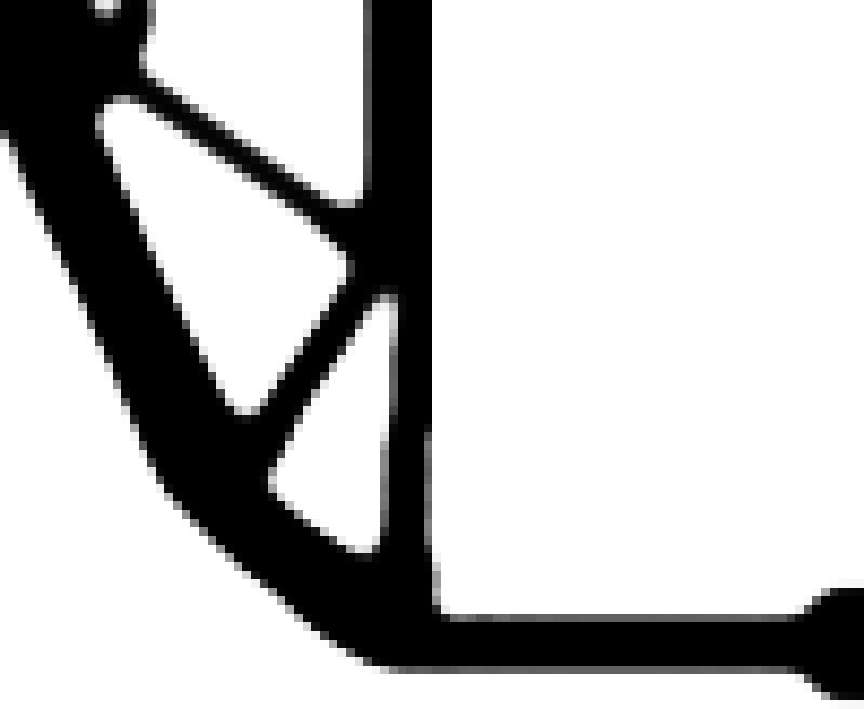}
         \caption{$C^b = {\CSingleHomB}$}
         \label{fig:LShape:Homogenized:Design:1}
     \end{subfigure}
     \hfill
     \begin{subfigure}[t]{0.29\textwidth}
     	\centering
         \includegraphics[height=0.8\linewidth,angle=90,origin=c]{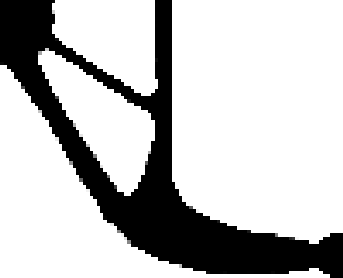}       
         \caption{$C^b = {\CSingleBuckHomB}$}
         \label{fig:LShape:Homogenized:Design:2}
     \end{subfigure}
     \hfill
     \begin{subfigure}[t]{0.29\textwidth}
         \centering
         \includegraphics[height=0.8\linewidth,angle=90,origin=c]{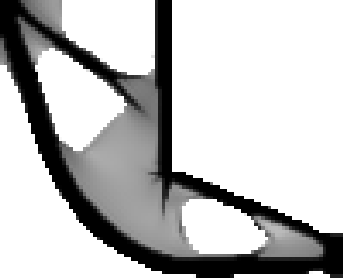} 
         \caption{$C^b = {\CMultiBuckHomBOriginal}$}
         \label{fig:LShape:Homogenized:Design:3}
     \end{subfigure}
     \hfill
     \begin{subfigure}[t]{0.1\textwidth}
		\centering
		\vspace{-4.4cm}
		\includegraphics[scale=1]{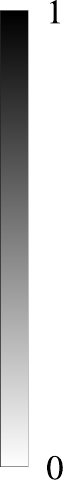} 
	\end{subfigure}
        \caption{Compliance and buckling optimized L-Beam designs using both single- and multiscale material.
        (\subref{fig:LShape:Homogenized:Design:1}) Singlescale compliance optimized design, (\subref{fig:LShape:Homogenized:Design:2}) Singlescale buckling optimized design, (\subref{fig:LShape:Homogenized:Design:3}) Multiscale buckling optimized design.}
        \label{fig:LShape:Homogenized:Design}
\end{figure}

\begin{figure}
     \centering
     \begin{tabular}{lcr}
     \begin{subfigure}[b]{0.3\textwidth}
     	\centering
         \includegraphics[height=0.8\linewidth,angle=90,origin=c]{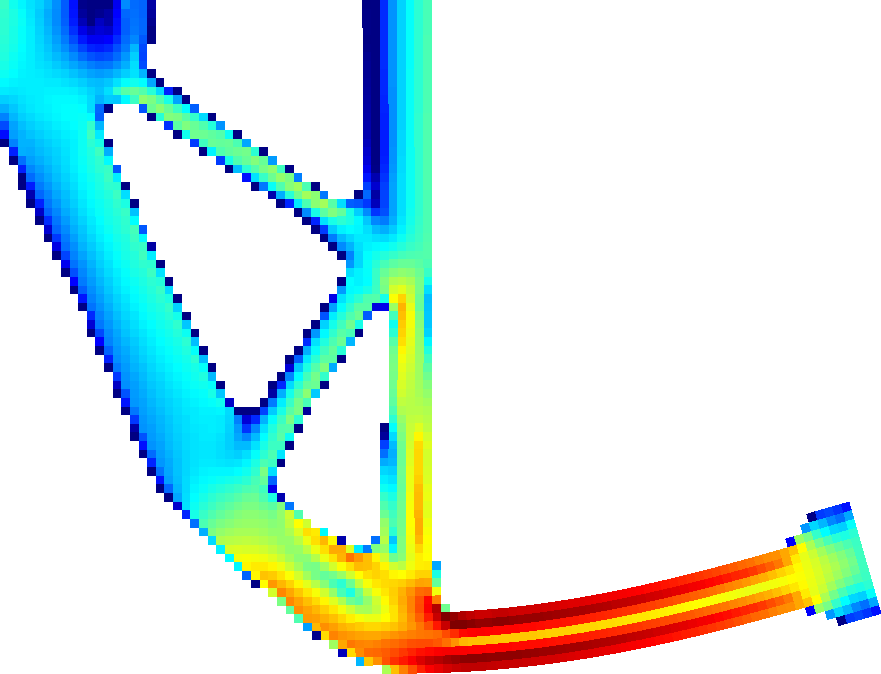}       
         \caption{$\lambda_1 = \lambdaSingleHomA$}
         \label{fig:LShape:Homogenized:Modes:4}
     \end{subfigure}
     &
     \begin{subfigure}[b]{0.3\textwidth}
         \centering
         \includegraphics[height=0.8\linewidth,angle=90,origin=c]{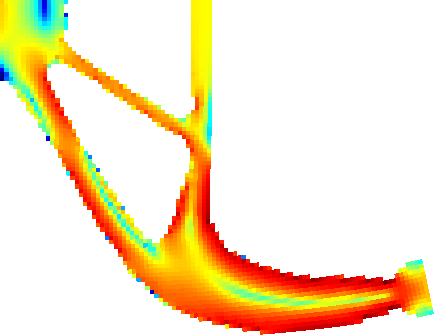}   
         \caption{$\lambda_1 = \lambdaSingleBuckHomA$}
         \label{fig:LShape:Homogenized:Modes:5}
     \end{subfigure}
     &
     \begin{subfigure}[b]{0.3\textwidth}
     	\centering
         \includegraphics[height=0.8\linewidth,angle=90,origin=c]{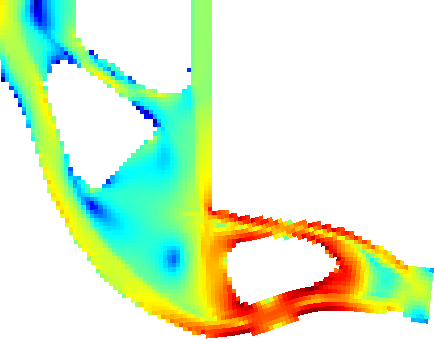}       
         \caption{$\lambda_1 = \lambdaMultiHomA$}
         \label{fig:LShape:Homogenized:Modes:6}
     \end{subfigure}
     \\
     \begin{subfigure}[b]{0.3\textwidth}
     	\centering
         \includegraphics[height=0.8\linewidth,angle=90,origin=c]{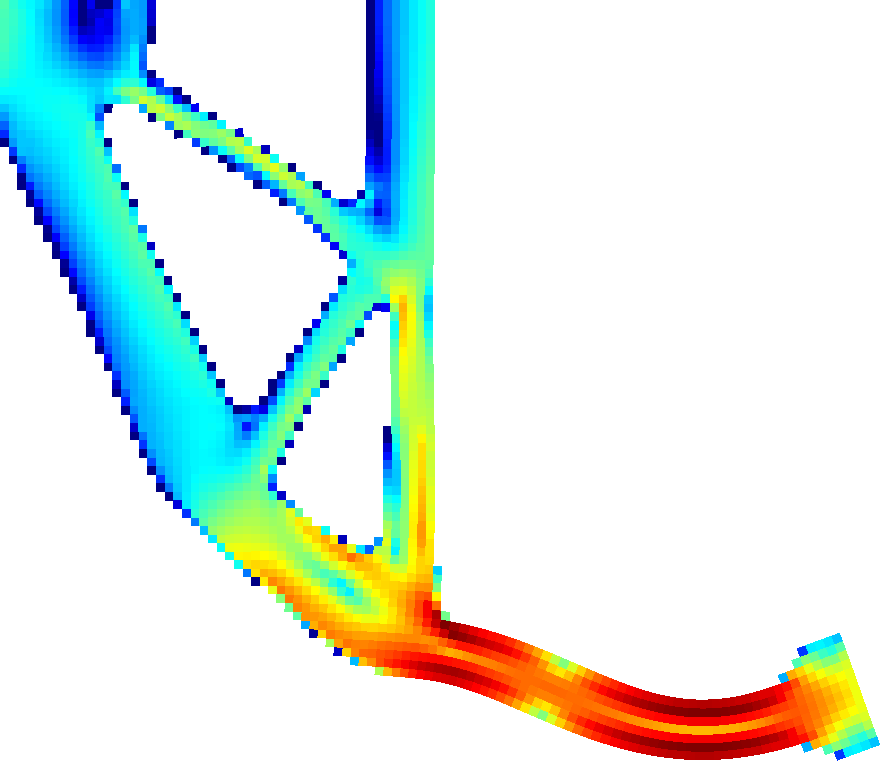}       
         \caption{$\lambda_2 = \lambdaSingleHomB$}
         \label{fig:LShape:Homogenized:Modes:7}
     \end{subfigure}
     &
     \begin{subfigure}[b]{0.3\textwidth}
         \centering
         \includegraphics[height=0.8\linewidth,angle=90,origin=c]{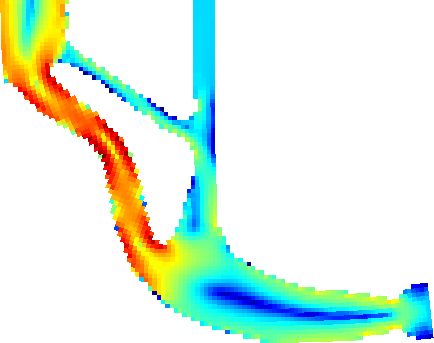}   
         \caption{$\lambda_2 = \lambdaSingleBuckHomB$}
         \label{fig:LShape:Homogenized:Modes:8}
     \end{subfigure}
     &
     \begin{subfigure}[b]{0.3\textwidth}
     	\centering
         \includegraphics[height=0.8\linewidth,angle=90,origin=c]{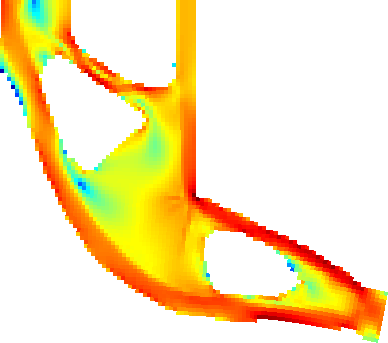}       
         \caption{$\lambda_2 = \lambdaMultiHomB$}
         \label{fig:LShape:Homogenized:Modes:9}
     \end{subfigure}
     \\
     \begin{subfigure}[b]{0.3\textwidth}
     	\centering
         \includegraphics[height=0.8\linewidth,angle=90,origin=c]{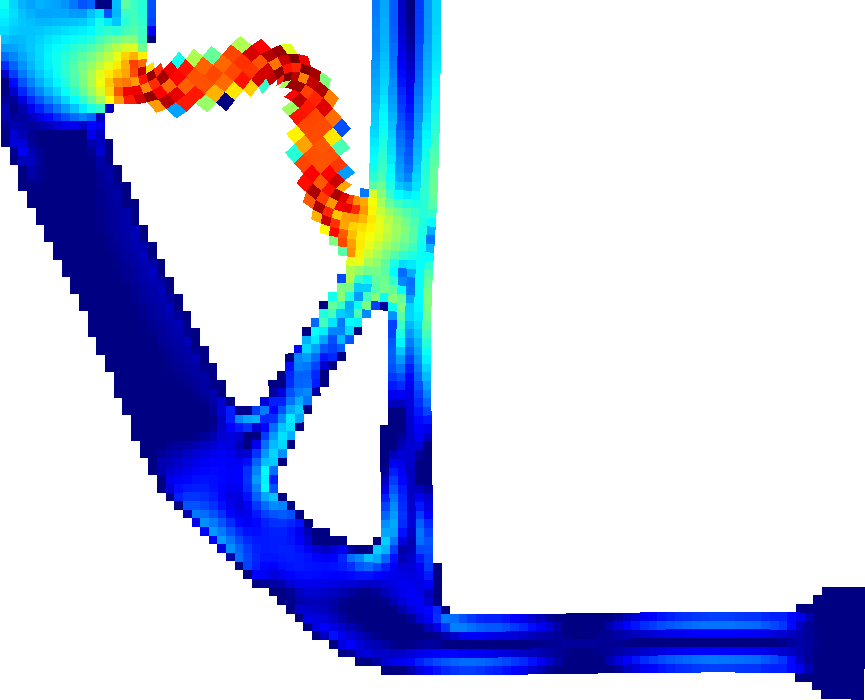}   
         \includegraphics[scale=1]{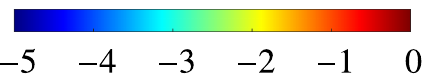}     
         \caption{$\lambda_3 = \lambdaSingleHomC$}
         \label{fig:LShape:Homogenized:Modes:10}
     \end{subfigure}
     &
     \begin{subfigure}[b]{0.3\textwidth}
         \centering
         \includegraphics[height=0.8\linewidth,angle=90,origin=c]{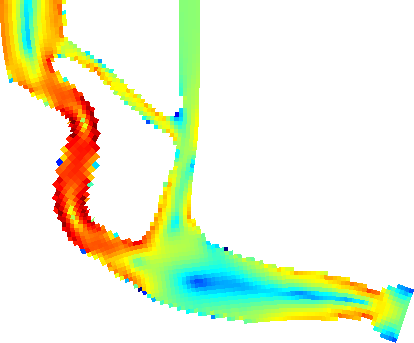}   
         \includegraphics[scale=1]{ColorBarJetHorizontal_converted2.pdf} 
         \caption{$\lambda_3 = \lambdaSingleBuckHomC$}
         \label{fig:LShape:Homogenized:Modes:11}
     \end{subfigure}
     &
     \begin{subfigure}[b]{0.3\textwidth}
     	\centering
         \includegraphics[height=0.8\linewidth,angle=90,origin=c]{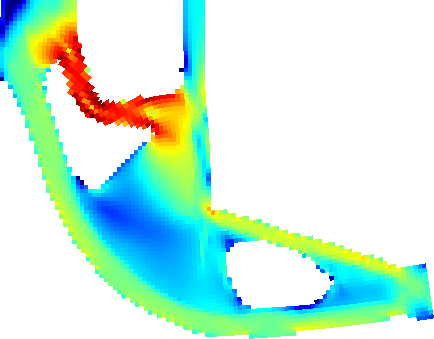} 
         \includegraphics[scale=1]{ColorBarJetHorizontal_converted2.pdf}  
         \caption{$\lambda_3 = \lambdaMultiHomC$}
         \label{fig:LShape:Homogenized:Modes:12}
     \end{subfigure}
     \end{tabular}
        \caption{Buckling modes for designs using both single- and multiscale material. The normalized strain energy density $\log(\phi_j/\phi_{max})$ is indicated by the color scale. (\subref{fig:LShape:Homogenized:Modes:4},\subref{fig:LShape:Homogenized:Modes:7},\subref{fig:LShape:Homogenized:Modes:10}) Singlescale buckling modes of compliance optimized design, 
(\subref{fig:LShape:Homogenized:Modes:5},\subref{fig:LShape:Homogenized:Modes:8},\subref{fig:LShape:Homogenized:Modes:11}) Singlescale buckling modes of buckling optimized design, 
(\subref{fig:LShape:Homogenized:Modes:6},\subref{fig:LShape:Homogenized:Modes:9},\subref{fig:LShape:Homogenized:Modes:12}) Multiscale buckling modes.}
        \label{fig:LShape:Homogenized:Modes}
\end{figure}

All the optimized homogenized designs are presented in Figure~\ref{fig:LShape:Homogenized:Design} with the corresponding buckling modes presented in Figure~\ref{fig:LShape:Homogenized:Modes}.
The homogenized design for the compliance minimization is presented in Figure~\ref{fig:LShape:Homogenized:Design:1}. The design has a characteristic vertical bar which obviously has low buckling resistance. This is visible in the first two buckling modes shown in Figure~\ref{fig:LShape:Homogenized:Modes:4} and \ref{fig:LShape:Homogenized:Modes:7} where the critical buckling mode is clearly visible in the vertical bar. The critical BLF also confirms that the design is not buckling resistant with a value of $\lambda_1 = \lambdaSingleHomA$. The compliance of the \textit{eroded}, \textit{blue print} and \textit{dilated} designs are $C^m = [{\CSingleHom}]$, respectively.

Next a buckling optimization is performed using the singlescale material and a compliance constraint allowing \ComplianceRelaxation\% higher compliance on the \textit{eroded} design. The optimized design is presented in Figure~\ref{fig:LShape:Homogenized:Design:2}. The design is slightly different from the compliance optimized design. The single vertical bar has thickened to increase buckling stability. The critical buckling mode in Figure~\ref{fig:LShape:Homogenized:Modes:5} is distributed more evenly over the structure and as a result the critical BLF increases with  \pgfmathprintnumber[precision=0]{\SIMPtoSIMPBuck}\% to $\lambda_1 = \lambdaSingleBuckHomA$. The higher order modes are active at significantly higher BLFs. Furthermore, they are mainly located at the thinner inclined bars in the lower left part of the design as seen in Figure~\ref{fig:LShape:Homogenized:Modes:8} and \ref{fig:LShape:Homogenized:Modes:11}. The compliance values are $C^m = [\CSingleBuckHom]$, which for the \textit{eroded} design is an increase of \ComplianceRelaxation\% but for the \textit{blue print} and \textit{dilated} designs correspond to larger increases. 

Finally, the optimization is performed using the multiscale material with the compliance constraint allowing \ComplianceRelaxation\% higher compliance on the \textit{eroded} design. The optimized design is presented in Figure~\ref{fig:LShape:Homogenized:Design:3}. The design is visibly different from the singlescale designs. The single vertical bar has been split into two to increase buckling stability. The critical buckling mode in Figure~\ref{fig:LShape:Homogenized:Modes:6} is now located {in the two bars in the upper part of the domain.} As a result the BLF increases with a factor of approximately \pgfmathprintnumber[precision=0]{\SIMPtoHSHom} compared to the compliance optimized design, resulting in a critical BLF of $\lambda_1 = \lambdaMultiHomA$. {The second buckling mode in Figure~\ref{fig:LShape:Homogenized:Modes:9} is distributed over the entire structure. The third buckling mode in Figure~\ref{fig:LShape:Homogenized:Modes:12} shows buckling at the transverse bar originating from the lower left corner of the domain. This is very similar to the corresponding mode for the compliance optimized design.} Common for the two higher order modes presented are that their BLFs are closer to the critical BLF. The difference between the first and third BLF is only \pgfmathprintnumber[precision=3]{\HSBuckAtoHSBuckD}. The compliance values are $C^m = [\CMultiBuckHom]$ which again is a \ComplianceRelaxation\% increase for the \textit{eroded} design. The \textit{blue print} and \textit{dilated} designs have larger increases, just as it was the case for the singlescale buckling optimized design.

\subsubsection{De-Homogenized Buckling Analysis}
\label{sec:LShape:De_Homogenization}
%
%
Using multiscale material for any kind of optimization is only beneficial if the de-homogenized designs perform at the same level as the homogenization based designs. To verify that this is the case the multiscale design in Figure~\ref{fig:LShape:Homogenized:Design:3} is de-homogenized. { The de-homogenization procedure is based on the approach described in \cite{Groen2020}. First the densities are mapped to a fine mesh.} In the current work a refinement ratio of $30$ from the coarse to fine mesh is used {resulting in a fine mesh using a total of $9\times 10^6$ elements. Then the densities are converted to widths $\boldsymbol{w}$ of the bars in the microstructure
\begin{equation}\label{eq:shellWidth}
w_j = \left(\dfrac{\sqrt{3}}{3}-\dfrac{\sqrt{3-3\rho_j^b}}{3}\right) \dfrac{2\sqrt{3}}{3}.
\end{equation}
This is used to create an implicit geometry description $\tilde{\boldsymbol{\rho}}_i, ~i\in[1,3] $ for each of the three layers
\begin{equation}
\tilde{\boldsymbol{\rho}}_i = H\left(\left(\dfrac{1}{2} +\dfrac{1}{2} \mathcal{S}\Bigl\{ P \left( \cos(\theta_i)\boldsymbol{x}_1 + \sin(\theta_i)\boldsymbol{x}_2\right) \Bigr\} \right) - \boldsymbol{w} \right),
\end{equation}
%
%
where $H$ is the Heaviside function, $\mathcal{S}\in [-1,1]$ is a triangular wave function and $\boldsymbol{x}_1$ and $\boldsymbol{x}_2$ are the spacial coordinates. The cell size is determined through the periodicity $P$
\begin{equation}
P = \dfrac{2\pi}{\epsilon},
\end{equation}
where $\epsilon$ is the absolute length describing the periodicity between two parallel bars in the microstructure as illustrated in Figure~\ref{fig:cellSize}.
\begin{figure}
\centering
\includegraphics[scale=1]{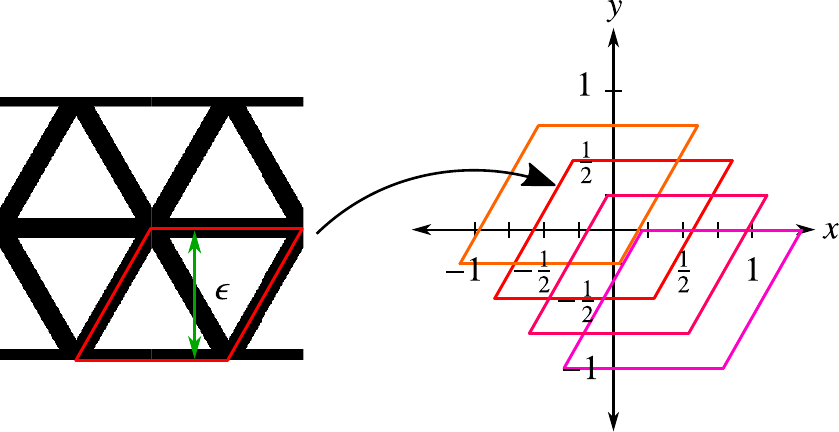}
\caption{Illustration of the cell size (right) and position (left). The positions is defined at the center of the cell. Here the red cell is located at $(0,0)$ and three remaining position are illustrated at $(-0.25,0.25)$, $(0.25,-0.25)$ and $(0.5,-0.5)$.}
\label{fig:cellSize}
\end{figure}
The orientation of the individual layers is determined through $\theta_i$. Given the triangular microstructure the layers are positioned in intervals of $60^{\circ}$. Finally, the layers are combined to create an implicit geometry of the de-homogenized structure
\begin{equation}
\tilde{\boldsymbol{\rho}} = \min\left(1, \sum_{i=1}^3 \tilde{\boldsymbol{\rho}}_i\right).
\end{equation}
}

\begin{figure}
     \centering
     \begin{subfigure}[t]{0.24\textwidth}
         \centering
         \includegraphics[scale=1,angle=90,origin=c]{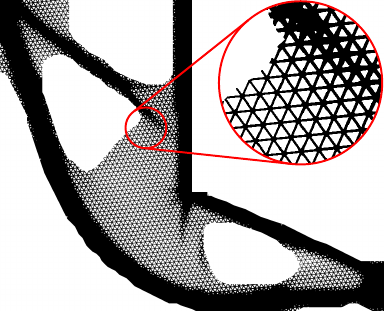}
         \caption{$\epsilon = 0.01$}
         \label{fig:LShape:Dehomogenization:1}
     \end{subfigure}
     \begin{subfigure}[t]{0.24\textwidth}
     	\centering
         \includegraphics[scale=1,angle=90,origin=c]{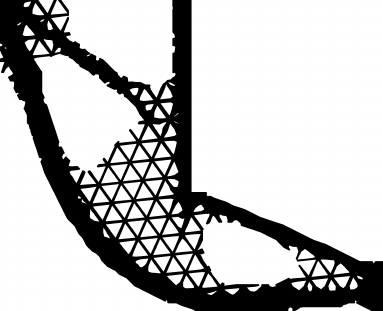}       
         \caption{$\epsilon = 0.05$}
         \label{fig:LShape:Dehomogenization:2}
     \end{subfigure}
     \begin{subfigure}[t]{0.24\textwidth}
         \centering
         \includegraphics[scale=1,angle=90,origin=c]{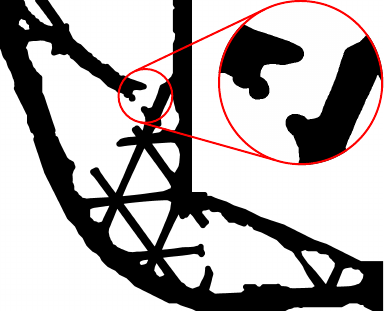}   
         \caption{$\epsilon = 0.15$}
         \label{fig:LShape:Dehomogenization:3}
     \end{subfigure}
     \begin{subfigure}[t]{0.24\textwidth}
         \centering
         \includegraphics[scale=1,angle=90,origin=c]{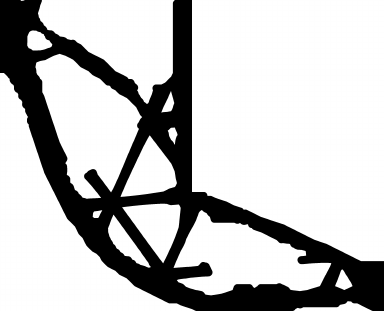}   
         \caption{$\epsilon = 0.2$}
         \label{fig:LShape:Dehomogenization:4}
     \end{subfigure}
        \caption{De-homogenization of the multiscale buckling optimized design at different unit cell sizes $\epsilon$.}
        \label{fig:LShape:Dehomogenization}
\end{figure}

Examples of de-homogenization using four different cell sizes are presented in Figure~\ref{fig:LShape:Dehomogenization}.
The example using $\epsilon = 0.01$ has very fine microstructures which visually looks similar to the homogenized design in Figure~\ref{fig:LShape:Homogenized:Design:3}. For the higher values of $\epsilon$ the size of the microstructures increases to a level where only a few triangles are present. This obviously introduces a risk of loosing connectivity throughout the structure as cell sizes exceed the length scale used on the homogeneous design. Nevertheless, it is interesting to look at large cell sizes to gain insight into the validity of the homogenization assumptions. An example is presented in Figure~\ref{fig:LShape:Dehomogenization:3} where a cell size of $\epsilon = 0.15$ is used. The region in the red circle is disconnected as a result of the coarse microstructure. This will severely reduce stiffness and stability of the structure. Additionally, coarse microstructures will result in free hanging members which are not contributing to neither stiffness nor stability. This is especially pronounced in Figure~\ref{fig:LShape:Dehomogenization:3} and \ref{fig:LShape:Dehomogenization:4} where $\epsilon = 0.15$ and $\epsilon = 0.2$ is used.

The influence of both cell size and position is examined using 20 different cell sizes $\{\epsilon \in {\mathbb{R}} \ | \ 0.01\leq \epsilon \leq 0.2\}$ at 16 different cell positions {  $\{(x, y) \in \mathbb{R} \ | \ -0.25\epsilon\leq x \leq 0.5\epsilon, \ -0.25\epsilon\leq y \leq 0.5\epsilon\}$ as illustrated in Figure~\ref{fig:cellSize}}. This results in a total of 320 test cases that are used for the buckling post validation. The first buckling mode of four selected cell sizes are presented in Figure~\ref{fig:LShape:DeHomogenized}.
\newcommand\Myscale{0.5}
\newcommand\MyVSpace{-0.052}
\begin{figure}
     \centering
     \begin{subfigure}[b]{0.33\textwidth}
         \centering
         \includegraphics[scale=\Myscale]{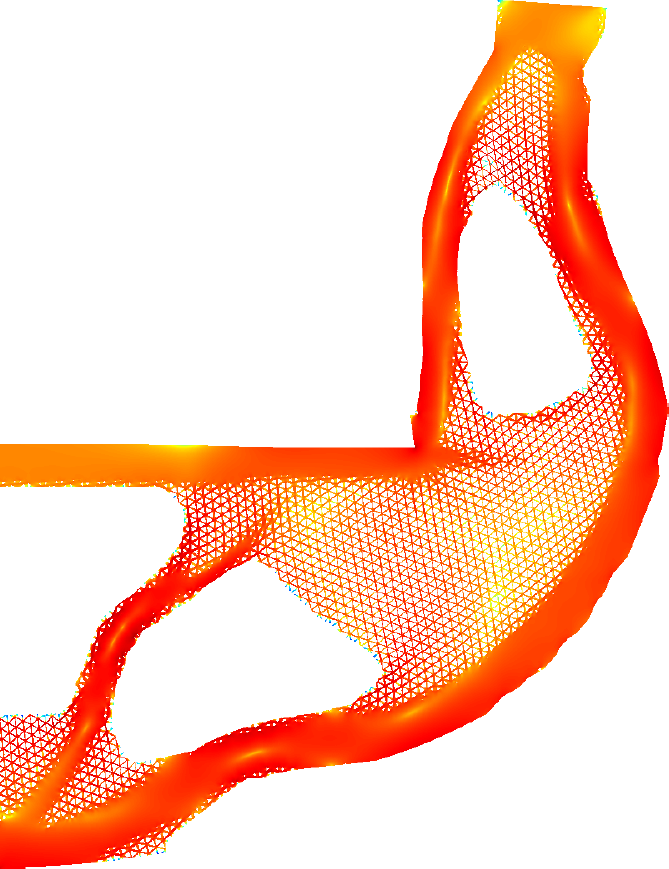}
         \vspace*{\MyVSpace cm} 
         \caption{$\epsilon = \CellSizeA$, $\min(\lambda_1) = \lambdaMultiDeHomAMin$}
         \label{fig:LShape:DeHomogenized:1}
     \end{subfigure}
     \hfill
     \begin{subfigure}[b]{0.33\textwidth}
         \centering
         \includegraphics[scale=\Myscale]{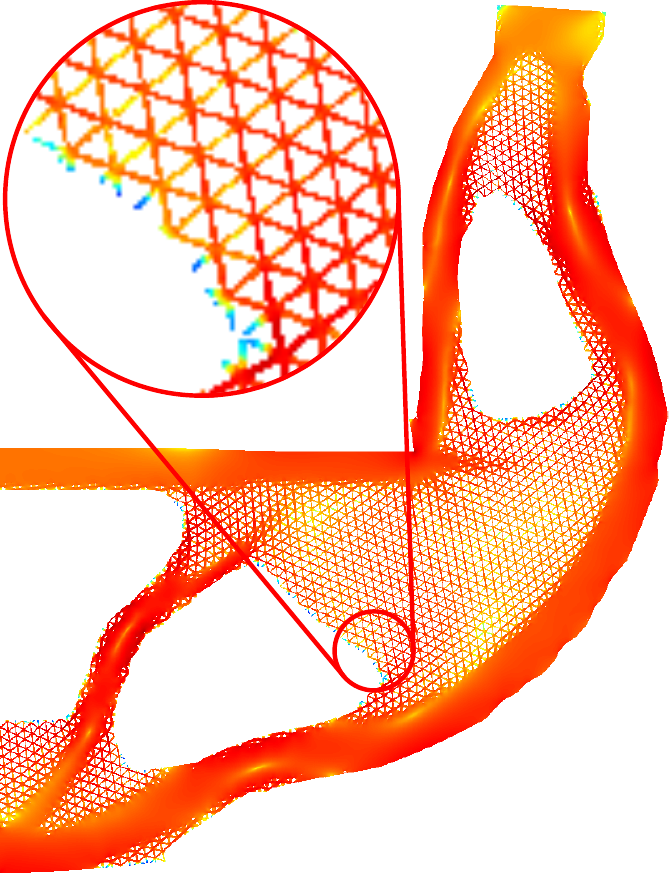}
         \vspace*{\MyVSpace cm}    
         \caption{$\epsilon = \CellSizeA$, $\lambda_1 = \lambdaMultiDeHomAUser$}
         \label{fig:LShape:DeHomogenized:2}
     \end{subfigure}
     \hfill
     \begin{subfigure}[b]{0.33\textwidth}
     	\centering
         \includegraphics[scale=\Myscale]{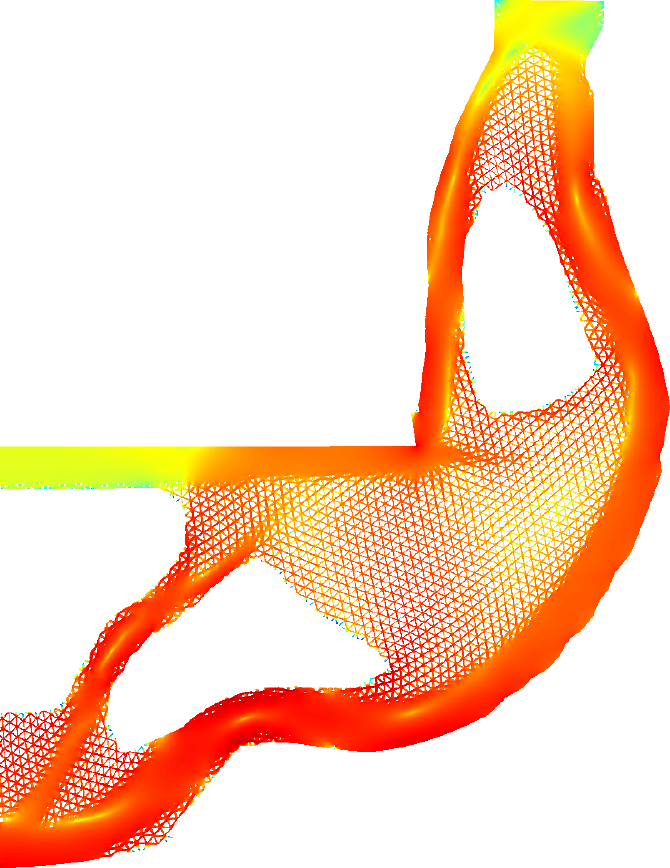}     
		 \vspace*{\MyVSpace cm}        
         \caption{$\epsilon = \CellSizeA$, $\max(\lambda_1) = \lambdaMultiDeHomAMax$}
         \label{fig:LShape:DeHomogenized:3}
     \end{subfigure}
     \\
          \begin{subfigure}[b]{0.33\textwidth}
         \centering
         \includegraphics[scale=\Myscale]{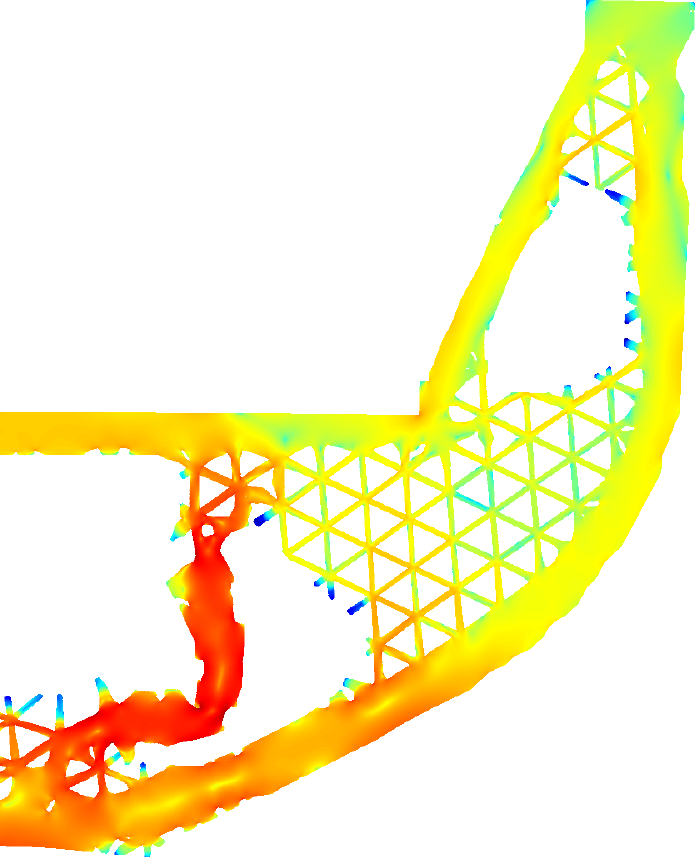}
         \vspace*{\MyVSpace cm} 
         \caption{$\epsilon = \CellSizeC$, $\min(\lambda_1) = \lambdaMultiDeHomCMin$}
         \label{fig:LShape:DeHomogenized:7}
     \end{subfigure}
     \hfill
     \begin{subfigure}[b]{0.33\textwidth}
         \centering
         \includegraphics[scale=\Myscale]{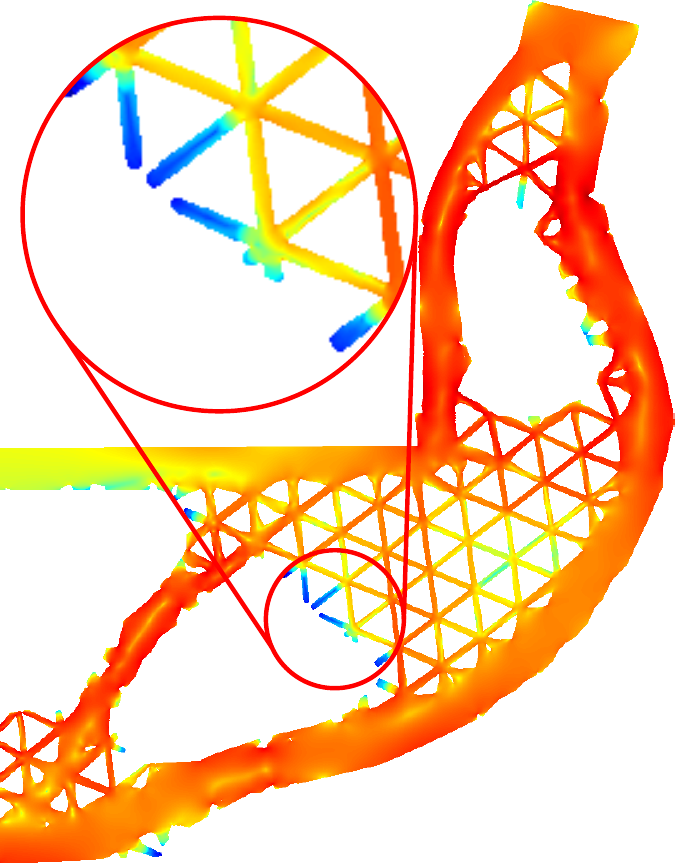}   
         \vspace*{\MyVSpace cm}
         \caption{$\epsilon = \CellSizeC$, $\lambda_1 = \lambdaMultiDeHomCUser$}
         \label{fig:LShape:DeHomogenized:8}
     \end{subfigure}
     \hfill
     \begin{subfigure}[b]{0.33\textwidth}
     	\centering
         \includegraphics[scale=\Myscale]{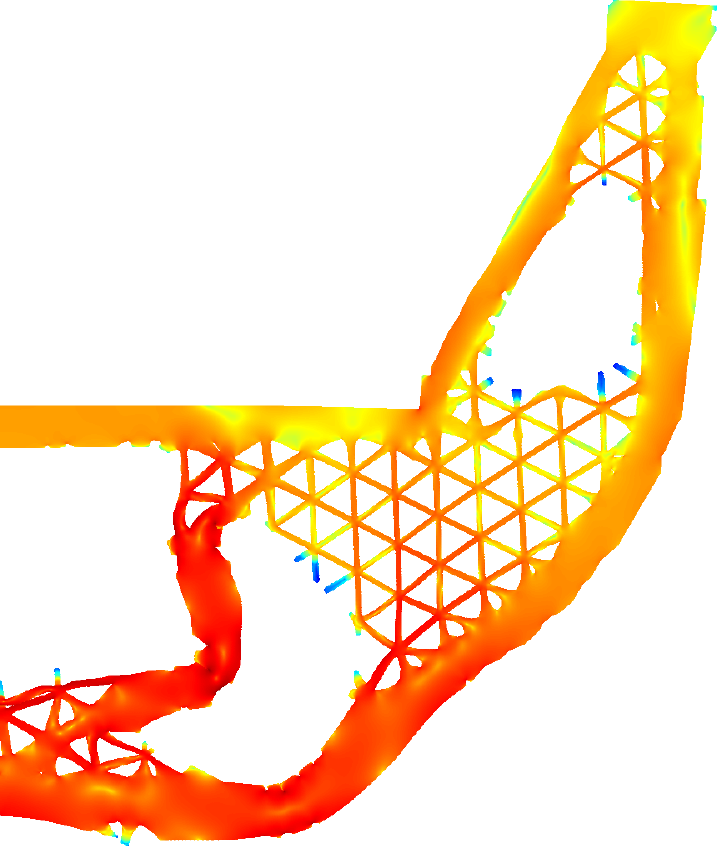} 
         \vspace*{\MyVSpace cm}    
         \caption{$\epsilon = \CellSizeC$, $\max(\lambda_1) = \lambdaMultiDeHomCMax$}
         \label{fig:LShape:DeHomogenized:9}
     \end{subfigure}
     \\
     \begin{subfigure}[b]{0.33\textwidth}
         \centering
         \includegraphics[scale=\Myscale]{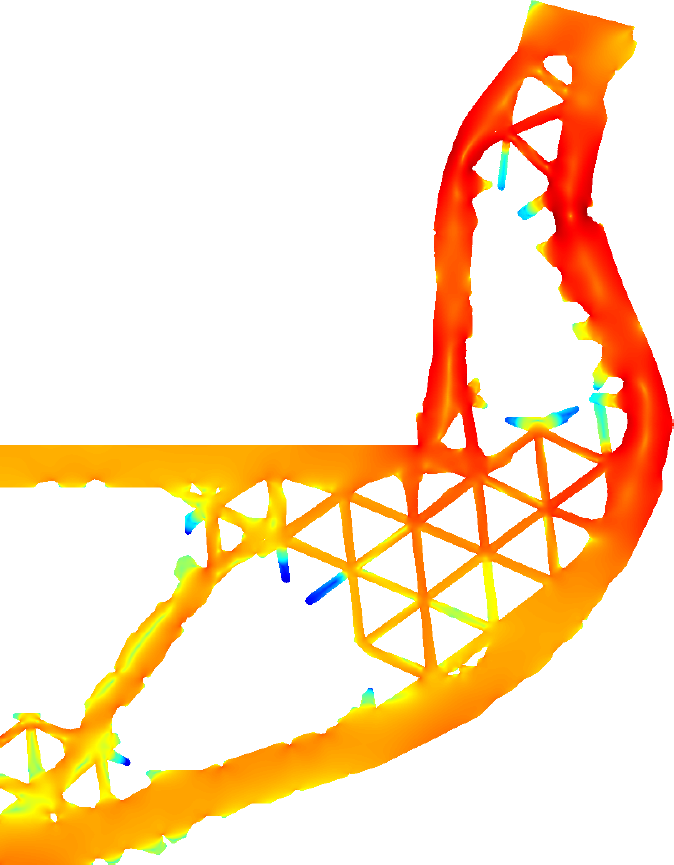}
         \vspace*{\MyVSpace cm}
         \caption{$\epsilon = \CellSizeB$, $\min(\lambda_1) = \lambdaMultiDeHomBMin$}
         \label{fig:LShape:DeHomogenized:4}
     \end{subfigure}
     \hfill
     \begin{subfigure}[b]{0.33\textwidth}
         \centering
         \includegraphics[scale=\Myscale]{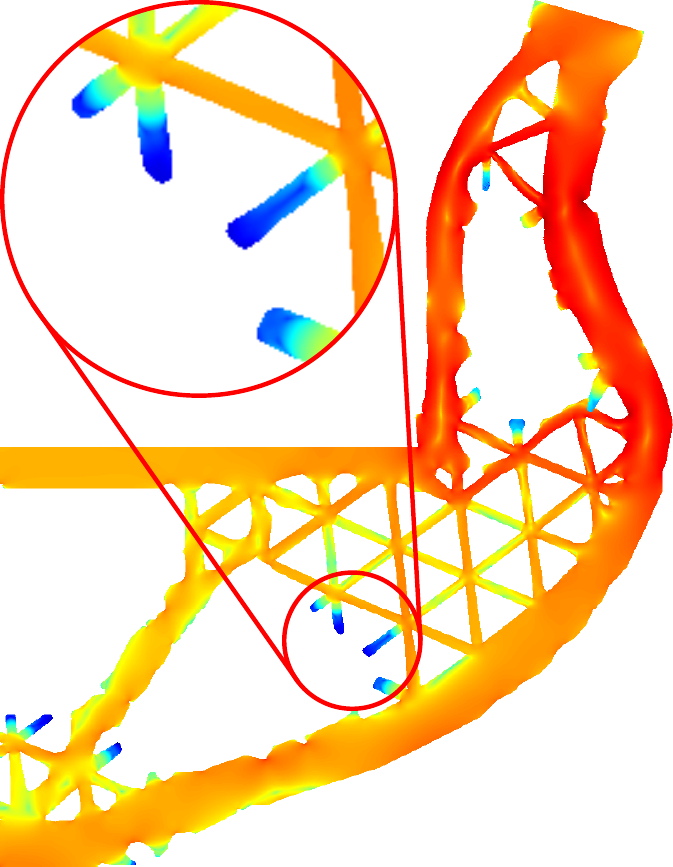}   
         \vspace*{\MyVSpace cm}
         \caption{$\epsilon = \CellSizeB$, $\lambda_1 = \lambdaMultiDeHomBUser$}
         \label{fig:LShape:DeHomogenized:5}
     \end{subfigure}
     \hfill
     \begin{subfigure}[b]{0.33\textwidth}
     	\centering
         \includegraphics[scale=\Myscale]{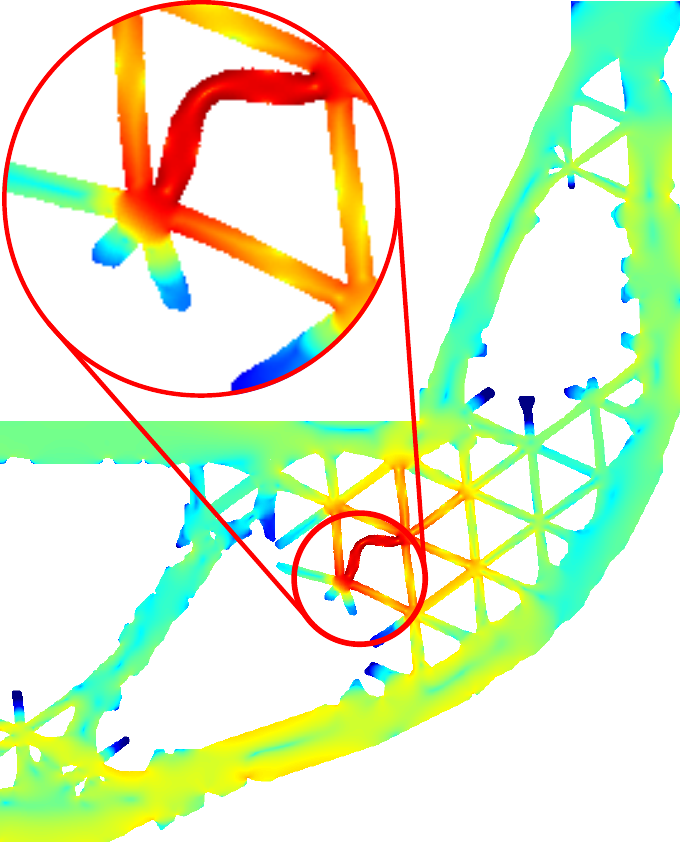}
         \vspace*{\MyVSpace cm}     
         \caption{$\epsilon = \CellSizeB$, $\max(\lambda_1) = \lambdaMultiDeHomBMax$}
         \label{fig:LShape:DeHomogenized:6}
     \end{subfigure}
     \\
     \begin{subfigure}[b]{0.33\textwidth}
         \centering
         \includegraphics[scale=0.5]{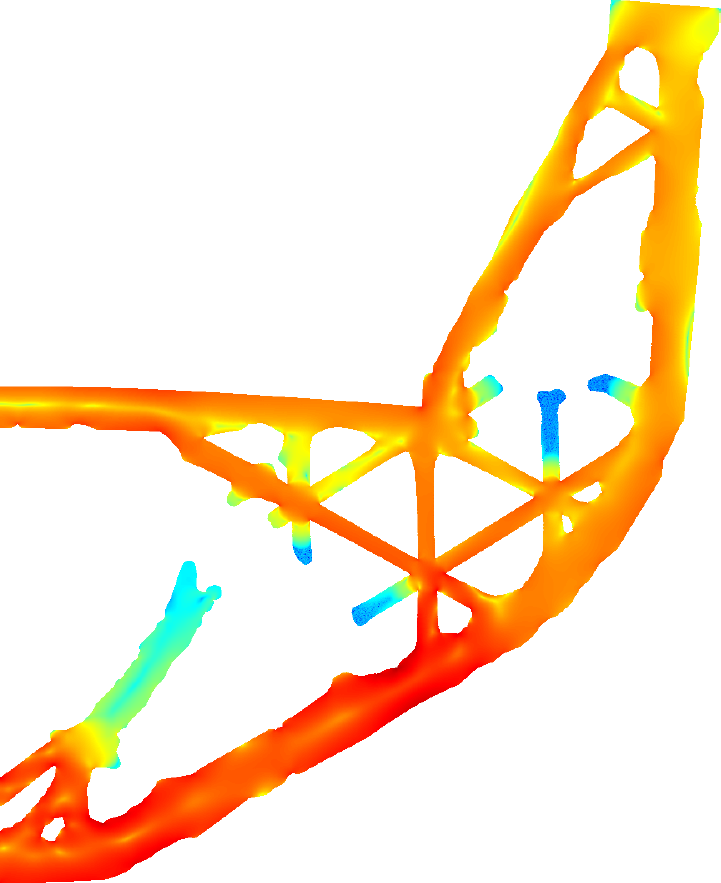}
         \includegraphics[scale=1]{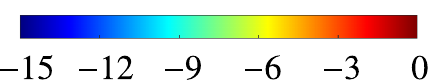} 
         \vspace*{\MyVSpace cm} 
         \caption{$\epsilon = \CellSizeD$, $\min(\lambda_1) = \lambdaMultiDeHomDMin$}
         \label{fig:LShape:DeHomogenized:10}
     \end{subfigure}
     \hfill
     \begin{subfigure}[b]{0.33\textwidth}
         \centering
         \includegraphics[scale=0.5]{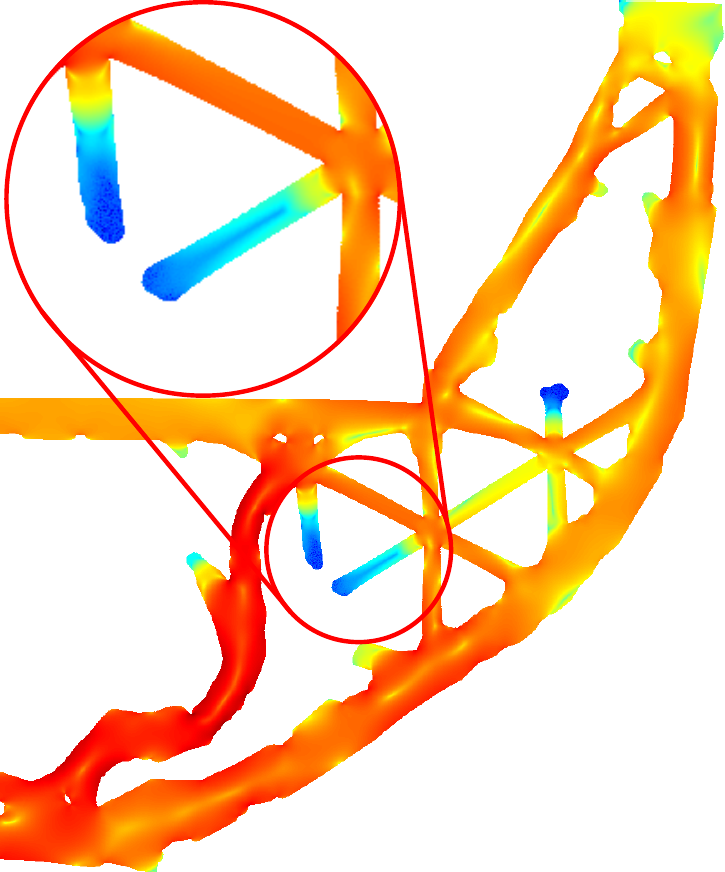} 
         \includegraphics[scale=1]{colorBarJetHorizontal_converted.pdf}          
         \vspace*{\MyVSpace cm} 
         \caption{$\epsilon = \CellSizeD$, $\lambda_1 = \lambdaMultiDeHomDUser$}
         \label{fig:LShape:DeHomogenized:11}
     \end{subfigure}
     \hfill
     \begin{subfigure}[b]{0.33\textwidth}
     	\centering
         \includegraphics[scale=0.5]{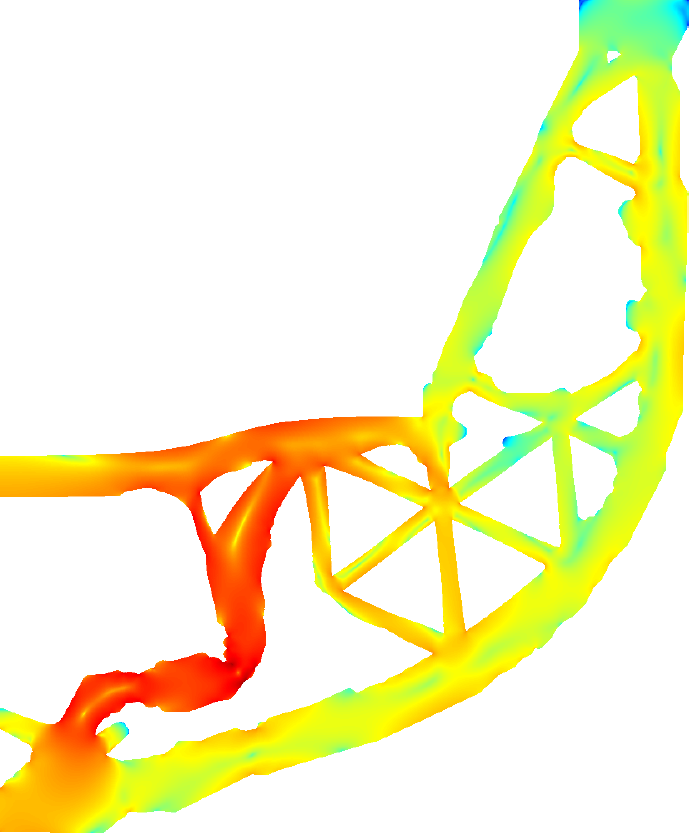} 
         \includegraphics[scale=1]{colorBarJetHorizontal_converted.pdf}     
         \vspace*{\MyVSpace cm}
         \caption{$\epsilon = \CellSizeD$, $\max(\lambda_1) = \lambdaMultiDeHomDMax$}
         \label{fig:LShape:DeHomogenized:12}
     \end{subfigure}
        \caption{Comparison of results of different cell sizes and positions used for de-homogenization of the multiscale design. The left column show $\min(\lambda_1)$ for the selected cell size. The right column shows $\max(\lambda_1)$ for the same cell size. The center column shows the design with $\lambda_1$ closest to $\text{mean}(\lambda_1)$. The color scale illustrates the normalized strain energy density $\log(\phi_j/\phi_{max})$.}
        \label{fig:LShape:DeHomogenized}
\end{figure}
For each cell size the minimum and maximum of the first buckling mode as well as the buckling mode with $\lambda_1$ closest to the average are presented.


Common for the structures using $\epsilon = \CellSizeA$ is that the BLFs are all within a very tight range. Furthermore, the buckling modes are very similar for all the cell positions. {Comparing the buckling modes to those of the homogenized design it is seen that they do not correspond exactly to the first buckling mode. The upper part of the structure experiences similar buckling compared to the homogenized design. However, the de-homogenized design also experiences buckling of the two bars in the lower left part of the structure. The average BLF of the de-homogenized structures is $\lambda_1 = \lambdaMultiDeHomAMean$. This is slightly higher than the BLF of the homogenized design which was $\lambda_1 = \lambdaMultiHomA$. The reason for not seeing the same first buckling mode in the de-homogenized designs as the homogenized design is most likely found in the de-homogenization of the thinner bars. A small dilation or erosion to the width of these bars will change their stiffness significantly.} Even though the buckling modes of the de-homogenized designs are not identical to those of the homogenized design the critical BLF is still higher. This means that for $\epsilon = \CellSizeA$ the de-homogenized design performs better than the homogenized design. { In general the homogenized designs should act as an upper bound. However, added volume and too stiff finite elements in the coarsely discretized microstructure as well as possible local thickening of critical features can increase the buckling stability of the de-homogenized structures.}


Looking at the coarser cell sizes reveals additional different buckling modes and an increasing range of BLF for each cell size. In general the critical BLF decreases in value resulting in a worse performance than the homogenized design. This is largely due to the size of the effective region of the microstructure being reduced. The center column of Figure~\ref{fig:LShape:DeHomogenized} highlights how smaller cell sizes are able to effectively use a larger part of the structural domain. The problem is that larger cell sizes risk producing larger free hanging members at the boundary of the microstructure. These members will not carry any load and therefore not contribute to stability. This is seen by the blue color in these members indicating low strain energy density. The worst case scenario is that the effective region of the microstructure is reduced in all direction by the width of the cell $\epsilon$. Therefore for $\epsilon \rightarrow 0$ the effective load carrying region of the de-homogenized design will be identical to the homogenized region. This effect highlights the assumption that the homogenized multiscale method only holds in the limit of infinitely periodic microstructure. However, based on the presented example it seems like a cell size of $\epsilon = 0.01$, corresponding to $1/100$ of the largest dimension of the design domain is small enough to provide consistent results with less than $4$\% deviation from the homogenized result.

In some cases the structure is subject to local buckling of the microstructure before global buckling occurs. { An example of this is presented in Figure~\ref{fig:LShape:DeHomogenized:6}. Here buckling occurs in a single bar in the microstructure at the BLF $\lambda_1 = \lambdaMultiDeHomBMax$.} The reason for this local buckling can be found in the open microstructure at the boundary next to the bar experiencing local buckling. The fact that the microstructure is incomplete weakens the surrounding microstructure. This leads to local buckling even though the homogenized local buckling stress constraint is satisfied.

As discussed earlier there is a risk of having disconnected structures when $\epsilon$ is increased to much. An example of this using $\epsilon = 0.15$ was shown in Figure~\ref{fig:LShape:Dehomogenization:3} and the critical buckling mode for that design is presented in Figure~\ref{fig:LShape:DeHomogenized:10}. The structure has a critical BLF of $\lambda_1 = \lambdaMultiDeHomDMin$ which is even lower than the buckling optimized single scale design. The color scheme clearly visualizes how the transverse bar is not carrying any load and therefore not contributing to the stability.

\begin{figure}
     \centering
     \begin{subfigure}[]{1\textwidth}
         \centering
         \includegraphics[scale=1]{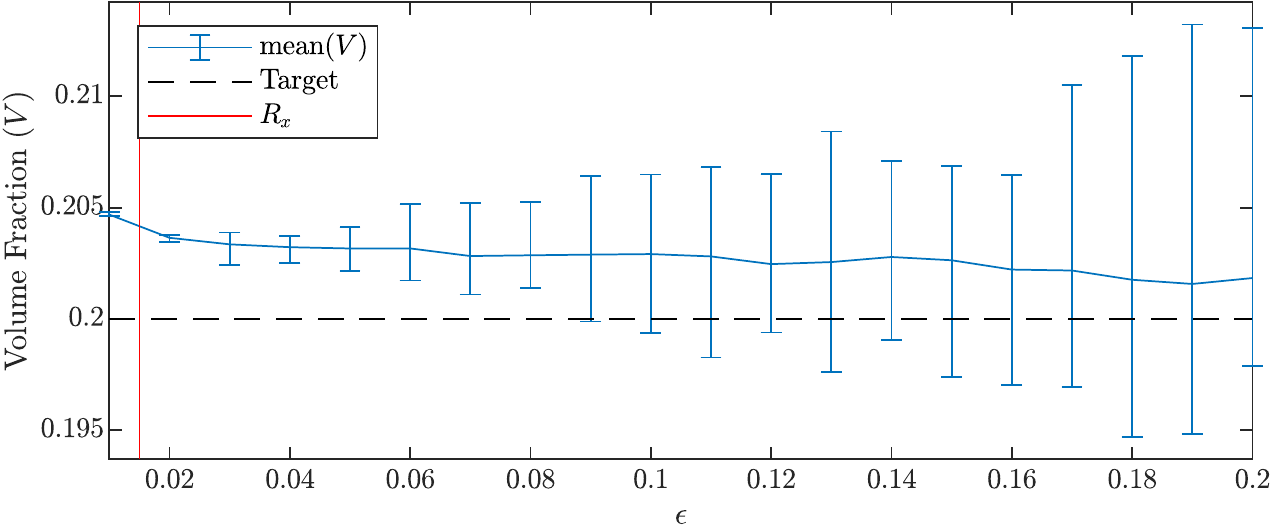}
         \caption{}
         \label{fig:LShape:DeHomogenized_Convergence:1}
     \end{subfigure}
     \\
     \begin{subfigure}[]{1\textwidth}
         \centering
         \includegraphics[scale=1]{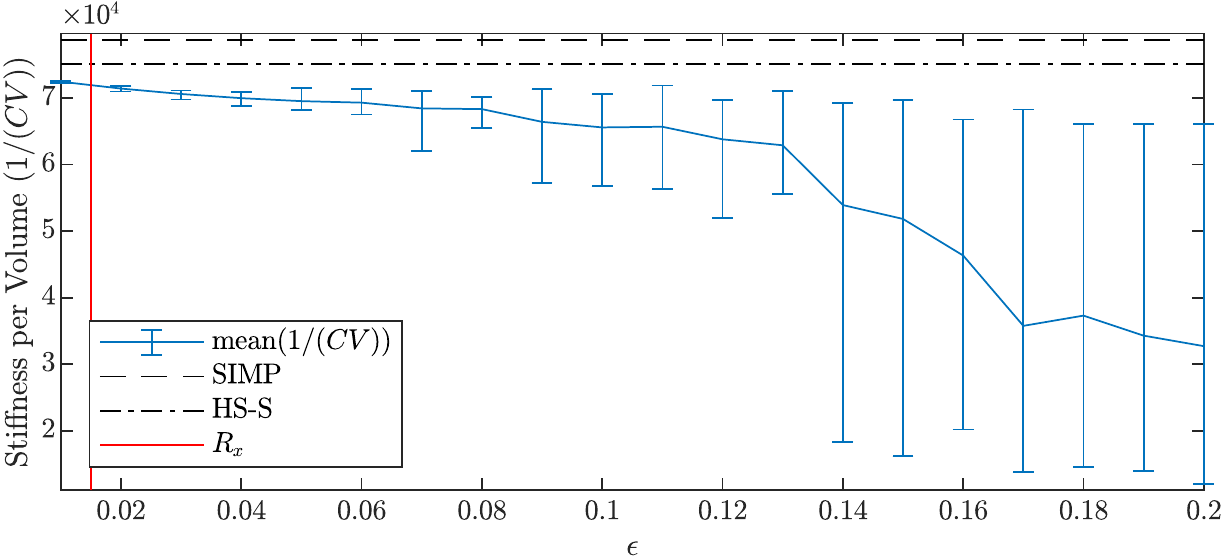}
         \caption{}
         \label{fig:LShape:DeHomogenized_Convergence:2}
     \end{subfigure}
     \\
     \begin{subfigure}[]{1\textwidth}
         \centering
         \includegraphics[scale=1]{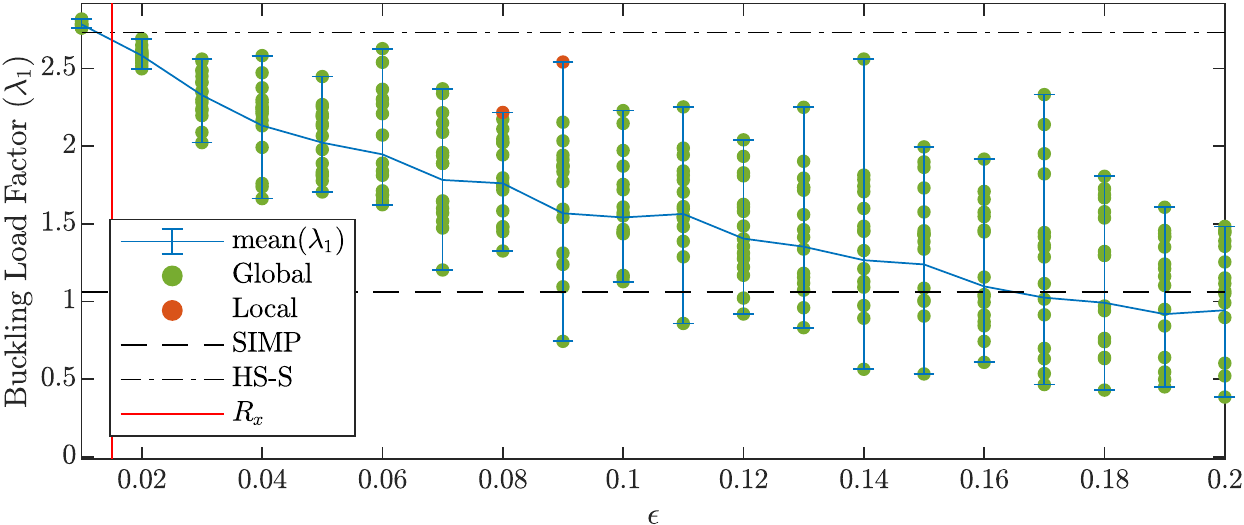}
         \caption{}
         \label{fig:LShape:DeHomogenized_Convergence:3}
     \end{subfigure}
        \caption{Convergence of the volume fraction, stiffness per volume and BLF for the de-homogenized structures relative to the cell size. SIMP and HS-S mark the values of the homogenized designs. The upper bound for the volume fraction on the \textit{blue print} design is marked as Target:  
(\subref{fig:LShape:DeHomogenized_Convergence:1}) Volume fraction,
(\subref{fig:LShape:DeHomogenized_Convergence:2}) Stiffness per volume,
(\subref{fig:LShape:DeHomogenized_Convergence:3}) BLF with colors indicating global (\textit{green}) and local (\textit{red}) buckling modes.}
        \label{fig:LShape:DeHomogenized_Convergence}
\end{figure}

All the average values of the volume fractions, stiffness per volume and BLFs relative to the cell size are presented in Figure~\ref{fig:LShape:DeHomogenized_Convergence}.
The plots also indicate the minimum and maximum values obtained using the different cell positions at each cell size. The volume fractions of the de-homogenized designs are compared to the target for the homogenized design. The stiffness per volume and buckling values of the de-homogenized designs are compared to the single- and multiscale homogenized \textit{blue print} designs. The comparison of the volume fraction in Figure~\ref{fig:LShape:DeHomogenized_Convergence:1} shows that the de-homogenized volumefraction $V_b^*$ in general is higher than the target volume fraction on the homogenized design. As expected, the volume fraction varies increasingly when cell sizes are increased. 

To filter the effect of volume change out of the discussion, we also compute the stiffness per volume ratio as presented in Figure~\ref{fig:LShape:DeHomogenized_Convergence:2}. Not unexpected, the general tendency is that structures get softer when cell sizes increase. For the finest microstructure using $\epsilon = \CellSizeA$ the variation is very small and the  average value $\text{mean}(1/(CV)) = \cMultiDeHomAMean$ is very close to that of the homogenized design $1/(C^b V) = \CMultiBuckHomB$ with a difference of only \pgfmathprintnumber[precision=2]{\HomtoDehomA}\%.

The BLFs are presented in Figure~\ref{fig:LShape:DeHomogenized_Convergence:3}. Colored dots represent the individual critical BLFs { visually identified as global buckling (\textit{green}) or local buckling only present in the microstructure (\textit{red}).} The figure shows that for $\epsilon = \CellSizeA$ only global buckling modes are present. However for {$\epsilon = [0.08,0.09]$} local buckling modes occur at some cell positions. For $\epsilon \geq 0.1$ only global buckling modes are present. This is partly because the length scale of the microstructure enters global length scale and partly because the microstructure is so poorly connected that only solid regions are left to buckle. Figure~\ref{fig:LShape:DeHomogenized_Convergence:3} also indicates that for increasing cell sizes the BLF varies more between individual cell positions. This demonstrates that the homogenized multiscale method holds, not only in the limit of infinite periodicity, but at least up to the length scale imposed via the filter radius $R_x \approx \epsilon = 0.015$ on the homogenized infill density as indicated by the red line in Figure~\ref{fig:LShape:DeHomogenized_Convergence}. 

Based on Figure~\ref{fig:LShape:DeHomogenized} the cell size $\epsilon$ should be equal to the element {size} used for the homogenized optimization. 
Alternatively, it should be equivalent to the length scale {imposed on the density field $\tilde{\boldsymbol{x}}$} used for the homogenized optimization, but this is yet to be investigated. 

\subsubsection{Closing the Boundary Shell}
\begin{figure}
\centering
\includegraphics[scale=1]{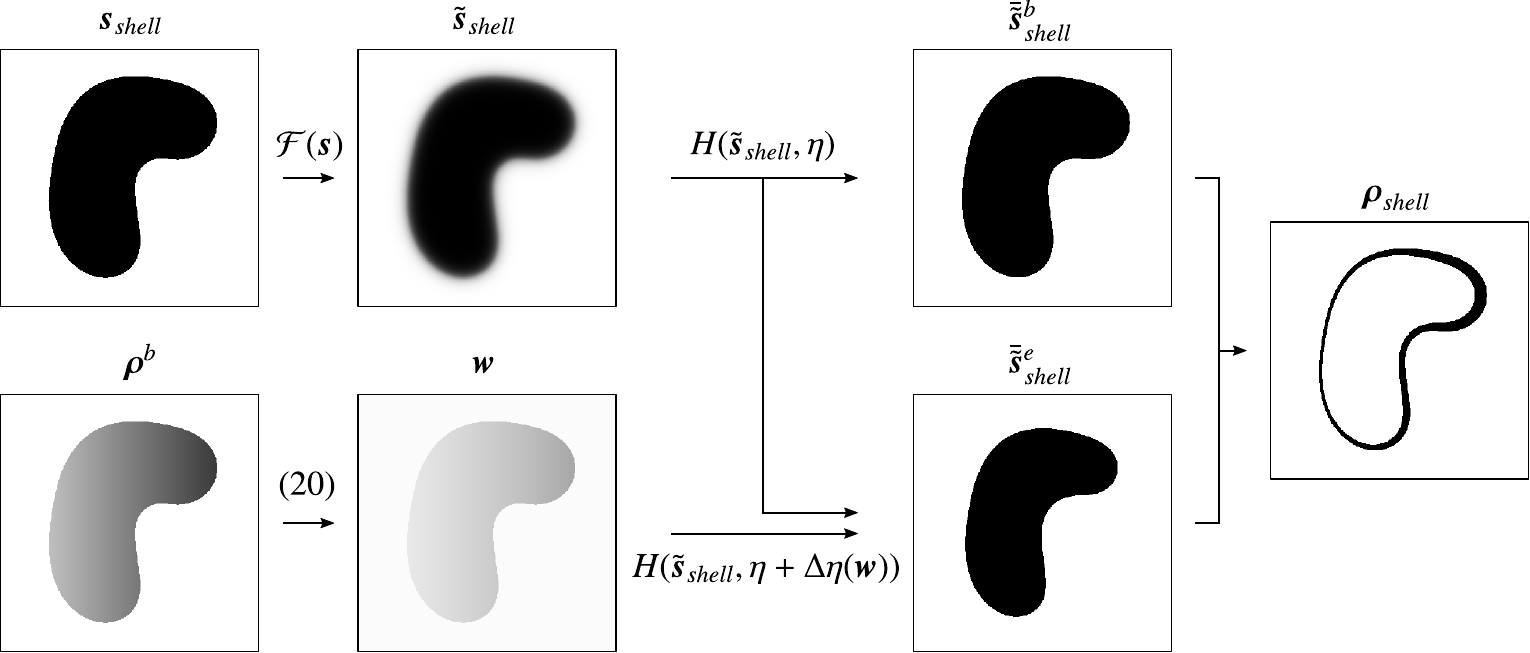}
\caption{Illustration of the erosion-based interface detection method used to generate the shells with variable thickness.}
\label{fig:ShellDerivation}
\end{figure}

The previous section showed instabilities mainly related to boundary effects at void interfaces due to incomplete microstructures. One way to overcome this problem and make the de-homogenized structures more stable and independent of the cell size and position is to close the microstructure at void interfaces. This is now done using an erosion-based interface detection \cite{Luo2019} with a local erosion threshold parameter $\Delta\eta_j$
\begin{equation}
\Delta\eta_j = \dfrac{1}{2} - \dfrac{1}{2} e^{-\dfrac{2 w_j \sqrt{3}}{R_{shell}}},
\end{equation}
where $w_j$ is the width of the shell and $R_{shell}$ is the filter radius in the standard filter. In this work the PDE filter based on a Helmholtz-type partial differential equation \cite{Lazarov2011} is used to filter the indicator field $\boldsymbol{s}_{shell} = \bar{\tilde{\boldsymbol{s}}}^b$
\begin{equation}
-r_{shell}^2\nabla^2\tilde{\boldsymbol{s}}_{shell} +\tilde{\boldsymbol{s}}_{shell} = \boldsymbol{s}_{shell}.
\end{equation}
The relation between $R_{shell}$ and $r_{shell}$ is
\begin{equation}
r_{shell} = \dfrac{R_{shell}}{2\sqrt{3}}.
\end{equation}
The shell width $w_j$ is determined locally according to the densities using \eqref{eq:shellWidth}.
%
%
The shell emerges from the difference between the eroded and non-eroded indicator fields
\begin{equation}
\boldsymbol{\rho}_{shell} = {H}(\tilde{\boldsymbol{s}}_{shell},\eta) - {H}(\tilde{\boldsymbol{s}}_{shell},\eta+\Delta\boldsymbol{\eta})
\end{equation}
where $\eta = 0.5$ and ${H}$ is a sharp heaviside step function. An illustration of the erosion-based interface identification is presented in Figure~\ref{fig:ShellDerivation}.

\begin{figure}
     \centering
     \begin{subfigure}[t]{0.24\textwidth}
         \centering
         \includegraphics[width=1\textwidth,angle=90,origin=c]{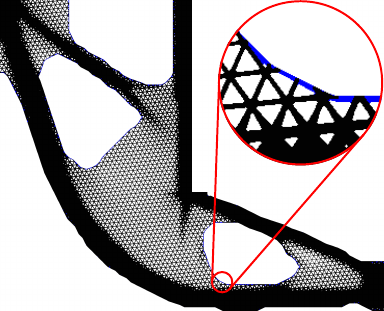}
         \caption{$\epsilon = 0.01$}
         \label{fig:LShape:DehomogenizationShell:1}
     \end{subfigure}
     \begin{subfigure}[t]{0.24\textwidth}
     	\centering
         \includegraphics[width=1\textwidth,angle=90,origin=c]{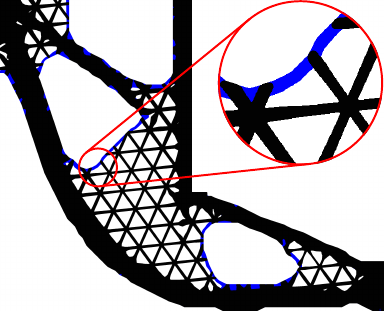}       
         \caption{$\epsilon = 0.05$}
         \label{fig:LShape:DehomogenizationShell:2}
     \end{subfigure}
     \begin{subfigure}[t]{0.24\textwidth}
         \centering
         \includegraphics[width=1\textwidth,angle=90,origin=c]{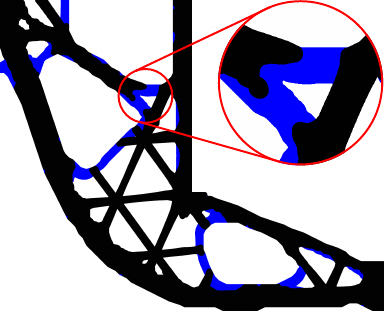}   
         \caption{$\epsilon = 0.15$}
         \label{fig:LShape:DehomogenizationShell:3}
     \end{subfigure}
     \begin{subfigure}[t]{0.24\textwidth}
         \centering
         \includegraphics[width=1\textwidth,angle=90,origin=c]{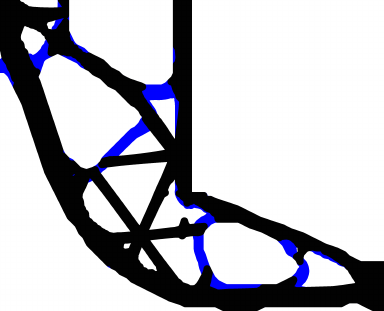}   
         \caption{$\epsilon = 0.2$}
         \label{fig:LShape:DehomogenizationShell:4}
     \end{subfigure}
        \caption{De-homogenization for different unit cell sizes $\epsilon$ with shells used to close off the microstructure. Black indicates the original structure while blue color represents the added shells.}
        \label{fig:LShape:DehomogenizationShell}
\end{figure}

The method is applied to the de-homogenized structures presented in the previous section. Illustrations of the structures shown in Figure~\ref{fig:LShape:Dehomogenization} with added shells are presented in Figure~\ref{fig:LShape:DehomogenizationShell}.
Here the original structure is black and the added shells are shown in blue. The structure using the smallest cell size i.e. $\epsilon = 0.01$, only has a very thin shell added. As a result the volume of the structure is only increased very little {i.e. 0.73\% at average}. However, for increasing cell sizes the thickness of the shell increases and the added shell has a larger impact on the volume of the structure. 

The case using $\epsilon = 0.15$ presented in Figure~\ref{fig:LShape:Dehomogenization:3} had a region where the structure was disconnected prior to the addition of the shell. This resulted in very poor stiffness and buckling performance as shown in Figure~\ref{fig:LShape:DeHomogenized:10} and \ref{fig:LShape:DeHomogenized_Convergence}. The addition of the shell helps reconnect the otherwise disconnected region as shown in Figure~\ref{fig:LShape:DehomogenizationShell:3}. This correction should result in increased stiffness and buckling stability.

The free hanging members, which were not contributing to neither stiffness nor stability before the shell was added, are now all connected to the shells. This helps activate them such that they can contribute to the performance of the structure. Especially for larger cell sizes, as illustrated in Figure~\ref{fig:LShape:DehomogenizationShell:3} and \ref{fig:LShape:DehomogenizationShell:4}, this will reactivate a large region of the microstructure and thereby have a significant impact on the stiffness and buckling stability.

\begin{figure}
     \centering
     \begin{subfigure}[b]{0.3\textwidth}
         \centering
         \includegraphics[scale=0.5]{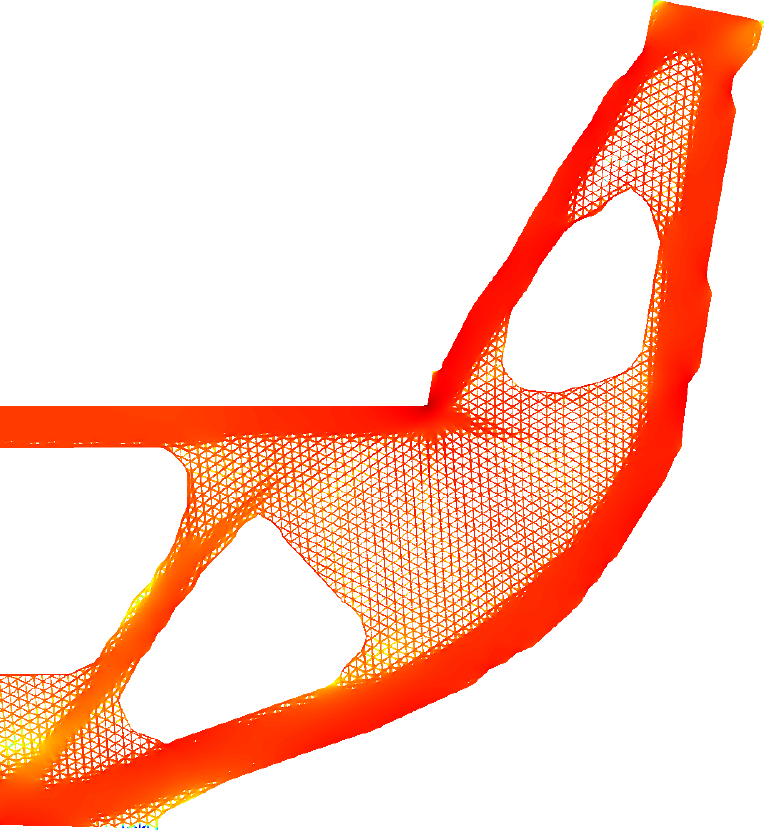}
         \caption{$\epsilon = \CellSizeAShell$, $\lambda_1 = \lambdaMultiDeHomAShell$}
         \label{fig:LShape:DeHomogenized_Shell:1}
     \end{subfigure}
     \hfill
     \begin{subfigure}[b]{0.3\textwidth}
     	\centering
         \includegraphics[scale=0.5]{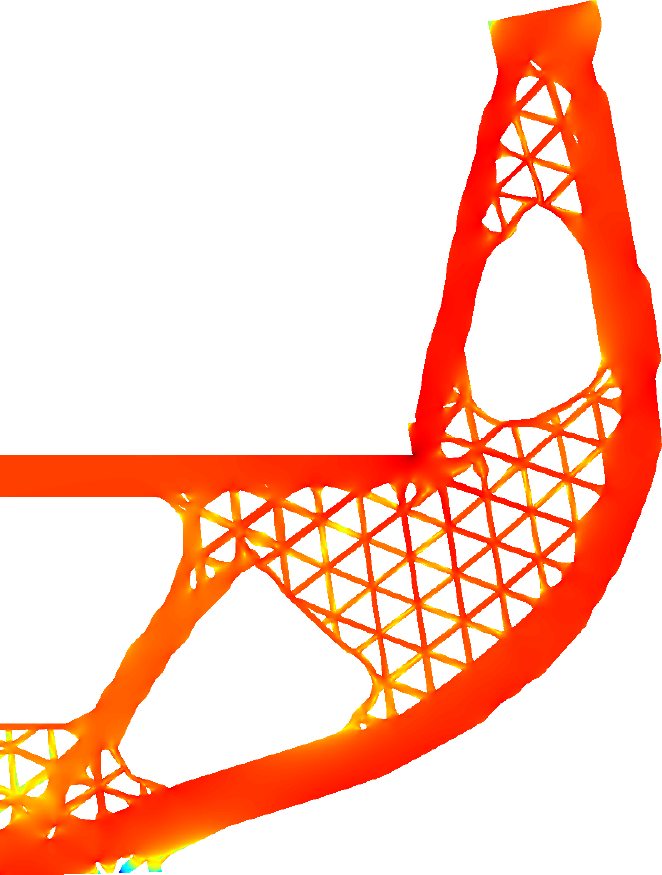}     
         \caption{$\epsilon = \CellSizeCShell$, $\lambda_1 = \lambdaMultiDeHomCShell$}
         \label{fig:LShape:DeHomogenized_Shell:3}
     \end{subfigure}
     \hfill
     \begin{subfigure}[b]{0.3\textwidth}
         \centering
         \includegraphics[scale=0.5]{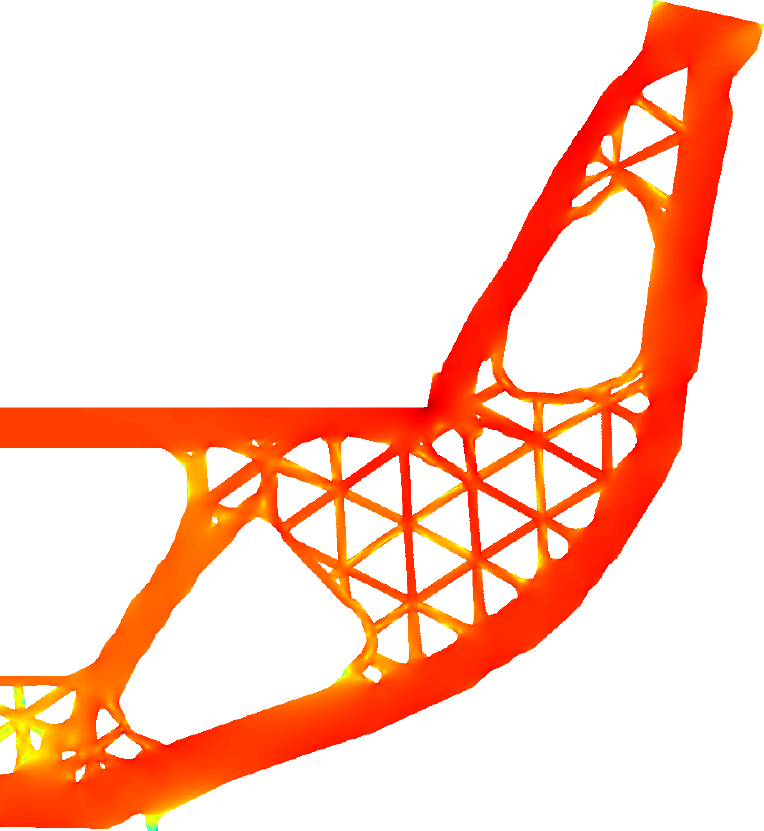}   
         \caption{$\epsilon = \CellSizeBShell$, $\lambda_1 = \lambdaMultiDeHomBShell$}
         \label{fig:LShape:DeHomogenized_Shell:2}
     \end{subfigure}
     \\
     \begin{subfigure}[b]{0.3\textwidth}
         \centering
         \includegraphics[scale=0.5]{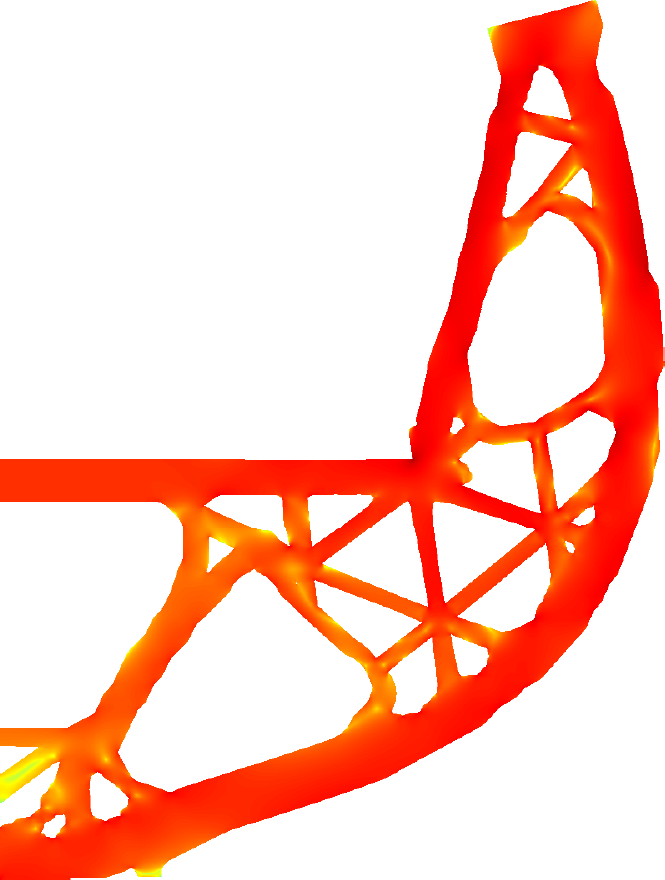}
         \includegraphics[scale=1]{colorBarJetHorizontal_converted.pdf} 
         \caption{$\epsilon = \CellSizeDShell$, $\lambda_1 = \lambdaMultiDeHomDShell$}
         \label{fig:LShape:DeHomogenized_Shell:4}
     \end{subfigure}
     \hfill
     \begin{subfigure}[b]{0.3\textwidth}
         \centering
         \includegraphics[scale=0.5]{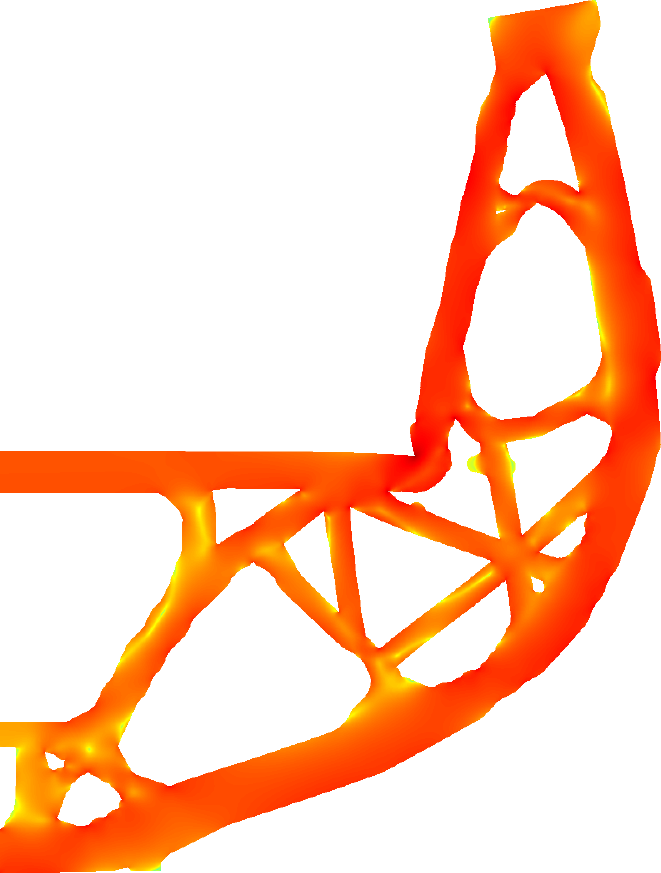}  
         \includegraphics[scale=1]{colorBarJetHorizontal_converted.pdf}  
         \caption{$\epsilon = \CellSizeEShell$, $\lambda_1 = \lambdaMultiDeHomEShellMin$}
         \label{fig:LShape:DeHomogenized_Shell:5}
     \end{subfigure}
     \hfill
     \begin{subfigure}[b]{0.3\textwidth}
     	\centering
         \includegraphics[scale=0.5]{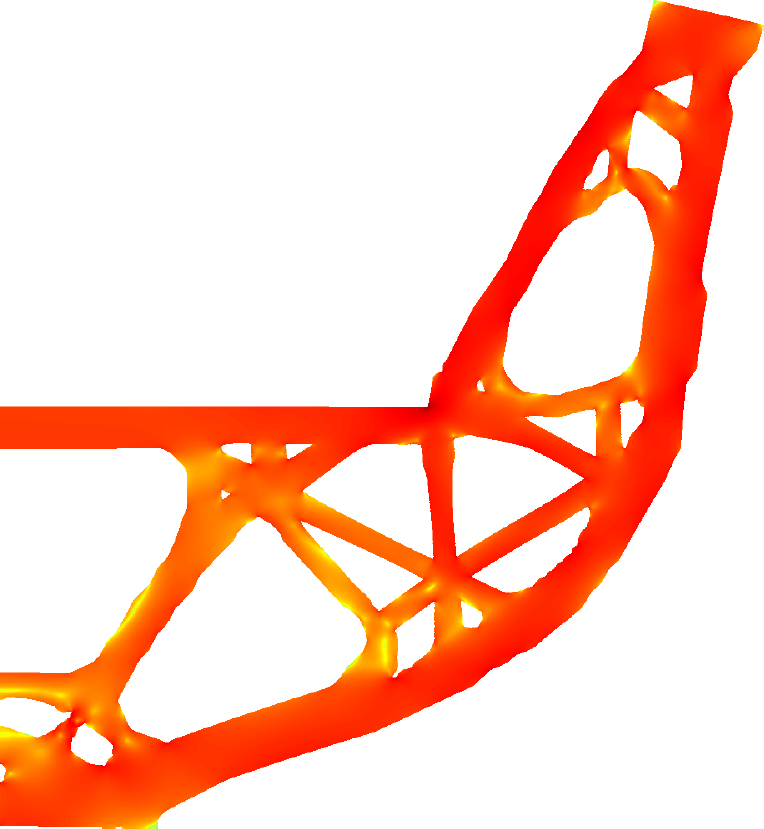}  
         \includegraphics[scale=1]{colorBarJetHorizontal_converted.pdf}    
         \caption{$\epsilon = \CellSizeEShell$, $\lambda_1 = \lambdaMultiDeHomEShellMax$}
         \label{fig:LShape:DeHomogenized_Shell:6}
     \end{subfigure}
        \caption{Comparison of buckling modes obtained using different cell sizes and positions during the de-homogenization including added shell. The color scale illustrates the normalized strain energy density $\log(\phi_j/\phi_{max})$.}
        \label{fig:LShape:DeHomogenized_Shell}
\end{figure}

The same 320 test cases that were used in Section~\ref{sec:LShape:De_Homogenization} are evaluated with the de-homogenized structures where shells are added. Figure~\ref{fig:LShape:DeHomogenized_Shell} shows the first buckling mode of six selected cases using five different cell sizes.
Common for all six example is that they experience global buckling. The example in Figure~\ref{fig:LShape:DeHomogenized_Shell:1} uses $\epsilon = \CellSizeAShell$ and has a critical BLF of $\lambda_1 = \lambdaMultiDeHomAShell$ which is an \pgfmathprintnumber[precision=2]{\DeHomtoDehomShellA}\% increase compared to the same structure without the shell. The buckling mode of the structure with the shell also appears to be more global than the one for the structure without the shell. { These results are mainly due to two factors. First, the shell adds material to the structure, which increases stiffness and stability. Second, the finite element discretization of the microstructure can increase stiffness and stability artificially.}
 
{The global buckling mode presented in \ref{fig:LShape:DeHomogenized_Shell:2}, using $\epsilon = \CellSizeBShell$, is a clear testimony to the improved stability obtained using the shells.} The corresponding structure without the shell experienced local buckling of the microstructure due to incomplete microstructures resulting in weakening near void interfaces as seen in Figure~\ref{fig:LShape:DeHomogenized:6}. The added shell closes off the microstruture which can then exploit the full potential of the multiscale material all the way to the void interface. For this very reason the de-homogenized structures experience better stability and are less sensitive to the position of the microstructure.

For very coarse microstructures it was seen that a risk of having disconnected regions in the structure was introduced, see Figure~\ref{fig:LShape:Dehomogenization:3} and \ref{fig:LShape:DeHomogenized:10}. Adding the shell along void interfaces helped reconnect these regions as shown in Figure~\ref{fig:LShape:DehomogenizationShell:3}. The first buckling mode of this structure is presented in Figure~\ref{fig:LShape:DeHomogenized_Shell:4}. The mode is global and active at a BLF of $\lambda_1 = \lambdaMultiDeHomDShell$. This is a significant improvement of \pgfmathprintnumber[precision=0]{\DeHomtoDehomShellD}\% highlighting the stability gained by using the shell.

Even for the coarsest case using $\epsilon = \CellSizeEShell$, where the microstucture is almost entirely incomplete, stability is improved significantly using the shell. Figure~\ref{fig:LShape:DeHomogenized_Shell:5} and \ref{fig:LShape:DeHomogenized_Shell:6} show the buckling modes with the minimum and maximum BLFs respectively. With a minimum BLF of $\lambda_1 = \lambdaMultiDeHomEShellMin$ and a maximum BLF of $\lambda_1 = \lambdaMultiDeHomEShellMax$ the difference is only \pgfmathprintnumber[precision=1]{\ShellMinToMaxE}\%. This is not negligible but still a significant improvement compared to the structure without the shell.

\begin{figure}
     \centering
     \begin{subfigure}[]{1\textwidth}
         \centering
         \includegraphics[scale=1]{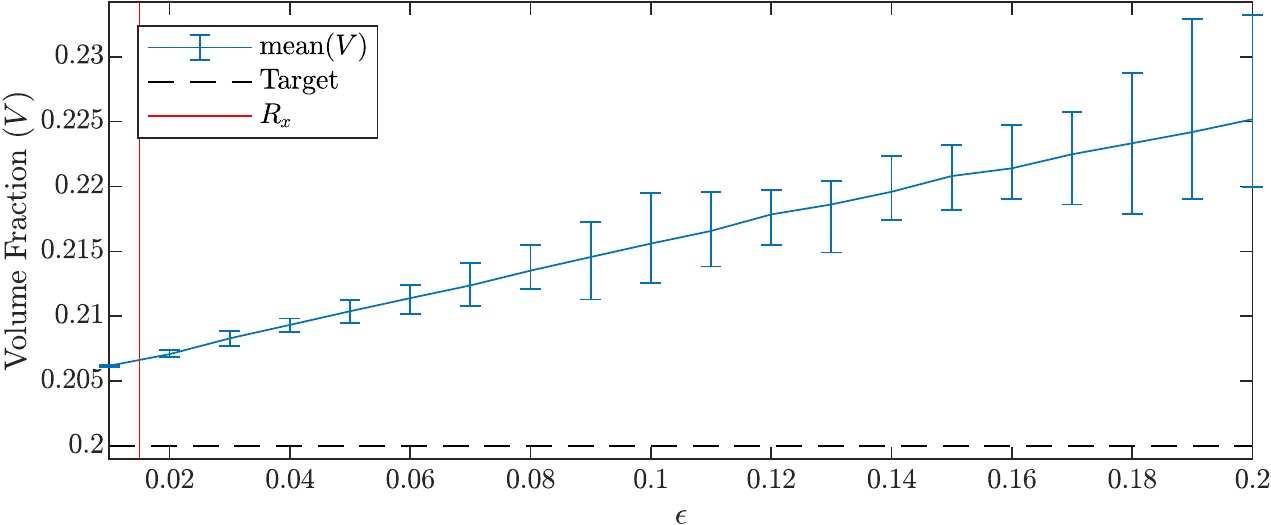}
         \caption{}
         \label{fig:LShape:DeHomogenized_Convergence_Shell:1}
     \end{subfigure}
     \\
     \begin{subfigure}[]{1\textwidth}
         \centering
         \includegraphics[scale=1]{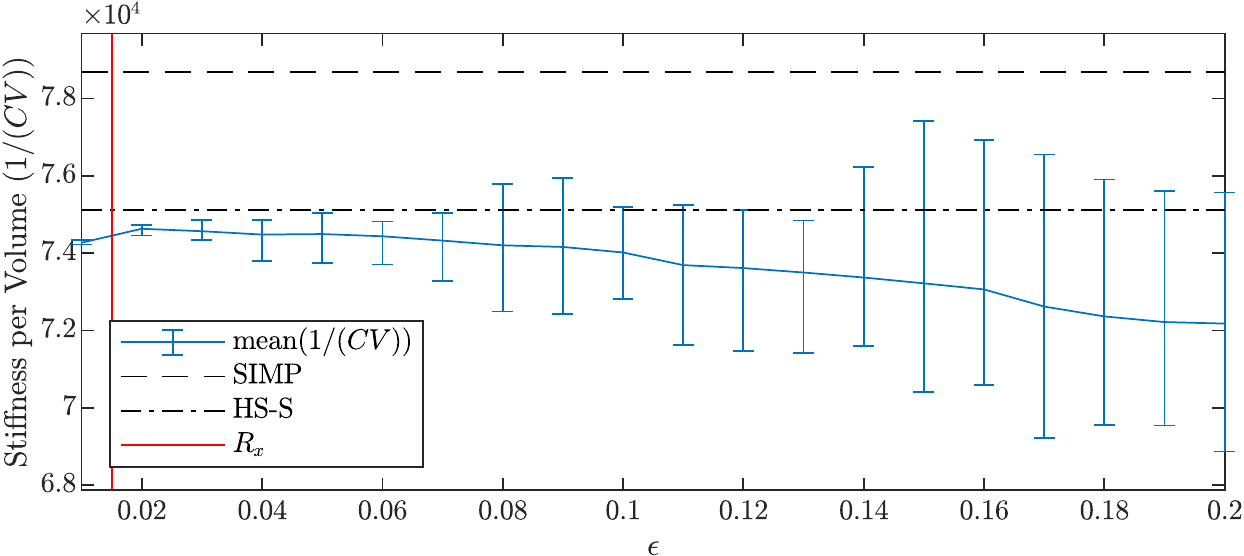}
         \caption{}
         \label{fig:LShape:DeHomogenized_Convergence_Shell:2}
     \end{subfigure}
     \\
     \begin{subfigure}[]{1\textwidth}
         \centering
         \includegraphics[scale=1]{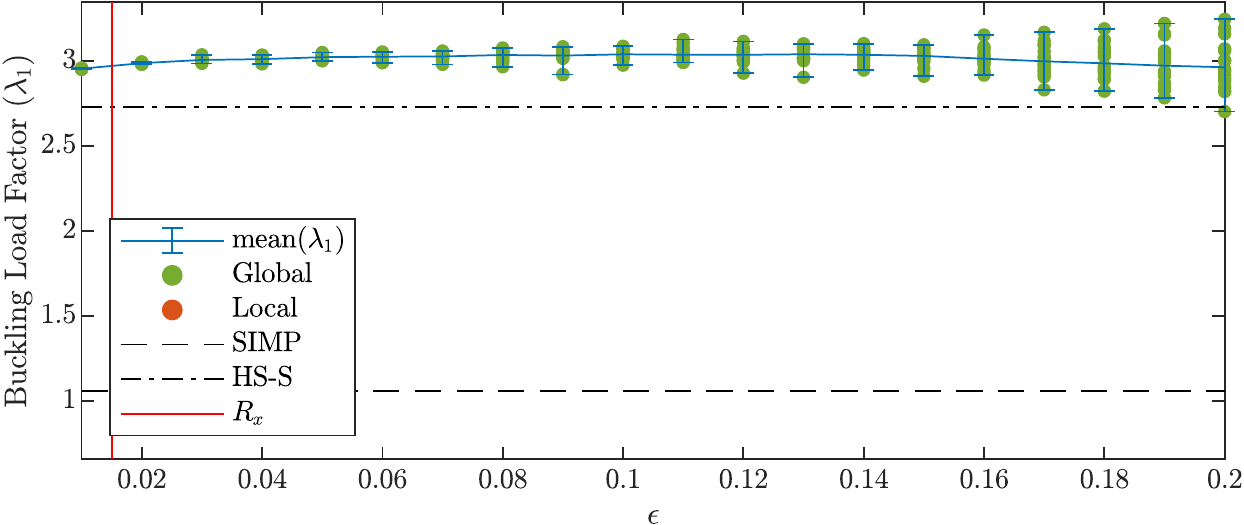}
         \caption{}
         \label{fig:LShape:DeHomogenized_Convergence_Shell:3}
     \end{subfigure}
        \caption{Convergence of the volume fraction, stiffness per volume and BLF for the de-homogenized structures using a shell to close off the microstructure. SIMP and HS-S mark the values of the homogenized designs. The upper bound for the volume fraction on the \textit{blue print} design is marked as Target:  
(\subref{fig:LShape:DeHomogenized_Convergence_Shell:1}) Volume fraction,
(\subref{fig:LShape:DeHomogenized_Convergence_Shell:2}) Stiffness per volume,
(\subref{fig:LShape:DeHomogenized_Convergence_Shell:3}) BLF with colors indicating global (\textit{green}) and local (\textit{red}) buckling modes.}
        \label{fig:LShape:DeHomogenized_Convergence_Shell}
\end{figure}

The volume fraction, stiffness per volume and BLFs for all 320 cases are summarized in Figure~\ref{fig:LShape:DeHomogenized_Convergence_Shell}.
The volume constraint is increasingly violated with increasing cell sizes as seen in Figure~\ref{fig:LShape:DeHomogenized_Convergence_Shell:1}. For the finest case the average volume fraction is $V = \volMultiDeHomAShellMean$ which is an increase of only \pgfmathprintnumber[precision=2]{\volDehomtoDehomAShell}\% compared to the structure without the shell. However, for the coarsest cell size the average volume fraction is $V = \volMultiDeHomEShellMean$ resulting in a significantly higher increase of \pgfmathprintnumber[precision=2]{\volDehomtoDehomEShell}\%. Furthermore, the variation of the volume fraction increases making it difficult to predict the volume fraction after de-homogenization.

With the added volume comes a benefit of added stiffness. However, this is only the case for cell sizes up to {$\epsilon = 0.07$} as shown in Figure~\ref{fig:LShape:DeHomogenized_Convergence_Shell:2}. For larger cell sizes the stiffness per volume measure decreases below that of the smallest cell size. Nevertheless, the average stiffness per volume never exceeds a variation above {$2.82$\%} even for the largest cell size. The variation in stiffness per volume relative to cell position increases in correlation with cell size. However, this is not as significant as it was the case when the shell was not added. 

Finally, the critical BLF is presented in Figure~\ref{fig:LShape:DeHomogenized_Convergence_Shell:3}. Compared to the results without the shell the stability is significantly improved. The average BLF is almost constant for all cell sizes. Furthermore, the variation is significantly reduced compared to Figure~\ref{fig:LShape:DeHomogenized_Convergence:3}. Looking at the critical buckling modes shows that all 320 cases experience global buckling before local. The case using $\epsilon = \CellSizeAShell$ has an average BLF of $\lambda_1 = \lambdaMultiDeHomAShellMean$ which is a \pgfmathprintnumber[precision=2]{\DeHomtoDehomShellAMean}\% improvement compared to the structure without the shell. This is a significant improvement considering the small price of only adding \pgfmathprintnumber[precision=2]{\volDehomtoDehomAShell}\% material. Furthermore, the compliance is reduced by \pgfmathprintnumber[precision=2]{\cDehomtoDehomAShell}\% meaning that there is a stiffness gain associated with the shell.

%% file: Conclusion.tex
This paper has presented an approach to obtain topology optimized {multiscale} designs that takes both local and global buckling stability into account. This was done using homogenization based topology optimization with isotropic multiscale material modeled by the Hashin-Shtrikman upper bound. Local buckling stability was ensured using a local buckling stress constraint. The constraint was based on the buckling stress limit obtained using Bloch-Floquet-based cell analysis on eight different relative cell densities. These limits where used to determine the shape of the Willam-Warnke yield surface to formulate an effective buckling stress surface. The combination of the local buckling stress surface and the global multiscale formulation provided a method capable of obtaining topology optimized designs that takes both local and global buckling stability into account. 

The method was demonstrated on two examples. First a uni-axial compression setup was evaluated using both singlescale and multiscale material. This example confirmed, what has been postulated for years, that isotropic microstructures enhance structural stability and are superior to singlescale structures when buckling is considered. The example also showed that even though stability was improved using the isotropic microstructure, there is still room for further improvements. Here the use of hierarchical microstructures with enhanced buckling stability \cite{thomsen2018a} is an obvious choice to examine.

The second example was the modified L-beam. A significant factor $2.5$ improvement of the BLF was observed when using the multiscale instead of the singlescale design. The de-homogenization of the multiscale design without adding a shell showed very good performance in compliance and BLF with only a small increase in volume fraction for the finest microstructure. However, for the larger cell sizes growing variations in performance related to cell positions were observed for both volume fraction, compliance and BLF. These instabilities were reduced significantly by adding a shell at the void interfaces to the de-homogenized designs. No local buckling modes were observed in any of the 320 test cases and the variation related to cell position was reduced drastically for the BLF. In fact, all the de-homogenized structures perform better than the homogenized design in terms of BLF. Regarding the stiffness per volume measure all cell positions for cell sizes up to {$\epsilon = 0.14$} are within $5$\% of the homogenized design. Finally, the volume fraction of the de-homogenized designs increase with cell size. However, all cell positions for cell sizes up to {$\epsilon = 0.04$, i.e. more than} twice the length scale of the infill in the homogenized design, are within $5$\% of the limit imposed on the homogenized design. Even for extremely large cells sizes, errors are moderate and confirm immense computational savings by formulating the complex buckling optimization problem as a multiscale problem that takes local buckling effects into account.

%% file: APP_Constraint.tex
The stress constraint based on the Willam-Warnke yield surface presented in Section~\ref{sec:WillamWarnkeModel} consists of several parameters. The shape of the Willam-Warnke yield surface defined in (\ref{eq:alpha}) consists of the design dependent parameters $A(\tilde{x}_j)$, $B(\tilde{x}_j)$, $C(\tilde{x}_j)$, $D(\tilde{x}_j)$ and $E(\tilde{x}_j)$ which are defined as
\begin{equation} \label{eq:ABCDE}
\begin{split}
A(\tilde{x}_j) &= \frac{4}{\sigma_c(\tilde{x}_j)} \sqrt{\frac{2}{15}} \left( r_c(\tilde{x}_j)^2 -r_t(\tilde{x}_j)^2 \right),
\quad
B(\tilde{x}_j) = \frac{1}{\sigma_c(\tilde{x}_j)} \sqrt{\frac{2}{15}} \left(r_c(\tilde{x}_j) - 2 r_t(\tilde{x}_j) \right)^2, 
\\
C(\tilde{x}_j) &= 2 r_c(\tilde{x}_j) \left( r_c(\tilde{x}_j)^2 - r_t(\tilde{x}_j)^2 \right) ,  
\quad
D(\tilde{x}_j) = 4 r_c(\tilde{x}_j)^2 \left( r_c(\tilde{x}_j) - 2 r_t(\tilde{x}_j) \right)^2 \left(r_c(\tilde{x}_j)^2 - r_t(\tilde{x}_j)^2 \right),
\\
E(\tilde{x}_j) &= r_c(\tilde{x}_j)^2 \left(r_c(\tilde{x}_j)-2 r_t(\tilde{x}_j) \right)^2 \left(5 r_t(\tilde{x}_j)^2 - 4 r_t(\tilde{x}_j) r_c(\tilde{x}_j) \right),
\end{split}
\end{equation}
where $r_c(\tilde{x}_j)$ and $r_t(\tilde{x}_j)$ are defined as
\begin{equation}
r_c(\tilde{x}_j) = \sqrt{\frac{6}{5}} \frac{\sigma_b(\tilde{x}_j) \sigma_t(\tilde{x}_j)}{3 \sigma_b(\tilde{x}_j) \sigma_t(\tilde{x}_j) + \sigma_c(\tilde{x}_j) \left(\sigma_b(\tilde{x}_j)-\sigma_t(\tilde{x}_j) \right)}, \quad
r_t(\tilde{x}_j) = \sqrt{\frac{6}{5}} \frac{\sigma_b(\tilde{x}_j) \sigma_t(\tilde{x}_j)}{\sigma_c(\tilde{x}_j) \left(2 \sigma_b(\tilde{x}_j) +\sigma_t(\tilde{x}_j) \right)}.
\end{equation}
The modified Lode angle $\hat{\theta}(\rho_j^m)$ in (\ref{eq:alpha}) is defined as
\begin{equation}\label{eq:theta_hat}
\hat{\theta}(\rho_j^m) = \frac{1}{3} \sin^{-1} \left(\sin(3\theta(\rho_j^m))\right) + \frac{\pi}{6},
\end{equation}
with $\theta(\rho_j^m)$ being the Lode angle \cite{Lode1926} defined as
\begin{equation}
\theta(\rho_j^m) = \frac{1}{3} \sin^{-1} \left( -\frac{3 \sqrt{3}}{2} \frac{J_3(\rho_j^m)}{J_2(\rho_j^m)^{3/2}} \right),
\end{equation}
where the third invariant of the diviatoric stress is introduced. The defineition of the term $G(I_1(\rho_j^m),\tilde{x}_j)$ in (\ref{eq:sigmaEQ}) depends on the yields surface. Here the Willam-Warnke yield surface is used but definitions for other yield surfaces can be found in \cite{Giraldo-Londono2020}. The definition of $G(I_1(\rho_j^m),\tilde{x}_j)$ is
\begin{equation}\label{eq:G}
G(I_1(\rho_j^m),\tilde{x}_j) = {\beta}(\tilde{x}_j) I_1(\rho_j^m),
\end{equation}
where the parameter $\beta(\tilde{x}_j)$ is defined as
\begin{equation}
\beta(\tilde{x}_j) = \frac{\sigma_b(\tilde{x}_j)-\sigma_t(\tilde{x}_j)}{3 \sigma_b(\tilde{x}_j) \sigma_t(\tilde{x}_j)}.
\end{equation}
The three invariants are defined as
\begin{equation}
I_1 = \mathrm{tr}(\boldsymbol{\sigma}), \quad J_2 = \frac{1}{2}\mathrm{tr}(\boldsymbol{s} \cdot \boldsymbol{s}), \quad J_3 = \det(\boldsymbol{s}),
\end{equation}
with the Cauchy stress calculated using the HS upper bound for interpolation defined in \eqref{eq:stiffnessInterpolation}. The diviatoric stress tensor is defined as
\begin{equation}
\boldsymbol{s} = \boldsymbol{\sigma} - \frac{I_1}{3}\boldsymbol{I}.
\end{equation}

%% file: APP_Sensitivity.tex
The optimization problem \eqref{eq:optProblem} is solved by updating the design iteratively using the Method of Moving Asymptotes (MMA) \cite{Svanberg1987} based on the gradients of the objective and constraints. The sensitivity of the objective is determined assuming a distinct eigenvalue, $\gamma_i$  and that the eigenvector $\boldsymbol{\varphi}_i$ is $\boldsymbol{K}$-normalized, the sensitivity of the eigenvalue $\gamma_i$ with respect to $\rho^m_j$ is defined as  
\begin{equation}\label{eq:Sensitivity:mu}
	\dfrac{\partial \gamma_i \left( \boldsymbol{\rho}^m \right)  }{\partial \rho^m_j} = 
	\left( \boldsymbol{\varphi}^j_i\right) ^T \left(
	\dfrac{\partial \boldsymbol{G}^j_\sigma\left( \boldsymbol{\rho}^m \right) }{\partial \rho^m_j}
	- \gamma_i \dfrac{\partial \boldsymbol{K}^j\left( \boldsymbol{\rho}^m\right) }{\partial \rho^m_j}
	\right)
	\boldsymbol{\varphi}^j_i
	-\left(  \boldsymbol{v}^j\right) ^T \dfrac{\partial \boldsymbol{K}^j\left( \boldsymbol{\rho}^m\right)}{\partial \rho^m_j} \boldsymbol{u}_j,
\end{equation}
where $\boldsymbol{v}$ is the adjoint vector which is determined solving
\begin{equation}
	\boldsymbol{K}\left( \boldsymbol{\rho}^m\right)\boldsymbol{v} = \sum_j\left( \boldsymbol{\varphi}^j_i\right)^T 
	\dfrac{\partial \boldsymbol{G}^j_\sigma\left( \boldsymbol{\rho}^m\right)}{\partial \boldsymbol{u}_j}
	\boldsymbol{\varphi}^j_i.
\end{equation}
The sensitivity of the KS function with respect to $\rho_j^m$ is 
\begin{equation} \label{eq:Sensitivity:KS}
\dfrac{\partial g_{\lambda}}{\partial \rho_j^m} =
\dfrac{\partial J^{KS}}{\partial \rho_j^m} = 
\dfrac{\partial \gamma_1}{\partial \rho_e^m}
\left(1+ \dfrac{1}{P}
\ln\left(
\sum\limits_{i\in \mathcal{B}} e^{P\left(\dfrac{\gamma_i(\boldsymbol{\rho}^m)}{\gamma_1(\boldsymbol{\rho}^m)}-1\right)} \right) \right)
+
\left(\dfrac{\sum \limits_{i\in \mathcal{B}}
\left(e^{P\left(\dfrac{\gamma_i(\boldsymbol{\rho}^m)}{\gamma_1(\boldsymbol{\rho}^m)}-1\right)}
\left(\dfrac{\partial \gamma_i}{\partial \rho_e^m} \gamma_1(\boldsymbol{\rho}^m)-
\dfrac{\partial \gamma_1}{\partial \rho_e^m} \gamma_i(\boldsymbol{\rho}^m)\right)\right)}
{ \sum \limits_{i\in \mathcal{B}}
\left(\gamma_1(\boldsymbol{\rho}^m)*e^{P*\left(\dfrac{\gamma_i(\boldsymbol{\rho}^m)}{\gamma_1(\boldsymbol{\rho}^m)}-1\right)}\right)}\right),
\end{equation}
The sensitivities of the objective $g_{\lambda}$ with respect to ${x}_j$ and ${s}_j$ are determined using the chain rule
\begin{equation}\label{eq:objectiveChainRule}
	\dfrac{\partial g_{\lambda}}{\partial x_i} = 
	\sum\limits_{i\in N_e} \left(
	\dfrac{\partial g_{\lambda}}{\partial \rho_i^m}
	\frac{\partial \rho_i^m}{\partial \tilde{x}_i} 
	\frac{\partial \tilde{x}_i}{\partial x_j}
	\right), 
	\quad
	\dfrac{\partial g_{\lambda}}{\partial s_i} = 
	\sum\limits_{i\in N_e} \left(
	\dfrac{\partial g_{\lambda}}{\partial \rho_i^m}
	\frac{\partial \rho_i^m}{\partial \bar{\tilde{s}}_i^m}
	\frac{\partial \bar{\tilde{s}}_i^m}{\partial \tilde{s}^m_i}
	\frac{\partial \tilde{s}^m_i}{\partial s_j} 
	\right).
\end{equation}

The sensitivities of $g_c$ with respect to $\rho_j^m$ is calculated by
\begin{equation}
	\dfrac{\partial g_c}{\partial \rho_j^m} = 
	\dfrac{1}{C_e^*} \left( \boldsymbol{u}_j\right) ^T \dfrac{\partial \boldsymbol{K}^j\left( \boldsymbol{\rho}^e\right)}{\partial \rho^e_j} \boldsymbol{u}_j,
\end{equation}
and equivalent to \eqref{eq:objectiveChainRule} the chain rule is applied to get the sensitivity with respect to ${x}_j$ and ${s}_j$.

The gradients of the buckling stress constraint in \eqref{eq:StressConstraint} with respect to ${x}_j$ and ${s}_j$ are also determined using adjoint sensitivity
\begin{align}
	\frac{\partial g_s}{\partial x_j} &= \sum\limits_{i\in N_e} \left(
	\frac{1}{\tilde{\sigma}_{PN}^0}
	\left[
	c \left( 
	-\frac{1}{J^{KS}} 
	\sigma_{sum}^{\frac{1}{p}-1} \sigma_{eq,i}^{p-1} \frac{\partial \sigma_{eq,i}}{\partial \tilde{x}_i}
	+\frac{1}{(J^{KS})^2}
	\frac{\partial J^{KS}}{\partial \rho_i^m}
	\frac{\partial \rho_i^m}{\partial \tilde{x}_i} \sigma_{PN}
	\right) + 
	\boldsymbol{v}_i^{\sigma T} 
	\left(
	\frac{\partial \kappa}{\partial \rho_i^m} \boldsymbol{K}_{\kappa 0} +
	\frac{\partial \mu}{\partial \rho_i^m} \boldsymbol{K}_{\mu 0}
	\right)
	\boldsymbol{u}_i
	\frac{\partial \rho_i^m}{\partial \tilde{x}_i} 
	\right]
	\frac{\partial \tilde{x}_i}{\partial x_j}
	\right),
	\label{eq:dgsdx}
	\\
	\frac{\partial g_s}{\partial s_j} &= \sum\limits_{i\in N_e} \left(
	\frac{1}{\tilde{\sigma}_{PN}^0}
	\left[
	c \left( 
	-\frac{1}{J^{KS}} 
	\sigma_{sum}^{\frac{1}{p}-1} \sigma_{eq,i}^{p-1} \frac{\partial \sigma_{eq,i}}{\partial \bar{\tilde{s}}_i^m} 
	+ \frac{1}{(J^{KS})^2}
	\frac{\partial J^{KS}}{\partial \rho_i^m}
	\frac{\partial \rho_i^m}{\partial \bar{\tilde{s}}_i^m} \sigma_{PN}
	\right) +
	\boldsymbol{v}_i^{\sigma T} 
	\left( \frac{\partial \kappa}{\partial \rho_i^m} \boldsymbol{K}_{\kappa 0} +
	\frac{\partial \mu}{\partial \rho_i^m} \boldsymbol{K}_{\mu 0}
	\right)
	\boldsymbol{u}_i \frac{\partial \rho_i^m}{\partial \bar{\tilde{s}}_i^m} \right]
	\frac{\partial \bar{\tilde{s}}_i^m}{\partial \tilde{s}^m_i}
	\frac{\partial \tilde{s}^m_i}{\partial s_j} 
	\right), 
	\label{eq:dgsds}
\end{align}
where $\sigma_{sum}$ is the sum in the \textit{p}-norm such that $\sigma_{PN} = \sigma_{sum}^{1/p}$ and $\boldsymbol{v}_i^{\sigma}$ is the adjoint vector. The sensitivity of $J^{KS}$ is defined by \eqref{eq:Sensitivity:KS} and the sensitivities of $\sigma_{eq}$, $\kappa$, $\mu$, filters and projections are all presented in the following section. The element adjoint vector $\boldsymbol{v}_i^{\sigma}$ contains the degrees of freedom for each element $i$. The adjoint field is calculated by solving
\begin{equation}\label{eq:adjointProblem}
	\boldsymbol{K} \boldsymbol{v}^{\sigma} = -\frac{\partial g_s}{\partial \boldsymbol{u}}^T.
\end{equation}
where $\boldsymbol{K}$ is the global stiffness matrix. The derivative of $g_s$ with respect to $\boldsymbol{u}$ on element level is defined in the following section.

\subsection{Sensitivity of Stress Constraint Components}\label{app:ComponentSensitivity}
The sensitivities of $\sigma_{eq}$ with respect to $\tilde{x}_j$ and $\bar{\tilde{s}}_j^m$ used in (\ref{eq:dgsdx}) and (\ref{eq:dgsds}) are derived from (\ref{eq:sigmaEQ})
\begin{align}
\frac{\partial \sigma_{eq}}{\partial \tilde{x}_j} &= 
\frac{\partial \alpha}{\partial \tilde{x}_j} \sqrt{3 J_2} +
\frac{\alpha \sqrt{3}}{2 \sqrt{J_2}} \frac{\partial J_2}{\partial \rho_j^m} \frac{\partial \rho_j^m}{\partial \tilde{x}_j} +
\frac{\partial \beta}{\partial \tilde{x}_j} I_1 +
\beta \frac{\partial I_1}{\partial \rho_j^m} \frac{\partial \rho_j^m}{\partial \tilde{x}_j}, \label{eq:dSigmaEQdx}
\\
\frac{\partial \sigma_{eq,i}}{\partial \bar{\tilde{s}}_j^m} &= 
\frac{\partial \alpha}{\partial \bar{\tilde{s}}_j^m} \sqrt{3 J_2} +
\frac{\alpha \sqrt{3}}{2 \sqrt{J_2}} \frac{\partial J_2}{\partial \rho_j^m} \frac{\partial \rho_j^m}{\partial \bar{\tilde{s}}_j^m} +
\beta \frac{\partial I_1}{\partial \rho_j^m} \frac{\partial \rho_j^m}{\partial \bar{\tilde{s}}_j^m},
\label{eq:dSigmaEQds}
\end{align}
The gradients of $\alpha$ are defined as
\begin{align}
\begin{split}
\frac{\partial \alpha}{\partial \tilde{x}_j} = & 
\frac{\frac{\partial A}{\partial \tilde{x}_j} \cos^2(\hat{\theta}) 
- 2 A \cos(\hat{\theta}) \sin(\hat{\theta}) \frac{\partial \hat{\theta}}{\partial \rho_j^m} \frac{\partial \rho_j^m}{\partial \tilde{x}_j}
+ \frac{\partial B}{\partial \tilde{x}_j}}{C \cos(\hat{\theta}) 
+ \sqrt{D \cos^2(\hat{\theta}) +E}}
- 
\frac{A \cos^2(\hat{\theta}) + B}{\left( C \cos(\hat{\theta}) 
+ \sqrt{D \cos^2(\hat{\theta}) + E} \right)^2} \dots
\\ &
\left(
\frac{\partial C}{\partial \tilde{x}_j} \cos(\hat{\theta}) 
- C \sin(\hat{\theta}) \frac{\partial \hat{\theta}}{\partial \rho_j^m} \frac{\partial \rho_j^m}{\partial \tilde{x}_j}
+ \frac{\frac{\partial D}{\partial \tilde{x}_j} \cos^2(\hat{\theta}) 
- 2 D \cos(\hat{\theta}) \sin(\hat{\theta}) \frac{\partial \hat{\theta}}{\partial \rho_j^m} \frac{\partial \rho_j^m}{\partial \tilde{x}_j} 
+ \frac{\partial E}{\partial \tilde{x}_j}}{2 \sqrt{D \cos^2(\hat{\theta}) + E}}
\right),
\end{split} \label{eq:DalpahaDx}
\\
\begin{split}
\frac{\partial \alpha}{\partial \bar{\tilde{s}}_j^m} = & 
\frac{- 2 A \cos(\hat{\theta}) \sin(\hat{\theta}) \frac{\partial \hat{\theta}}{\partial \rho_j^m} \frac{\partial \rho_j^m}{\partial \bar{\tilde{s}}_j^m}}{C \cos(\hat{\theta}) 
+ \sqrt{D \cos^2(\hat{\theta}) +E}}
- 
\frac{A \cos^2(\hat{\theta}) + B}{\left( C \cos(\hat{\theta}) 
+ \sqrt{D \cos^2(\hat{\theta}) + E} \right)^2} \dots
\\ &
\left(
- C \sin(\hat{\theta}) \frac{\partial \hat{\theta}}{\partial \rho_j^m} \frac{\partial \rho_j^m}{\partial \bar{\tilde{s}}_j^m}
+ \frac{- 2 D \cos(\hat{\theta}) \sin(\hat{\theta}) \frac{\partial \hat{\theta}}{\partial \rho_j^m} \frac{\partial \rho_j^m}{\partial \bar{\tilde{s}}_j^m}}{2 \sqrt{D \cos^2(\hat{\theta}) + E}}
\right).
\end{split} \label{eq:DalpahaDs}
\end{align}
The derivatives of the coefficients $A$, $B$, $C$, $D$ and $E$ with respect to $\tilde{x}_j$ are derived from (\ref{eq:ABCDE}), i.e.
\begin{equation}
\begin{split}
\frac{\partial A}{\partial \tilde{x}_j} &= 
\frac{4}{15} \sqrt{30} 
\frac{2 \sigma_c r_c \frac{\partial r_c}{\partial \tilde{x}_j} 
- 2 \sigma_c r_t \frac{\partial r_t}{\partial \tilde{x}_j} 
- r_c^2 \frac{\partial \sigma_c}{\partial \tilde{x}_j} 
+ r_t^2 \frac{\partial \sigma_c}{\partial \tilde{x}_j}}{\sigma_c^2}
, \\
\frac{\partial B}{\partial \tilde{x}_j} &= 
\frac{\sqrt{30}}{15} 
\frac{-(r_c-2 r_t)
\left(
(r_c-2 r_t) \frac{\partial \sigma_c}{\partial \tilde{x}_j} - 2 \sigma_c
\left(
\frac{\partial r_c}{\partial \tilde{x}_j} - 2 \frac{\partial r_t}{\partial \tilde{x}_j}
\right)
\right)}{\sigma_c^2}
, \\
\frac{\partial C}{\partial \tilde{x}_j} &= 
6 r_c^2 \frac{\partial r_c}{\partial \tilde{x}_j}
- 2 r_t^2 \frac{\partial r_c}{\partial \tilde{x}_j}
- 4 r_c r_t \frac{\partial r_t}{\partial \tilde{x}_j}
, \\
\frac{\partial D}{\partial \tilde{x}_j} &= 
24 r_c (r_c -2 r_t)
\left(
\left(
r_c^3 -\frac{4}{3} r_c^2 r_t - \frac{2}{3} r_c r_t^2 + \frac{2}{3} r_t^3
\right)
\frac{\partial r_c}{\partial \tilde{x}_j}
- \frac{2}{3} r_c 
\left(
r_c^2 + \frac{r_c r_t}{2} - 2 r_t^2
\right)
\frac{\partial r_t}{\partial \tilde{x}_j}
\right)
, \\
\frac{\partial E}{\partial \tilde{x}_j} &= 
-4 r_c (r_c -2 r_t)
\left(
\left(
5 r_c^2 r_t - 11 r_c r_t^2 +5 r_t^3
\right)
\frac{\partial r_c}{\partial \tilde{x}_j}
+ r_c
\left(
r_c^2 - \frac{17}{2} r_c r_t +10r_t^2
\right)
\frac{\partial r_t}{\partial \tilde{x}_j}
\right).
\end{split}
\end{equation}
The derivative of the parameters $r_c$ and $r_t$ are
\begin{equation}
\begin{split}
\frac{\partial r_c}{\partial \tilde{x}_j} &= 
\frac{\sqrt{30}}{5}
\frac{-\sigma_b \sigma_t (\sigma_b - \sigma_t) \frac{\partial \sigma_c}{\partial \tilde{x}_j} 
+ \sigma_c
\left(
\frac{\partial \sigma_t}{\partial \tilde{x}_j} \sigma_b^2
- \sigma_t^2 \frac{\partial \sigma_b}{\partial \tilde{x}_j}
\right)}{
\left(
(\sigma_c + 3 \sigma_t) \sigma_b -\sigma_c \sigma_t
\right)^2}
, \\
\frac{\partial r_t}{\partial \tilde{x}_j} &= 
\frac{\sqrt{30}}{5}
\frac{\sigma_c \sigma_t^2 \frac{\partial \sigma_b}{\partial \tilde{x}_j} 
+ 2 \sigma_c \sigma_b^2 \frac{\partial \sigma_t}{\partial \tilde{x}_j} 
- \sigma_t^2 \sigma_b \frac{\partial \sigma_c}{\partial \tilde{x}_j}
- 2 \sigma_t \sigma_b^2 \frac{\partial \sigma_c}{\partial \tilde{x}_j}}{\sigma_c^2 (2 \sigma_b + \sigma_t)^2}.
\end{split}
\end{equation}
The gradient of $\beta$ used in (\ref{eq:dSigmaEQdx}) is
\begin{equation}
\frac{\partial \beta}{\partial \tilde{x}_j} =
\frac{\frac{\partial \sigma_b}{\partial \tilde{x}_j} \sigma_t^2 
- \sigma_b^2 \frac{\partial \sigma_t}{\partial \tilde{x}_j}}{3 \sigma_b^2 \sigma_t^2}
\end{equation}
The gradients of the three stress limit fits are derived from (\ref{eq:FitEquation}) as
\begin{equation}
\frac{\partial \bar{\sigma}_k}{\partial \tilde{x}_j} = 
E_0 \left(n_0 b_{0,k} \tilde{x}_j^{n_0-1} + (n_0+1) b_{1,k} \tilde{x}_j^{n_0} \right)
, \quad k \in \{c,t,b\}.
\end{equation}
Furthermore, the gradient of the relaxed stress limits are defined as
\begin{equation}
\frac{\partial \sigma_k}{\partial \tilde{x}_j} = 
\frac{\partial \bar{\sigma}_k}{\partial \tilde{x}_j} +
\psi \dfrac{\partial \mathcal{H}(\tilde{x}_j,\eta,\beta)}{\partial \tilde{x}_j}
, \quad k \in \{c,t,b\}.
\end{equation}
The gradient of the modified Lode angle $\hat{\theta}$ used in (\ref{eq:DalpahaDx}) and (\ref{eq:DalpahaDs}) is derived from (\ref{eq:theta_hat}) as
\begin{equation}
\frac{\partial \hat{\theta}}{\partial \rho_j^m} = 
\frac{\cos(3 \theta)}{\sqrt{-\sin^2(3 \theta)+1}} \frac{\partial \theta}{\partial \rho_j^m},
\end{equation}
where the derviative of the Lode angle $\theta$ is
\begin{equation}
\frac{\partial \theta}{\partial \rho_j^m} = 
\frac{-2 
}{3 \sqrt{-27 \frac{J_3^2}{J_2^3} + 4}} 
\left(
\frac{3 \sqrt{3}}{2 J_2^{3/2}} \frac{\partial J_3}{\partial \rho_j^m}
- \frac{9 \sqrt{3} J_3}{4 J_2^{5/2}} \frac{\partial J_2}{\partial \rho_j^m}
\right).
\end{equation}
The gradients of the stress inveriants are defined as
\begin{equation}
\frac{\partial I_1}{\partial \rho_j^m} = \frac{\partial I_1}{\partial \boldsymbol{\sigma}} \frac{\partial \boldsymbol{\sigma}}{\partial \rho_j^m}
= \boldsymbol{I} \frac{\partial\boldsymbol{\sigma}}{\partial \rho_j^m} 
, \quad
\frac{\partial J_2}{\partial \rho_j^m} = \frac{\partial J_2}{\partial \boldsymbol{\sigma}} \frac{\partial \boldsymbol{\sigma}}{\partial \rho_j^m}
= \boldsymbol{s} \frac{\partial\boldsymbol{\sigma}}{\partial \rho_j^m}
, \quad
\frac{\partial J_3}{\partial \rho_j^m} = \frac{\partial J_3}{\partial \boldsymbol{\sigma}} \frac{\partial \boldsymbol{\sigma}}{\partial \rho_j^m}
= \left(\boldsymbol{s} \cdot \boldsymbol{s} - \frac{2 J_2}{3} \boldsymbol{I} \right) \frac{\partial\boldsymbol{\sigma}}{\partial \rho_j^m},
\end{equation}
with the gradient of $\sigma$ being determined by the gradients of the HS interpolation in \eqref{eq:stiffnessInterpolation}. 

The derivative of $g_s$ with respect to $\boldsymbol{u}_j$ used on the right hand side of \eqref{eq:adjointProblem} is calculated on element level as 
\begin{equation}
\frac{\partial g_s}{\partial \boldsymbol{u}_j}^T = - \dfrac{c \sigma_{sum}^{\frac{1}{p}-1} \sigma_{eq}^{p-1}}{J^{KS}} \dfrac{\partial \sigma_{eq}}{\partial \boldsymbol{u}_j},
\end{equation}
with the gradient of $\sigma_{eq}$ being
\begin{equation}
\frac{\partial \sigma_{eq}}{\partial \boldsymbol{u}_j}  = 
\frac{\partial \alpha}{\partial \boldsymbol{u}_j} \sqrt{3 J_2} + \alpha \frac{\sqrt{3}}{2\sqrt{J_2}} \frac{\partial J_2}{\partial \boldsymbol{u}_j}
+ \beta \frac{\partial I_1}{\partial \boldsymbol{u}_j}.
\end{equation}
The gradient of $\alpha$ with respect to $\boldsymbol{u}_j$ is
\begin{equation}
\frac{\partial \alpha}{\partial \boldsymbol{u}_j} = 
\left(
\frac{-2 A \cos(\hat{\theta}) \sin(\hat{\theta})  }{C \cos(\hat{\theta}) + \sqrt{D \cos^2(\hat{\theta}) + E}}
-  \frac{A \cos^2(\hat{\theta}) + B}{
\left(
C \cos(\hat{\theta}) + \sqrt{D \cos^2(\hat{\theta}) + E}
\right)^2} 
\left(
-C \sin(\hat{\theta}) 
- \frac{D \cos(\hat{\theta}) \sin(\hat{\theta})}{\sqrt{D \cos^2(\hat{\theta}) + E}} 
\right)\right) \frac{\partial \hat{\theta}}{\partial \boldsymbol{u}_j},
\end{equation}
where the gradient of $\hat{\theta}$ is
\begin{equation}
\frac{\partial \hat{\theta}}{\partial \boldsymbol{u}_j} = 
\frac{\cos(3 \theta)}{\sqrt{-\sin^2(\theta) +1}} 
\frac{\partial {\theta}}{\partial \boldsymbol{u}_j}.
\end{equation}
The gradient of $\theta$ is
\begin{equation}
\frac{\partial {\theta}}{\partial \boldsymbol{u}_j} = 
\frac{-2 
\left(
\frac{3 \sqrt{3}}{2 J_2^{3/2}} \frac{\partial J_3}{\partial \boldsymbol{u}_j}
- \frac{9 \sqrt{3} J_3}{4 J_2^{5/2}} \frac{\partial J_2}{\partial \boldsymbol{u}_j}
\right)
}{3 \sqrt{-27 \frac{J_3^2}{J_2^3} +4}}.
\end{equation}
Finally, the gradients of $I_1$, $J_2$ and $J_3$ are defined as
\begin{equation}
\frac{\partial I_1}{\partial \boldsymbol{u}_j} = \frac{\partial I_1}{\partial \boldsymbol{\sigma}} \frac{\partial \boldsymbol{\sigma}}{\partial \boldsymbol{u}_j}
= \boldsymbol{I} \frac{\partial\boldsymbol{\sigma}}{\partial \boldsymbol{u}_j} 
, \quad
\frac{\partial J_2}{\partial \boldsymbol{u}_j} = \frac{\partial J_2}{\partial \boldsymbol{\sigma}} \frac{\partial \boldsymbol{\sigma}}{\partial \boldsymbol{u}_j}
= \boldsymbol{s} \frac{\partial\boldsymbol{\sigma}}{\partial \boldsymbol{u}_j}
, \quad
\frac{\partial J_3}{\partial \boldsymbol{u}_j} = \frac{\partial J_3}{\partial \boldsymbol{\sigma}} \frac{\partial \boldsymbol{\sigma}}{\partial \boldsymbol{u}_j}
= \left(\boldsymbol{s} \cdot \boldsymbol{s} - \frac{2 J_2}{3} \boldsymbol{I} \right) \frac{\partial\boldsymbol{\sigma}}{\partial \boldsymbol{u}_j},
\end{equation}
where the derivative of the Cauchy stress is
\begin{equation}
\frac{\partial\boldsymbol{\sigma}}{\partial \boldsymbol{u}_j} = 
\left(
\kappa \boldsymbol{C}_{0,\kappa} + \mu \boldsymbol{C}_{0,\mu}
\right) \boldsymbol{B}_0.
\end{equation}